\documentclass[12t,a4paper]{article}
\usepackage[margin=2.5cm]{geometry}
\usepackage{titlesec}
\usepackage{textcomp}
\usepackage{chngcntr}
\usepackage{parskip}
\usepackage{wrapfig}
\usepackage{geometry}
\usepackage{multicol}
\usepackage{adjustbox}
\usepackage[round]{natbib}
\usepackage{float}
\usepackage[utf8]{inputenc}
\usepackage[T1]{fontenc} 					
\usepackage{graphicx}
\usepackage{caption}
\usepackage{subcaption}
\usepackage{graphicx}       					
\usepackage{amsmath,amssymb} 
\newcommand{\bs}[1]{\boldsymbol{#1}}
\newcommand{\mc}[1]{\mathcal{#1}}
\renewcommand{\S}{\hat{\mathbb{S}}}

\usepackage{color}
\usepackage{ulem}
\usepackage{subcaption}

\newcommand{\Shat}[1]{\hat{\mathbb{S}}_{sy}^{#1}(t)}
\usepackage{authblk}
\usepackage{caption}
\usepackage{hyperref}
\usepackage{subcaption}

\title{Probabilistic prediction of the time to hard freeze using seasonal weather forecasts and survival time methods}

\author[1]{Thea Roksv{\aa}g \footnote{roksvag@nr.no}}
\author[1]{Alex Lenkoski}
\author[1]{Michael Scheuerer}
\author[1]{Claudio Heinrich-Mertsching}
\author[1]{Thordis L. Thorarinsdottir}

\affil[1]{Norwegian Computing Center, Postboks 114, 0314 Blindern, Oslo, Norway}

\begin{document}

\maketitle

\begin{abstract}
\noindent
Agricultural food production and natural ecological systems depend on a range of seasonal climate indicators that describe seasonal patterns in climatological conditions. This paper proposes a probabilistic forecasting framework for predicting the end of the freeze-free season, or the time to a mean daily near-surface air temperature below 0 $^\circ$C (here referred to as hard freeze). The forecasting framework is based on the multi-model seasonal forecast ensemble provided by the Copernicus Climate Data Store and uses techniques from survival analysis for time-to-event data. The original mean daily temperature forecasts are statistically post-processed with a mean and variance correction of each model system before the time-to-event forecast is constructed. In a case study for a region in Fennoscandia covering Norway for the period 1993-2020, the proposed forecasts are found to outperform a climatology forecast from an observation-based data product at locations where the average predicted time to hard freeze is less than 40 days after the initialization date of the forecast on October 1.        
\end{abstract}

\section{Introduction}

Agricultural decisions such as planting, harvest timing and field fertilization, are in large steered by weather and climate conditions, including conditions on subseasonal to seasonal timescales. For surface air temperature, the relevant seasonal climate indicators, or agroclimatic indicators, are mainly related to heat accumulation and frost characteristics. The most common index for heat accumulation is usually referred to as growing degree days, while frost characteristics are described by the onset, end, and duration of the freeze-free season for various freezing thresholds \citep{Kukal&2018, Weltzin&2020}. Skillful predictions of these agroclimatic indicators are needed for both agricultural operations planning and for predicting crop growth and yield potential, especially in the light of worldwide observed changes in the indicators over time due to climate change \citep[e.g.][]{Kunkel&2004, Schwartz&2006, Liu&2008, Zhang&2014}.   

Seasonal forecast users in the agricultural sector have criticized available forecasting products for a lack of forecast skill and usability. In particular, they have requested derived forecast products that directly predict relevant agroclimatic indicators, information on uncertainty and comparisons to previous years, see \cite{KlemmMcPherson2017} and references therein. This contrasts current practice in subseasonal and seasonal forecasting to aggregate variables such as temperature and precipitation over months, seasons and/or spatial regions in order to achieve skill that outperforms climatological reference forecasts \citep[e.g.][]{hemri_et_2020, vanStraaten&2020}. 

Seasonal predictions of agroclimatic indicators are currently lacking and available seasonal information is commonly in the form of local or regional climatologcial information, see e.g. \cite{Angel&2017}. Similarly, approaches to predict agricultural outcomes on the seasonal scale, e.g. agricultural yield prediction, are  not based on seasonal predictions of relevant agroclimatic indicators. Rather, predictions are often based on seasonal predictions for large-scale climate drivers such as the El Ni{\~n}o Southern Oscillation \citep{Hammer&1996, An-Vo&2019, Lehmann&2020}. While such approaches work well in regions where the seasonal climate is heavily influenced by teleconnections, they may not generalize to other regions. In a perspective, \cite{FischerConnor2018} identify improved seasonal weather forecasts as a potential key factor for crop agronomy developments in the coming decades.  

In this paper, we investigate the potential of seasonal forecasts to skillfully predict an agroclimatic indicator for the end of the freeze-free season. Specifically, we aim to predict the time to the next occurrence of hard freeze from the initialization date of the forecast, where hard freeze refers to a mean daily near-surface air temperature below 0$^\circ$C. Using methods from survival analysis, a branch of statistics for analyzing the expected duration of time until an event occurs \citep{kalbfleisch_ross_2011}, we construct a probabilistic time-to-event forecast for the time to hard freeze based on a multi-model ensemble of seasonal temperature forecasts available from the Copernicus Climate Data Store (CDS). As suggested by e.g. \cite{hemri_et_2020} in the context of seasonal mean forecasts, the daily temperature forecasts are statistically post-processed to remove systematic biases and errors in the ensemble spread before they are processed further to construct the time-to-event forecast. The proposed methodology is applied to seasonal temperature forecasts for a region in Fennoscandia covering Norway for the period 1993-2020, where forecasts initialized on October 1 are used to predict time to hard freeze in the period October 1 to December 31.

While survival time analysis is a broad statistical field, such methods are rarely used in meteorology. We believe that this approach has great potential for use in a wide range of applications, including the modeling of the onset of the rainy season, or the time to the next drought. The current study forms an entry point showing how this theory can be combined with numerical weather predictions to provide probabilistic forecasts for the time to a specific event.

The remainder of the paper is organized as follows. The raw seasonal forecasts and the observation-based gridded data product used in the analysis are described in Section~\ref{sec:data_products}. The proposed statistical post-processing and survival analysis methods are described in Section~\ref{sec:methods}, followed by a description of forecast evaluation approaches in Section~\ref{sec:evaluation}. The results of the time to freeze forecasting are presented in Section~\ref{sec:results}, with conclusions and a discussion given in Section~\ref{sec:discussion}.

\section{Data products}\label{sec:data_products}
The time to frost forecasts proposed in this article are constructed based on seasonal subdaily near-surface air temperature forecasts from the Copernicus Climate Data Store (CDS). CDS provides a selection of seasonal forecast products from different weather centers around the globe\footnote{\url{ https://confluence.ecmwf.int/display/CKB/C3S+Seasonal+Forecasts\%3A+datasets+documentation}}. 
The seasonal forecasts are generated by predicting the change in slow-varying components of the earth system, such as the ocean temperature. Based on the current state of the system and the directed movements of large ocean currents, it is possible to predict both ocean and atmospheric temperature several weeks or months ahead using numerical weather prediction models \citep[NWPs;][]{Robertson&2018}. The resulting forecasts are given as an ensemble, where each ensemble member represents one possible future weather scenario. As there are complex, non-linear interactions at play, forecasts at long lead times are associated with large uncertainty.
At the typical seasonal lead times of one to three months, skillful forecasts can often only be obtained after considering monthly or weekly averages, often additionally averaged over large spatial regions \citep{vanStraaten&2020}. 

\begin{figure}
	\centering
	\begin{subfigure}[b]{0.32\textwidth}
		\includegraphics[width=1\linewidth]{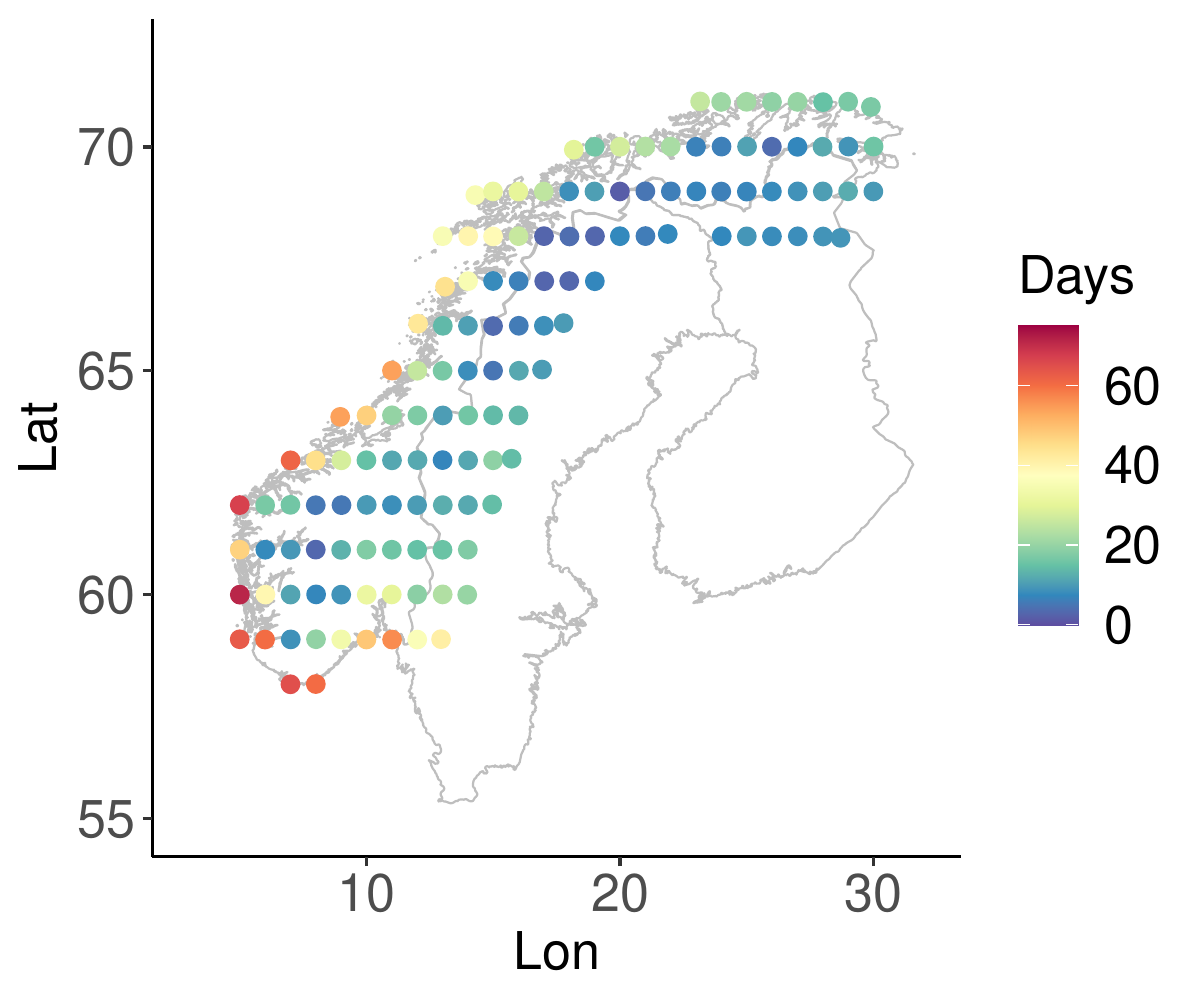}
		\caption{Mean days to hard freeze.}
		\label{fig:meanmap}
	\end{subfigure}
	\begin{subfigure}[b]{0.32\textwidth}
		\includegraphics[width=1\linewidth]{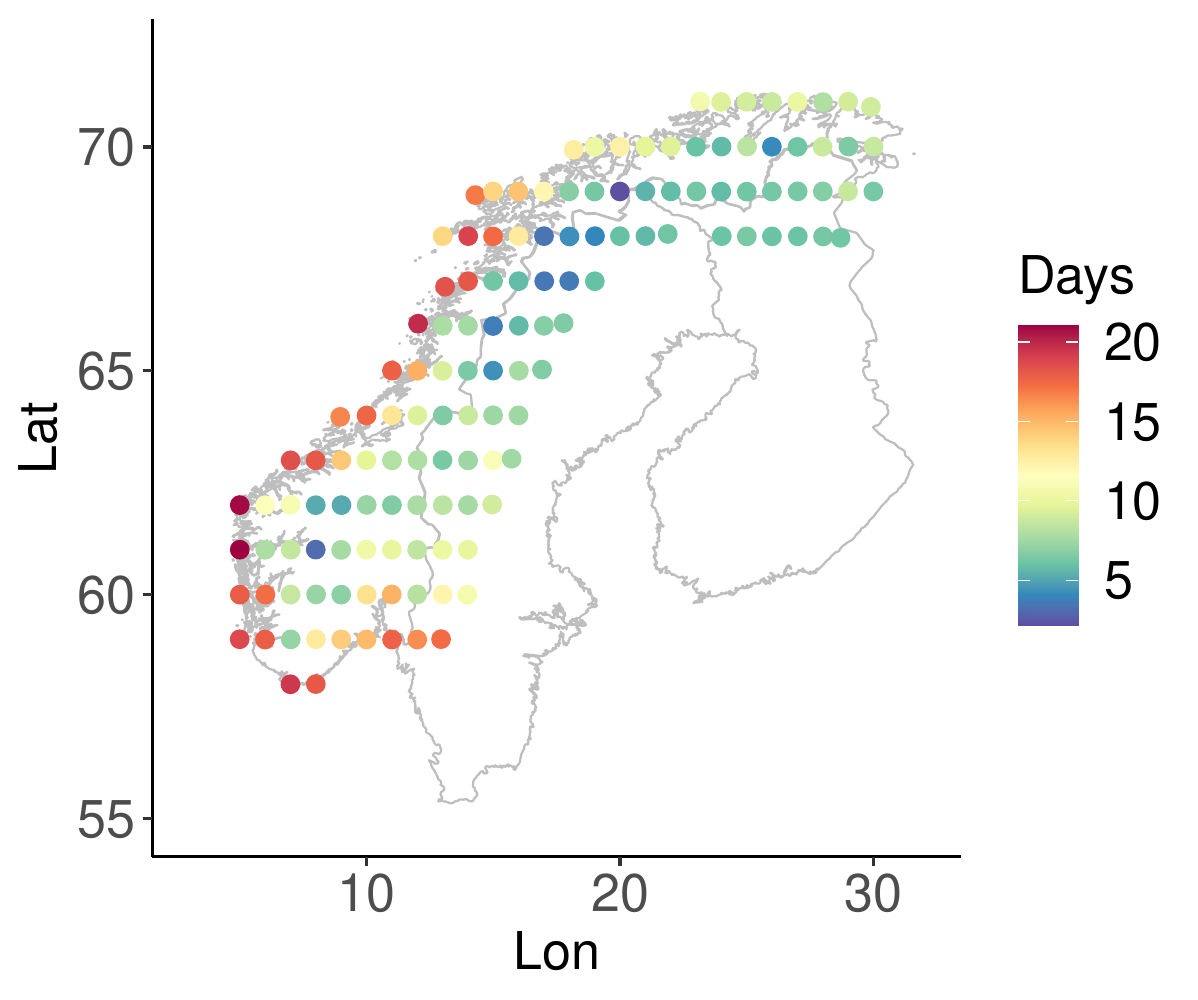}
		\caption{Standard deviation. }
		\label{fig:sdmap}
	\end{subfigure}
	\begin{subfigure}[b]{0.32\textwidth}
		\includegraphics[width=1\linewidth]{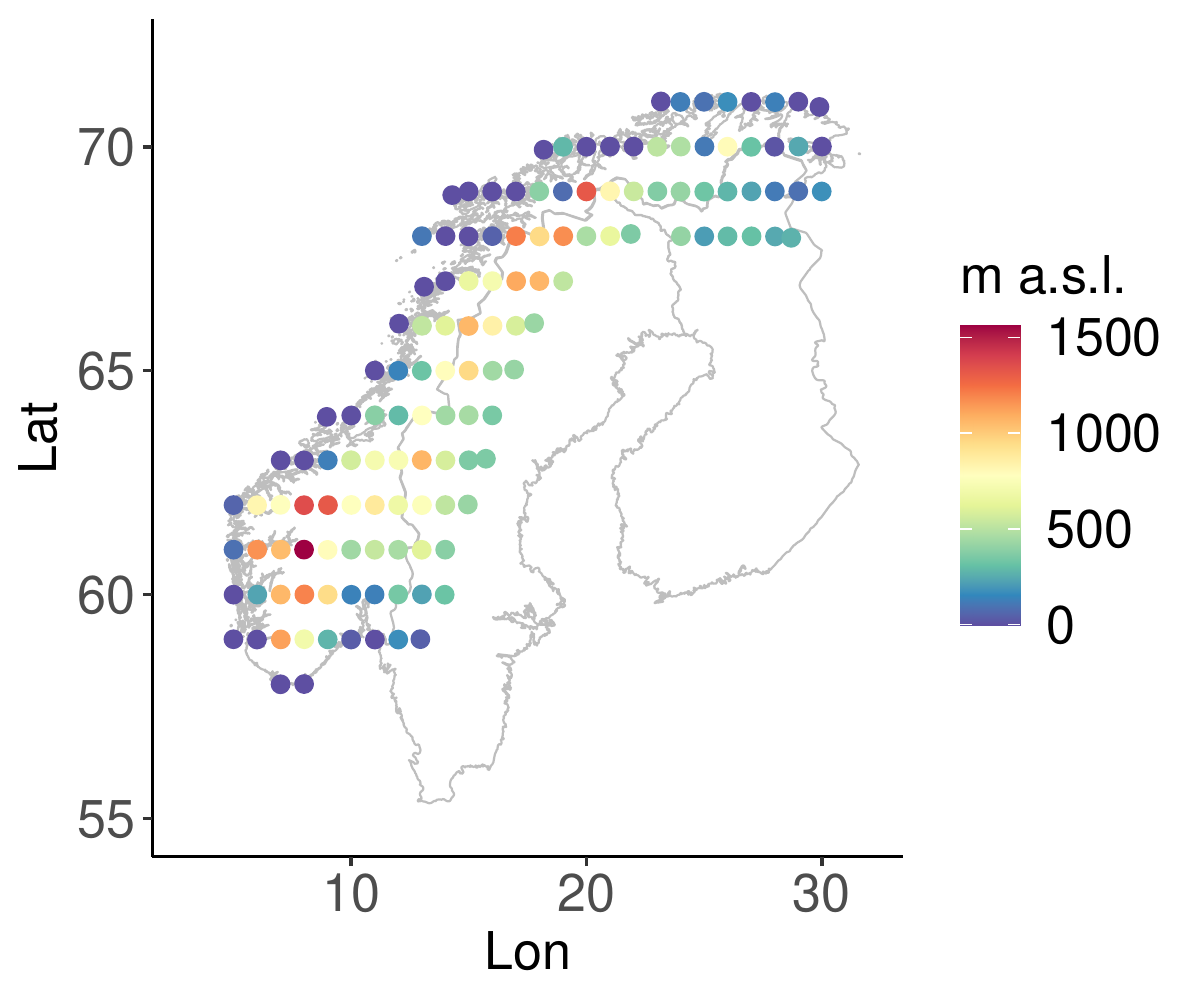}
		\caption{seNorge elevation. \\  }
		\label{fig:seNorge_elev}
	\end{subfigure}
	\caption{\small Mean number of days to hard freeze, from October 1st, for 1993-2020 for the seNorge product (left), the corresponding standard deviation (center) and seNorge's model elevation in meters above sea level (right). The study locations are in Norway, Sweden, Finland and Russia, on an approximate $1\times 1^\circ$ grid. }
	\label{fig:seNorge}
\end{figure}

We derive probabilistic forecasts for the time to hard freeze from NWP ensemble predictions with initialization date of October 1. October 1st was chosen since the first hard freeze typically occurs 0-3 months after this date within our study region in Fennoscandia, see Figure \ref{fig:meanmap}. On the CDS, hindcasts with initialization date of October 1 were, at the time we downloaded the data, available for four NWP systems for a total of approximately 100 ensemble members. The four centres that produce these forecasts are the European Centre for Medium-Range Weather Forecasts (ECMWF), the Euro-Mediterranean Center for Climate Change (CMCC), Météo France and the UK Met Office. For 1993-2016 and 2020, hindcasts with 6 hour temporal resolution and lead times from 0 to 5-7 months were available for all four systems. For 2017-2019, only hindcasts from ECMWF were available. See Table \ref{tab:dataprod} for further information on each NWP system and the data availability. 

As Table \ref{tab:dataprod} shows, the NWP forecasts are all  delivered on a $1\times1^\circ$ grid, but the grid specification varies between the systems. While the ECMWF forecasts are given on a grid centered at full-degree longitudes and latitudes (1$^\circ$, 2$^\circ$,...), the remaining forecasts are on a grid shifted by half a degree (0.5$^\circ$, 1.5$^\circ$, 2.5$^\circ$,...). Following the practice on the CDS website for monthly forecasts, we use a nearest neighbor approach to place the half degree products on the nearest full degree. This is assumed to be a reasonable approximation for our application, as the forecasts for each system is post-processed to reduce system specific, systematic errors. Furthermore, we consider mean daily temperatures when we derive the time to hard freeze forecasts. These are obtained by averaging the 6h subdaily temperature predictions for each forecast valid date.

\begin{table}[h!]\caption{\small Information on the data products used in the analysis. The forecasts have initialization date of October 1st. The table shows the number of ensemble members the hindcasts for each of the systems have for most years. From 2017-2020 the ECMWF forecast has 51 ensemble members, while the CMCC and Météo France forecasts have 50 members in 2020. The UK Met Office forecast only has 2 ensemble members in 2020. The $Degrees$ row indicates whether the forecasts are given for whole longitudes and latitudes  (1$^\circ$, 2$^\circ$,...) or half longitudes and latitudes (0.5$^\circ$, 1.5$^\circ$, 2.5$^\circ$,...).}\label{tab:dataprod}
	\begin{tabular}{llllll}\hline 
		NWP system	& ECMWF                          & CMCC                        & Météo France               & UK Met Office                        & seNorge2018            \\ \hline
		Spatial resolution              & $ 1^\circ \times 1^\circ$    & $ 1^\circ \times 1^\circ$ & $ 1^\circ \times 1^\circ$ & $ 1^\circ \times 1^\circ$ & 1 km $\times$ 1 km \\
		Degrees (half or whole) & Whole                          & Half                        & Half                        & Half                        & $\times$           \\
		Ensemble members        & 25  & 40                          & 25                          & 7                           & $\times$           \\
		Missing years           & None                           & 2017-2019              & 2017-2019              & 2017-2019              & None              \\
		\hline
	\end{tabular}
\end{table}

To validate and calibrate the seasonal forecasts, we use mean daily temperatures from the data product seNorge2018 for 1993-2020. The seNorge product is a gridded observation-based dataset developed at the Norwegian Meteorological Institute \citep[MET Norway;][]{Lussana2019} and freely available at \url{ https://thredds.met.no/thredds/catalog/senorge/catalog.html}. It is derived from observations of temperature from 200-450 meteorological stations, depending on year. These are interpolated to a 1$\times$1 km grid covering the Norwegian mainland and parts of Sweden, Finland and Russia. The interpolation method is based on optimal interpolation \citep[e.g.][]{Gandin,Kalnay}, which is a data assimilation method where observations are combined with a background field. The background field is here an estimated temperature field based on how temperature varies with elevation, see \cite{Lussana2018,Lussana2019} for details. The seNorge mean daily temperature dataset, and other seNorge products, have been used in a variety of studies \citep[e.g.][]{Erlandsen,seNorgeApplic,HANDELAND2021214,Lawrence}.

For agricultural applications, a hard freeze forecast is most useful if it provides information targeted towards a specific farm or field. In our experiments, we therefore aim to construct forecasts that predict the temperature locally at selected study locations that we can think of as farm locations. To select a set of locations we performed the following: Each whole longitude and latitude pair in the NWP grid was matched with the closest seNorge grid cell. Only NWP grid cells that were closer to a seNorge grid cell than 10 km were included in the analysis. This resulted in the 135 study locations shown in Figure \ref{fig:meanmap}. For these, we computed the number of days to hard freeze from October 1 from the seNorge mean daily temperature data, where a day with hard freeze is defined here as a day with mean daily temperature below 0. The historical mean number of days to hard freeze for 1993-2020 is visualized in Figure  \ref{fig:meanmap}. The mean ranges from 1 to 72 days depending on distance from coast, elevation, longitude and latitude. In Figure \ref{fig:sdmap} the corresponding standard deviation is presented, while the seNorge model elevation is shown in Figure \ref{fig:seNorge_elev} for reference.

\section{Predicting the time to hard freeze}\label{sec:methods}

\subsection{Post-processing daily NWP output}\label{sec:postprocessing}
Numerical weather predictions are often subject to systematic biases. This is especially true for near-surface temperature forecasts over complex terrain, where differences between the model grid elevation and the true elevation (or, as in the present setting, the elevation associated with the seNorge product) entail systematic differences in the respective near-surface temperature climatologies \citep{DabernigEA2017, KellerEA2021}. Statistical post-processing can remove these biases and thus lead to substantial improvements in forecast skill, even as NWP models continue to become better and their horizontal resolutions increase \citep{HemriEA2014}. Methods proposed for the statistical post-processing of near-surface temperature forecasts range from basic, regression-based bias correction \citep{Glahn+Lowry1972} over distributional regression methods, which also adjust the representation of uncertainty  \citetext{e.g. \citealp{GneitingEA2005}}, to advanced machine learning techniques which permit the use of multiple predictors and/or non-linear predictor-predictand relationships \citep{MessnerEA2017,TaillardatEA2016,Rasp+Lerch2018,Roebber+Crockett2019}.

Several applications require that the post-processed forecasts are provided in the form of an ensemble.  This ensemble must not only be unbiased and represent the forecast uncertainty, but in addition it must also exhibit realistic co-variability across space and time. This is important for ensuring that predictions of spatially and/or temporally aggregated quantities such as areal means, minima/maxima over a certain time period, etc.,  also are calibrated. In the present application, correlations across forecast lead times are important since temperature forecasts which oscillate wildly over time can be expected to fall below a given temperature threshold sooner than a forecast that is more persistent over time, but has otherwise similar statistical characteristics. Unfortunately, most statistical post-processing techniques focus on the marginal forecast distributions and lose the multivariate information encompassed in the ensemble. Techniques to construct a full multivariate predictive distribution have been proposed in the literature (see \citet{LerchEA2020} for a recent overview), but require additional modeling efforts. Alternatively, one can use methods like member-by-member post-processing approaches \citep{VanSchaeybroeck+Vannitsem2015} which retain the rank correlation structure of the original ensemble forecasts. The post-processing approach used in the present study falls into this category as it adjust the members individually and leaves their temporal (and spatial) structure intact.

Let $\mc{S}$ denote the study locations in Figure \ref{fig:seNorge} and $s \in \mc{S}$ a particular location.
Further, let $f_{s,t,y}^{k,m}$ be the mean daily temperature forecasted for day $t$ after the forecast initialization date, for year $y$, NWP system $k$, ensemble member $m$ and location $s$. We post-process the daily near surface air temperature forecasts in two steps. For each NWP system $k$, the forecasts are first standardized with respect to a system specific mean and standard deviation. The standardization is given by
\begin{equation}\label{eq:anomaly}
	\tilde{f}_{s,t,y}^{k,m}=\frac{(f_{s,t,y}^{k,m}-\tilde{\mu}^k_{s,t})}{\tilde{\sigma}^k_{s,t}},
\end{equation}
where $\tilde{\mu}_{s,t}^k$ is the mean daily temperature forecasted by system $k$ for day $t$ and location $s$, calculated based on historical ensemble members from this particular NWP system. Similarly, $\tilde{\sigma}_{s,t}^k$ is the historical standard deviation for system $k$, day $t$ and location $s$. In practice, when calculating $\tilde{\mu}^k_{s,t}$ and $\tilde{\sigma}^k_{s,t}$, we leave out all members of the forecast corresponding to the current year $y$.

The next step is to re-standardize $\tilde{f}_{s,t,y}^{k,m}$ with respect to historical observations from the target locations, in this case the seNorge locations in Figure \ref{fig:seNorge}. This is done as follows:
\begin{equation}\label{eq:SFE}
    \hat{f}_{s,t,y}^{k,m}=\tilde{f}_{s,t,y}^{k,m} \cdot \hat{\sigma}_{s,t}  + \hat{\mu}_{s,t},
\end{equation}
where $\hat{\mu}_{s,t}$ is the mean observed daily temperature for location $s$ on day $t$, and $\hat{\sigma}_{s,t}$ is the corresponding observed standard deviation. When calculating $\hat{\mu}_{s,t}$ and $\hat{\sigma}_{s,t}$, we leave out data from the current year $y$ as these data are unobserved at the forecast time. The result of equations \eqref{eq:anomaly} and \eqref{eq:SFE} is a post-processed ensemble  $\hat{f}_{s,t,y}^{k,m}$ where the mean and standard deviation of the marginal distribution of each member matches that of the observation for each target day and each target location. 

This form of post-processing by recalibrating the model climatology is frequently used for seasonal weather forecasts \citep[e.g.][]{Weigel&2009, Woldemeskel&2018}, and is similar in flavor to regression-based post-processing techniques. While regression-based techniques provide slightly more flexibility, 
\citet{hemri_et_2020} shows that they often perform poorly in related tasks in seasonal forecasting.  This is a result of an increased risk of overfitting in the context of seasonal forecasts, where observed time series are short and the signal-to-noise ratio is small, see also \citet{VanSchaeybroeck&2018} and \citet{Heinrich&2021}.

\subsection{Time-to-event forecasts with survival analysis}\label{sec:km}
We construct time to hard freeze forecasts by using methods from survival time analysis. In this subsection we provide a concise review of survival time modeling.  This topic has a substantial literature associated with it and a number of texts deal with the study in detail, see e.g. \citet{kalbfleisch_ross_2011} for a comprehensive introduction to the subject. While the methodology has become established in a diverse number of fields, it has gained most prominence in the field of biostatistics, where it is used to study the efficacy of treatments for disease, among other matters.  A key aspect of the appeal of the survival time approach is that it can address a broad array of complications that arise when conducting real-world experiments, especially covariate-dependent censoring. Many of the complications ecountered in medical studies are not present in our modeling and therefore we are able to specify a rather straightforward framework in this manuscript. 
Events other than the first hard freeze day could be considered with essentially no alteration to the framework, see Section \ref{sec:discussion}.

Let $\mc{T} = \{0, \dots, T\}$ denote a time sequence.  In our study it is sufficient for $\mc{T}$ to be discrete, but generalization to continuous time is straightforward.  A survival time model $\mathbb{S}:\mc{T} \to [0,1]$ is a monotonically decreasing function where, for our purposes $\mathbb{S}(0) = 1$ and $\mathbb{S}(t) = p$ implies that at time $t$, there is a probability $p$ that an event has occurred at some point in the interval $(0,t]$. In our hard freeze modeling $\mc{T} = \{1, \dots, 92\}$ denotes the number of days from October 1 of a given year, and the event in question is the first day that the mean daily temperature is below 0. We here use $\mc{T}=92$, corresponding to December 31, as our last possible hard freeze date. 

In the survival time framework, an observation $\bs{Y}_i$ consists of the pair $\bs{Y}_i = (T_i, d_i)$, where $T_i \in \mc{T}$ is the observed time to the event or the last observed time of the process and $d_i \in \{0,1\}$ indicates whether or not the event occurred at $T_i$.  In our example $T_i$ is either the number of days after October 1 on which hard freeze first occurred (in which case $d_i = 1$), or $T_i = 92$ and $d_i = 0$ indicating that hard freeze did not occur before December 31. See the attached files for example data.

Suppose we now have a collection of data $\mc{D} = \{\bs{Y}_1, \dots, \bs{Y}_n\}$.  There exist a panoply of methods for estimating $\mathbb{S}(t)$ using these data.  The Kaplan-Meier (KM) estimator \citep{kaplan_meier_1958} is the classic nonparametric estimator of $\mathbb{S}(t)$ and serves as the first step in any more detailed survival analysis.  Let $n_t = \sum_{i = 1}^n \mathbf{1}\{T_i \geq t\}$ denote the number of observations still at risk (i.e.\ the event has not occurred yet) at time $t$ and $e_t = \sum_{i = 1}^n \mathbf{1}\{T_i = t \cap d_i = 1\}$ denote the number of observations for which the event occurred at time $t$. The KM estimator  models the probability that the event of interest has not yet occurred at time $t$ and is formed as 
  \begin{equation}
  \hat{\mathbb{S}}(t| \mc{D}) = \prod_{l = 0}^t (1 - \lambda_l)\label{eq:kaplan_meier}
  \end{equation}
  where
  $$
  \lambda_l = e_l / n_l.
  $$ 
 
 Let $\mathbb{S}_{sy}(t)$ denote the year-$y$-specific survival curve for a location $s\in\mc{S}$. In our study, we will compare three estimators of $\mathbb{S}_{sy}(t)$, all of which use the KM estimator~(\ref{eq:kaplan_meier}) and differ solely on the basis of their input data.  The estimator $\hat{\mathbb{S}}(t| \mc{D}_{s,-y}^{seNorge})$ constructs $\mc{D}$ from the seNorge data at $s$ over the years 1993-2020, but excluding year $y$ and serves as our leave-one-year-out hard freeze climatology.  As such, we will write this as $\hat{\mathbb{S}}_{sy}^C(t)$.  When calculating  $\hat{\mathbb{S}}_{sy}^C(t)$, we have one data pair $\boldsymbol{Y}_i=(T_i,d_i)$  per historical year. Each year is hence treated similarly as an individual in a classical survival analysis setting.

In addition to the climatology estimate, we consider two year-specific survival curves $\hat{\mathbb{S}}(t | \mc{D}_{sy}^{Raw})$ and $\hat{\mathbb{S}}(t | \mc{D}_{sy}^{Post})$ that represent our probabilistic time to hard freeze predictions for year $y$.  In this case $\mc{D}_{sy}^{Raw}$ is constructed from the raw ensemble forecast (or hindcast) for the year $y$ initialized on October 1 for the ensemble systems discussed in Section~\ref{sec:data_products}.   $\mc{D}_{sy}^{Post}$  consist of the ensemble estimates after performing the post-processing according to Equation~(\ref{eq:SFE}).  For brevity we write these as $\hat{\mathbb{S}}_{sy}^R(t)$ and $\hat{\mathbb{S}}_{sy}^{P}(t)$, respectively. When calculating $\hat{\mathbb{S}}_{sy}^R(t)$ and $\hat{\mathbb{S}}_{sy}^{P}(t)$, we have one data pair $\boldsymbol{Y}_i=(T_i,d_i)$ per ensemble member. Each ensemble member is hence treated similarly as an individual in a classical survival analysis setting.

\begin{figure}
	\centering
	\includegraphics[width=0.35\linewidth]{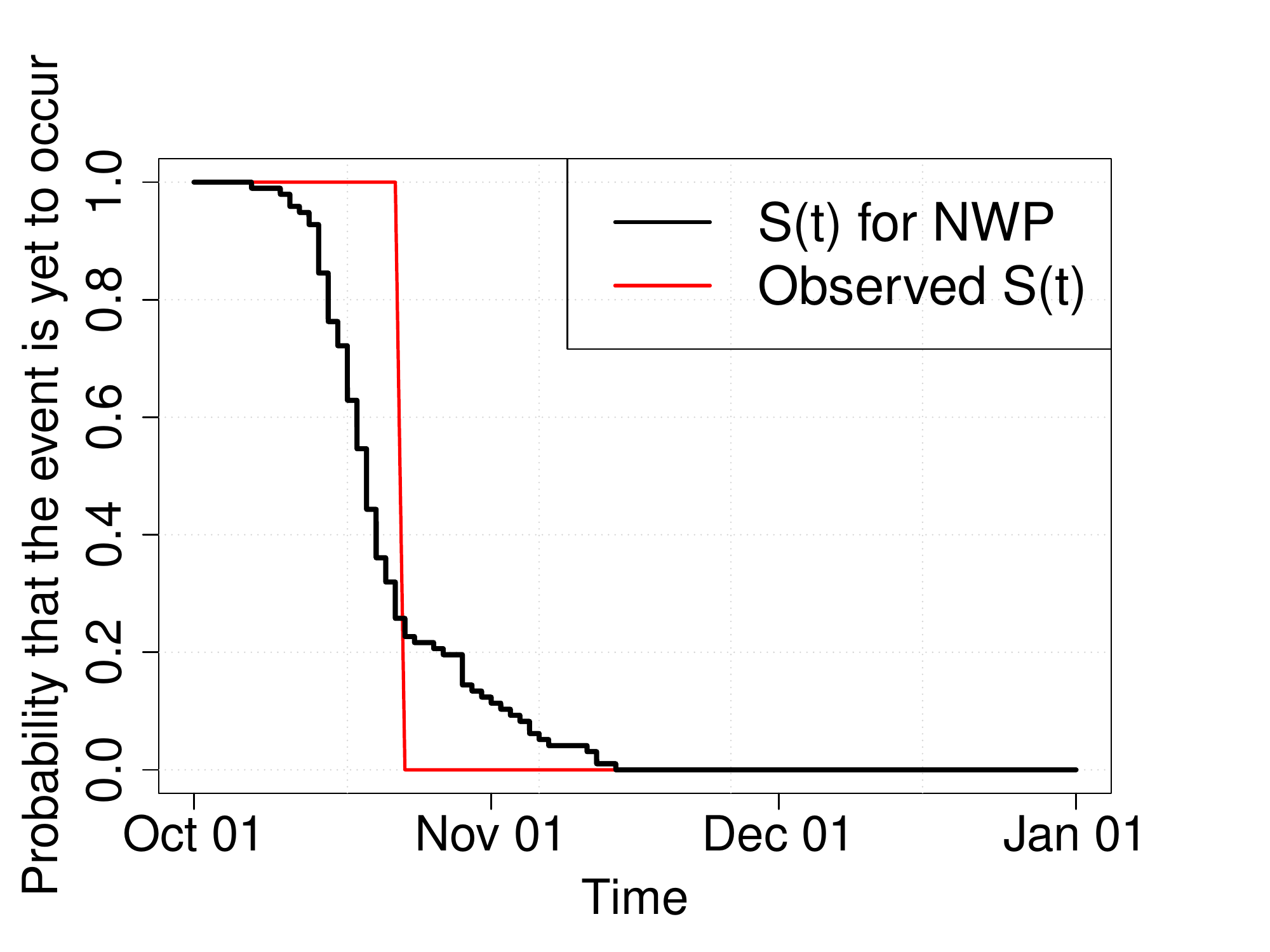}
	\caption{\small Example KM curve $\hat{\mathbb{S}}(t)$ for the time to hard freeze for a year and location based on the ensembles of several NWP systems (black). This is compared to the observed survival curve, which has probability 1 before hard freeze occurred historically for this year and location, and 0 afterwards.}\label{fig:motivex}
\end{figure}

In Figure \ref{fig:motivex}, an example forecast  $\mathbb{S}_{sy}(t)$ for the time to hard freeze is shown. The uncertainty of the NWP ensemble is represented in an intuitive way through the shape of $\mathbb{S}_{sy}(t)$ and how it decays over time. Each drop in $\mathbb{S}_{sy}(t)$ corresponds to a time when at least one ensemble member experienced the first hard freeze after the forecast initialization date. When $\hat{\mathbb{S}}(t)$ is equal to 0, the event has occurred for all ensemble members before time $t$.  If not all ensemble members predict hard freeze before the last day of the forecast, the $\hat{\mathbb{S}}(t)$ will simply not drop to zero.

\section{Forecast evaluation}\label{sec:evaluation}
We construct time to hard freeze forecast, $\hat{\mathbb{S}}_{sy}^C(t)$, $\hat{\mathbb{S}}_{sy}^R(t)$ and $\hat{\mathbb{S}}_{sy}^{P}(t)$, with initialization date October 1st for our 135 study locations and 28 years by using the methodology from Section \ref{sec:methods}. The time to freeze forecasts are evaluated under the paradigm that we aim to maximize the sharpness of the probabilistic forecasts subject to calibration \citep{Gneiting&2007}. Calibration (or reliability) refers to the property that predicted probabilities match observed frequencies. Sharpness is a property of the prediction only, a sharper forecast being more informative about future events. To this aim, we apply rank histograms and proper scoring rules as described in the following two subsections. 

In addition to evaluating time to hard freeze forecasts with initialization date of October 1, we evaluate survival curves for forecasts initialized on September 1. These are mainly used as a reference to the October forecasts to show how the forecast skill is affected when hard freeze occurs further into the future from the initialization date. The experimental set-up is otherwise identical to the one used for the October initializations.

\subsection{Calibration assessment}
Before evaluating the KM time-to-event forecasts, we perform a calibration assessment of the the post-processed ensembles. The goal is to explore whether post-processing of the coarse $1\times1^\circ$ NWP forecasts results in calibrated daily mean temperature forecasts at the $1\times1$ km scale and whether this translates into calibrated forecasts for the time to hard freeze. The calibration assessment is performed by calculating the rank of each seNorge observation within the corresponding forecast ensemble, and building rank histograms \citep{Hamill2001}. To facilitate comparison across forecast cases with varying ensemble size, we standardize the observed ranks to take values between 0 and 1 by subtracting 1 and then dividing by the ensemble size, with ties resolved at random \citep{Heinrich2021}. 

For a calibrated forecast, the observed standardized ranks should be uniformly distributed on [0,1]. If the histogram is skewed to the left or right, it means that the ensemble underestimates or overestimates the quantity of interest. A $\cup$- or $\cap$-shaped histogram indicates under- or over-dispersion. In our study, we only have 28 years of data for each location. To increase our sample size, we calculate standardized ranks for the daily mean temperature for each location, for a group of lead times at once. Calibration at all lead times in a group is thus evaluated simultaneously, and we wouldn't be able to tell whether e.g.\ an apparent bias applies to all or just a subset of those lead times. We summarize the resulting rank histograms by calculating the mean observed standardized ranks for each location. If the ensemble is unbiased, the mean should be close to 0.5. Smaller values of the mean observed standardized ranks indicate an overforecast bias, whereas values above 0.5 indicate an underforecast bias. We also compute the mean deviance from 0.5 for the observed standardized ranks. For a flat histogram we expect a value of 0.25. Smaller values indicate overdispersion of the forecast, whereas values above 0.25 indicate underdispersion, subject to the mean being close to 0.5.

We also consider rank histograms for our target variable, which is the predicted time to hard freeze. These can be derived in the same fashion as for the mean daily temperature. As the forecast and observations take discrete values between 1 and 92, we add random noise to the data to avoid ties when the observation rank within the ensemble is calculated. Furthermore, we cannot pool the observed ranks according to lead time here. Instead, we pool across all study locations, both for the raw and the post-processed ensemble forecasts.

\subsection{Evaluating probabilistic time-to-event forecasts}
To evaluate the KM predictions of time to hard freeze, the predicted survival curve  $\hat{\mathbb{S}}_{sy}^R(t)$, $\hat{\mathbb{S}}_{sy}^{P}(t)$ or $\hat{\mathbb{S}}_{sy}^{C}(t)$ is compared to the observed survival curve $\mathbb{S}_{sy}^{obs}(t)$ for $t=1,2,...,92$ for the target locations. For year $y$ and location $s$, the observed survival curve is 1 before frost occurred  and 0 afterwards, see Figure \ref{fig:motivex} for an example.

For a fixed time $t$, a survival model simply issues a probabilistic forecast for a categorical event that takes either 0 or 1 as values, namely whether hard freeze has occurred up to the time $t$ or not. Such predictions can be evaluated based on the average Brier score \citep{Brier1950}, which takes the form
\begin{equation*}
BS_{s,M}(t) := \frac{1}{28}\sum_{y = 1993}^{2020}(\mathbb{S}_{sy}^{obs}(t) - \hat{\mathbb{S}}_{sy}^{M}(t))^2,
\end{equation*}
where the model $\hat{\mathbb{S}}_{sy}^M$ is a placeholder for any of the models $\hat{\mathbb{S}}_{sy}^R$, $\hat{\mathbb{S}}_{sy}^P$ or $\hat{\mathbb{S}}_{sy}^C$ discussed in Section 2.3. We generally consider negatively oriented scores such that lower scores indicate better predictive performance. 

The score $BS_{s,M}(t)$ is a measure of the predictive skill solely focusing on the prediction of the occurrence of hard freeze up to time $t$. The overall fit of the survival curve at location $s$ can be assessed by summing (`integrating') the Brier scores over $t$, see \citet{Mogensen&2012},
\begin{equation}\label{eq:IBS}
IBS_{s,M} :=  \frac{1}{92}\sum_{t = 1}^{92}\bigg(\frac{1}{28}\sum_{y = 1993}^{2020}(\mathbb{S}_{sy}^{obs}(t) - \hat{\mathbb{S}}_{sy}^{M}(t))^2\bigg),
\end{equation}
where $T=92$ is the highest considered $t$. As a visual example, in Figure \ref{fig:motivex} the integrated Brier score would simply be the integrated squared difference between the two lines. The integrated Brier score is equivalent to the continuous ranked probability score \citep[CRPS;][]{Hersbach2000} for the time to hard freeze and thus a proper score.

A natural question for evaluating predictions is whether the forecast exhibits higher skill than a climatological forecast. To facilitate direct comparison, proper scores can be transformed into skill scores, see \citet{Wilks2011}. The integrated Brier skill score for model $M$ at location $s$ is defined as
\[
IBSS_{s,M} := \frac{IBS_{s,C} - IBS_{s,M}}{IBS_{s,C}},\]
where $IBS_{s,C}$ is the integrated Brier score of the climatological model $\hat{\mathbb{S}}_{sy}^C$.
Skill scores are positively oriented, such that  higher values indicate better predictive performance. They are normalized in the sense that a perfect forecast achieves a skill score of 1, and a forecast with the same skill as climatology achieves a skill score of 0. Positive skill scores thus indicate better predictive performance than a climatological forecast.

Above, we define location-specific evaluation scores. A year-specific integrated Brier (skill) score for the whole study area can be obtained similarly, by averaging over locations instead of years.

\section{Results}\label{sec:results}
In this section we present the results from the time to hard freeze analysis. We start by presenting the results from the calibration assessment, before considering the predictive performance of the time-to-event forecasts.

\subsection{Calibration assessment}
\begin{figure}[h!!!!!]
	\centering
	\begin{subfigure}[b]{0.32\textwidth}
		\includegraphics[width=1\linewidth]{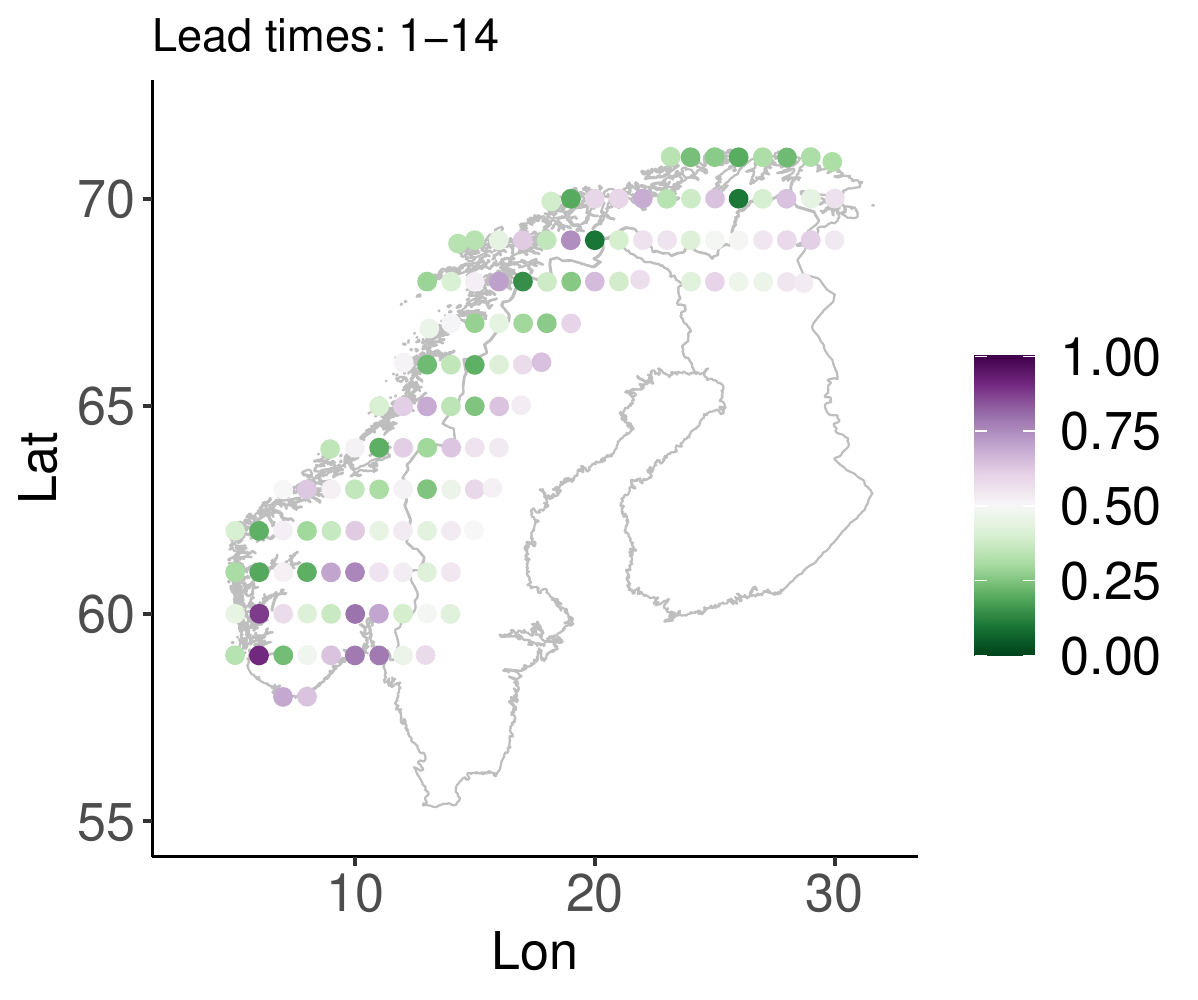}
		\caption{Raw: 1-14 days}
		\label{fig:PIT_daily_cds1}
	\end{subfigure}
		\begin{subfigure}[b]{0.32\textwidth}
		\includegraphics[width=1\linewidth]{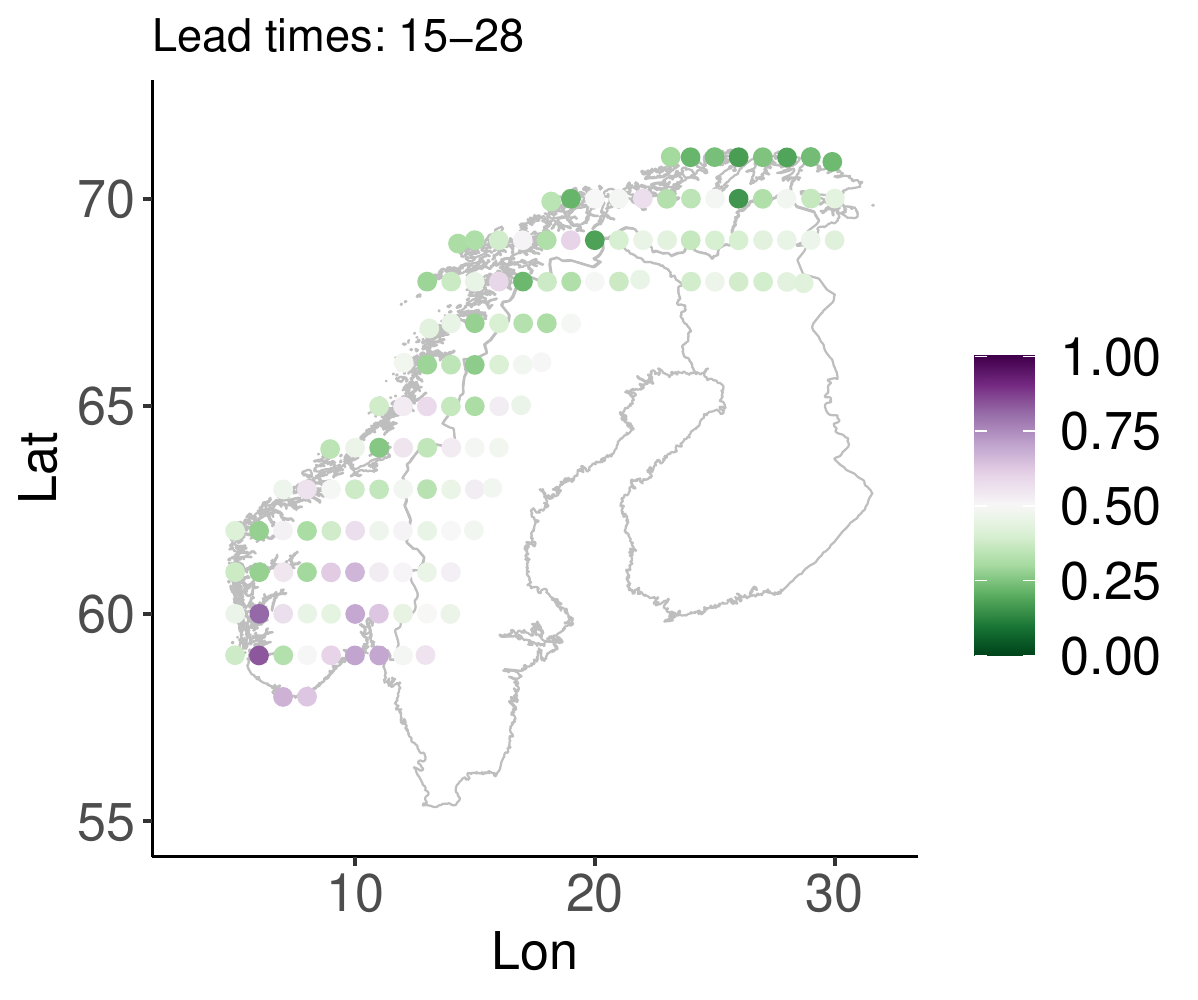}
		\caption{Raw: 15-28 days}
		\label{fig:PIT_daily_cds15}
	\end{subfigure}
			\begin{subfigure}[b]{0.32\textwidth}
		\includegraphics[width=1\linewidth]{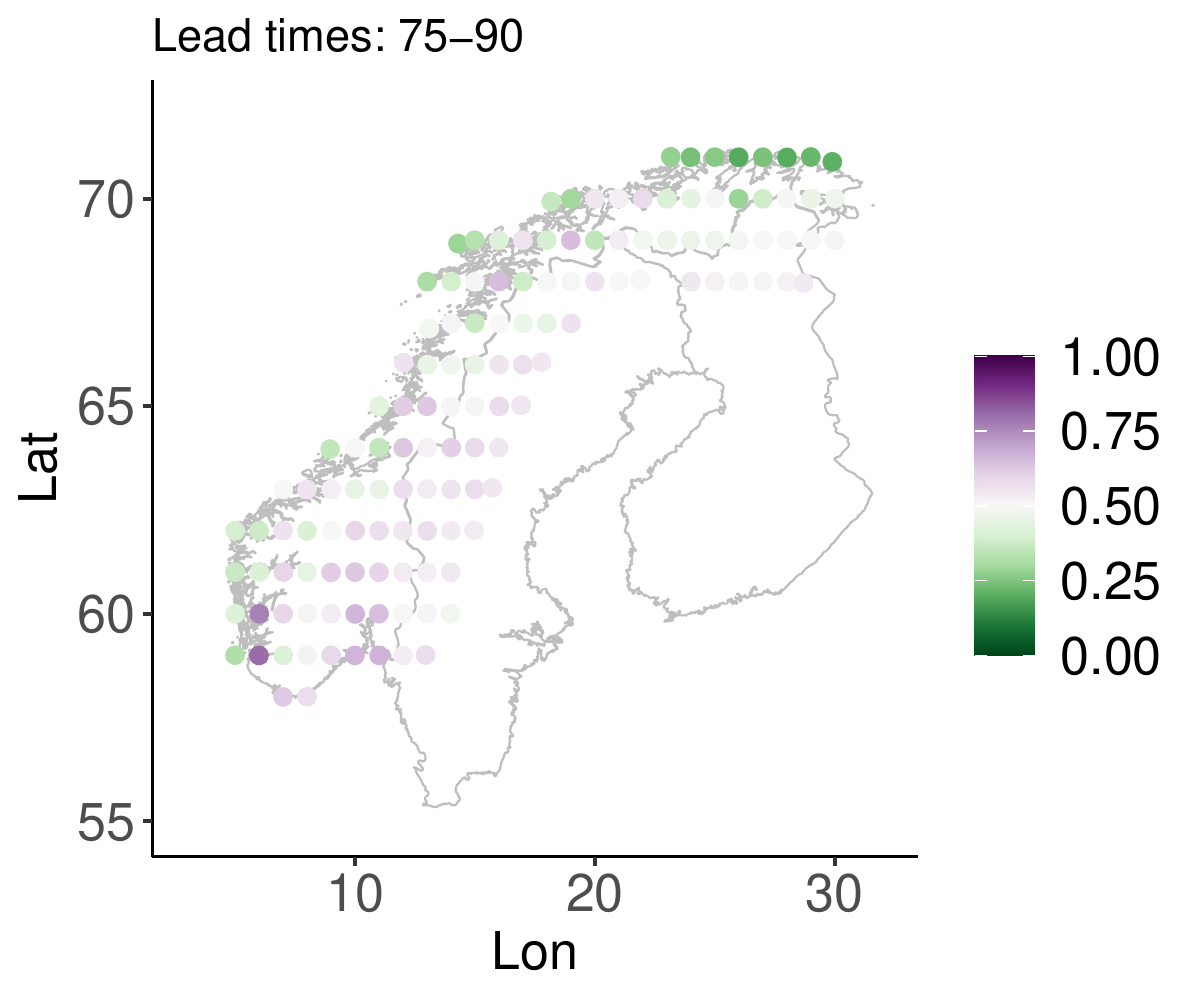}
		\caption{Raw: 75-90 days}
		\label{fig:PIT_daily_cds75}
	\end{subfigure}
	\begin{subfigure}[b]{0.32\textwidth}
		\includegraphics[width=1\linewidth]{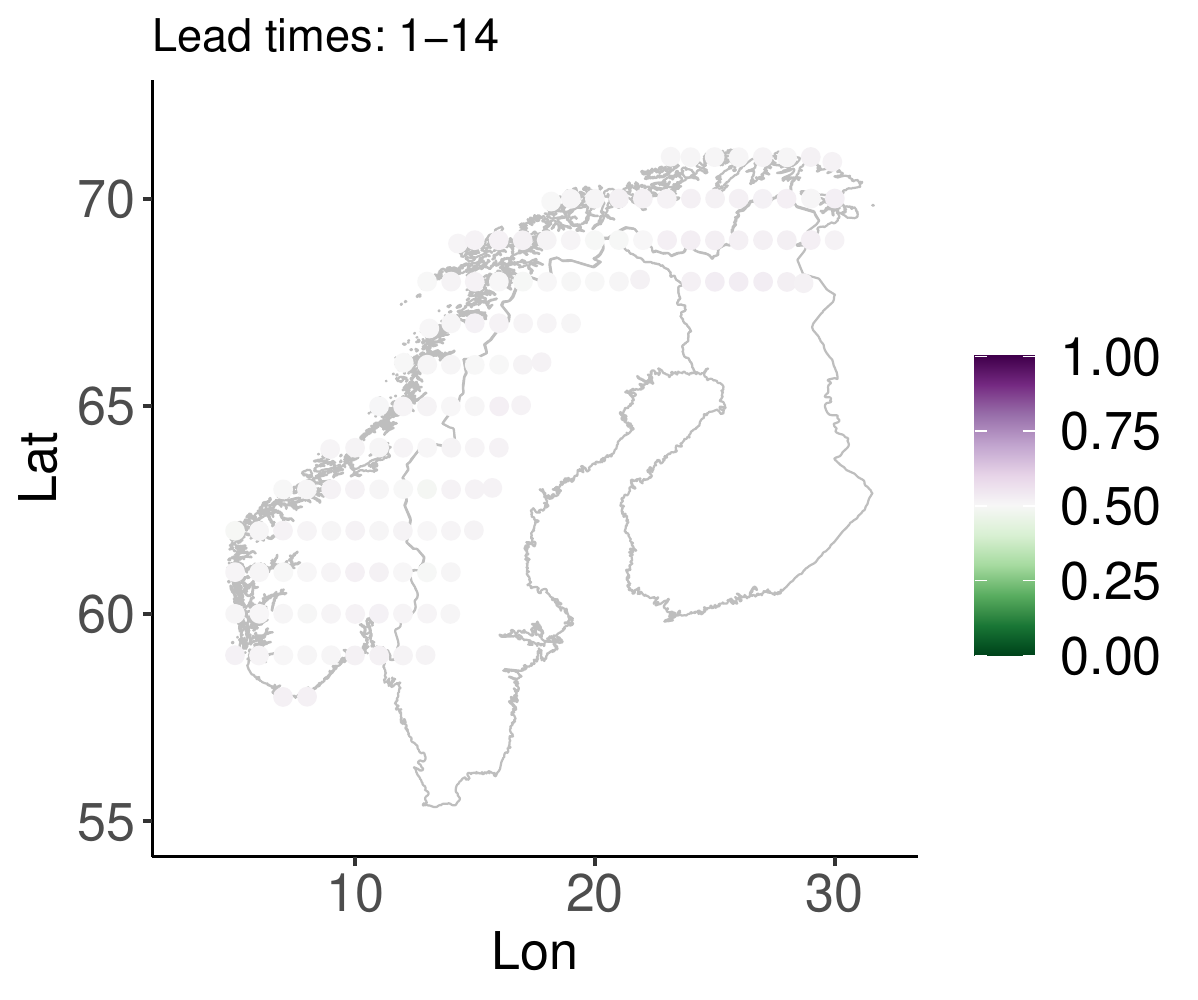}
		\caption{PP: 1-14 days}
		\label{fig:PIT_daily_sfe1}
	\end{subfigure}
		\begin{subfigure}[b]{0.32\textwidth}
		\includegraphics[width=1\linewidth]{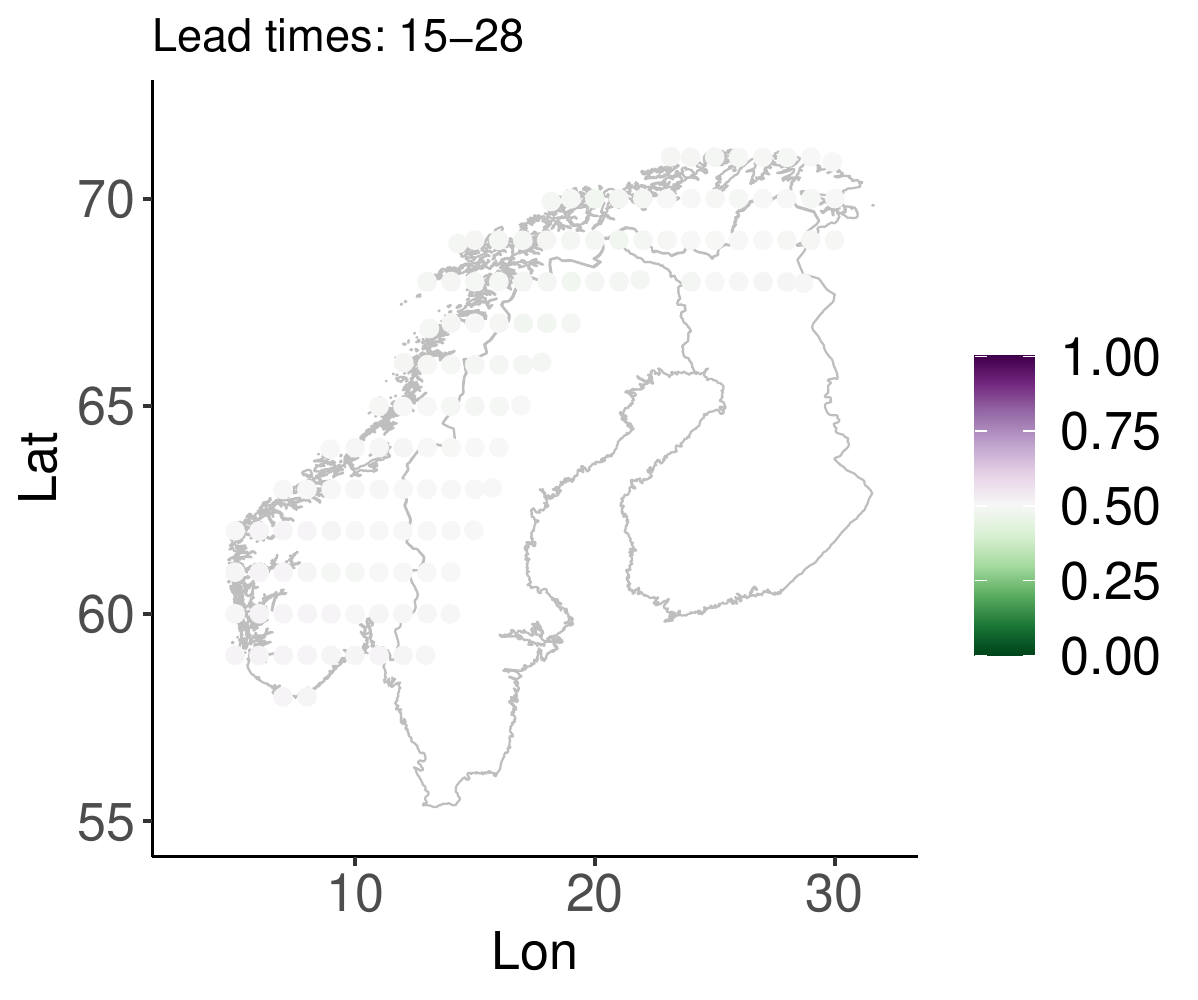}
		\caption{PP: 15-28 days}
		\label{fig:PIT_daily_sfe15}
	\end{subfigure}
			\begin{subfigure}[b]{0.32\textwidth}
		\includegraphics[width=1\linewidth]{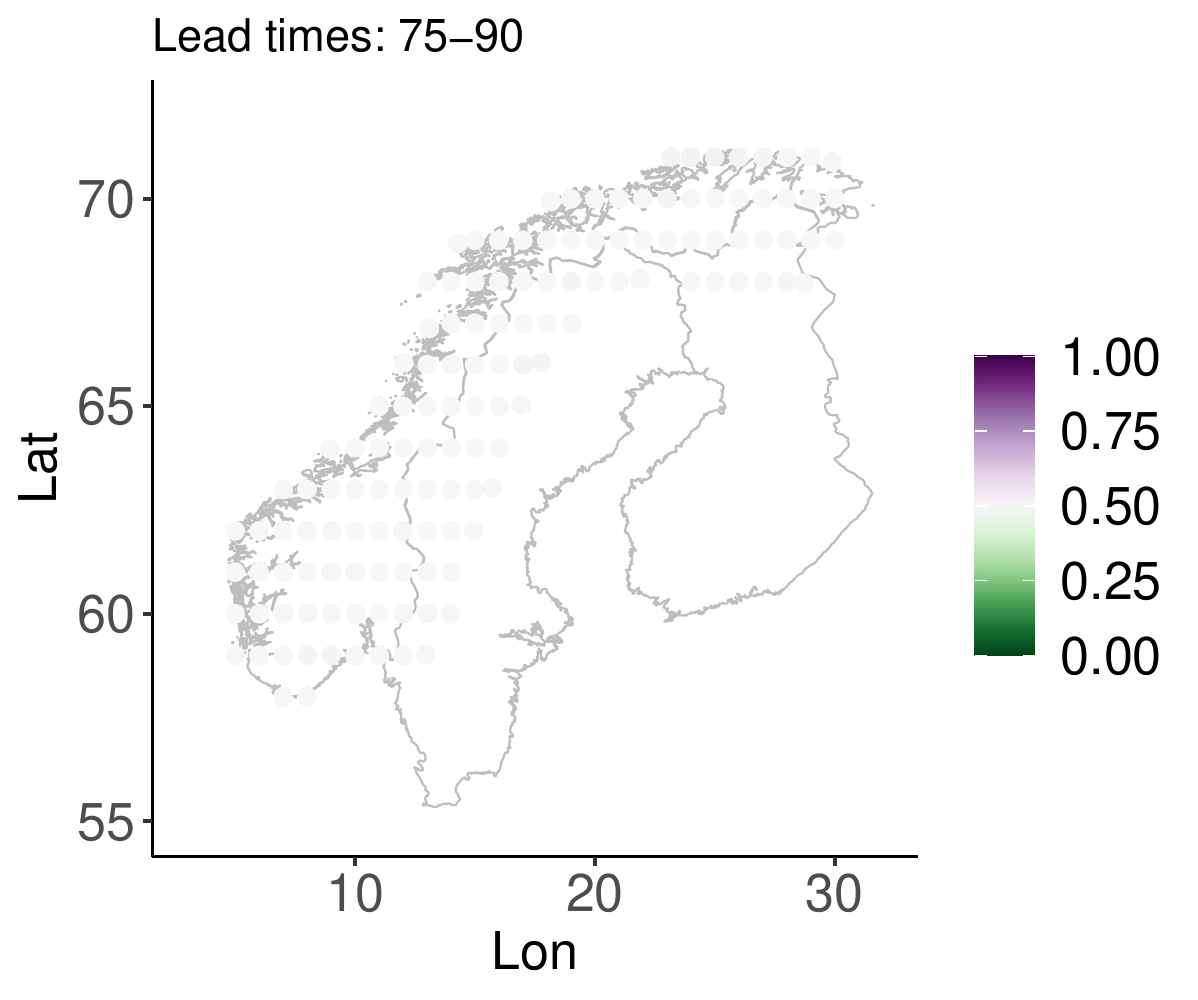}
		\caption{PP: 75-90 days}
		\label{fig:PIT_daily_sfe75}
	\end{subfigure}
	\caption{\small Mean observed standardized ranks for daily temperature forecasts initialized on October 1 for daily lead times 1-14 (leftmost column), 15-28 (middle column) and 75-90 (rightmost column) for the raw (top row) and the post-processed (bottom row) ensemble forecast. The results are aggregated over 1993-2020. If the forecast is calibrated, the mean observed standardized rank should be 0.5.}\label{fig:PITbias}
\end{figure}

The mean observed standardized ranks for the daily temperature forecasts are shown in Figure \ref{fig:PITbias} for some selected lead times for the raw and post-processed ensembles. Grid points colored in green indicate that the forecasts are too warm on average, whereas for purple grid points the forecasts are too cold on average. The results show that the raw ensemble is too warm at many grid points, in particular in the northern part of the study area and too cold at other grid points. We obtain a similar spatial distribution of mean ranks for all lead times, but the biases appear to be less extreme for the longest lead times (75-90 days). For the post-processed ensembles, the mean ranks are close to 0.5 for all study locations and lead times, indicating that the post-processed temperature forecasts are unbiased on average.

\begin{figure}[h!!!!]
	\centering
	\begin{subfigure}[b]{0.32\textwidth}
		\includegraphics[width=1\linewidth]{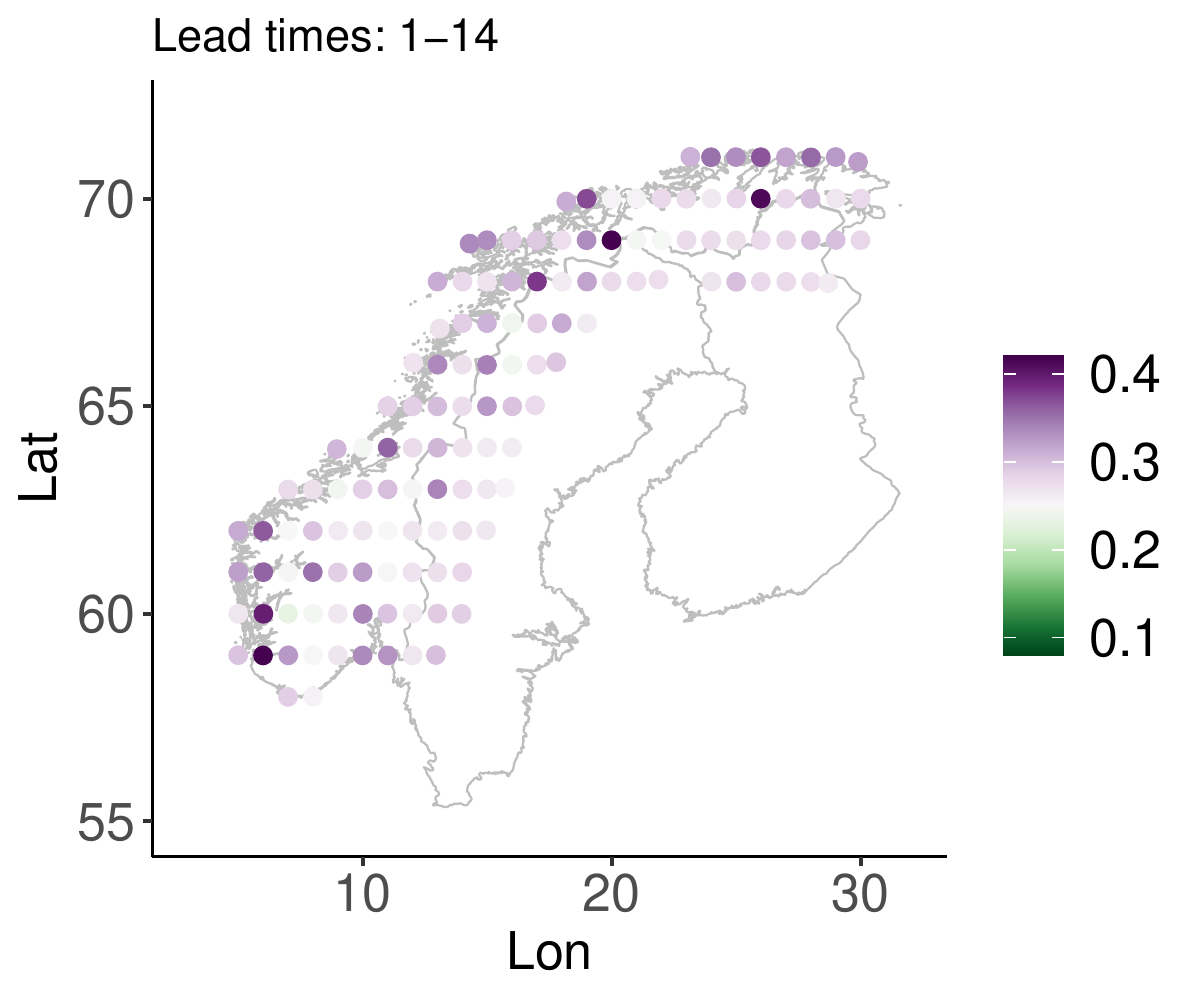}
		\caption{Raw: 1-14 days}
		\label{fig:PIT_error_cds1}
	\end{subfigure}
		\begin{subfigure}[b]{0.32\textwidth}
		\includegraphics[width=1\linewidth]{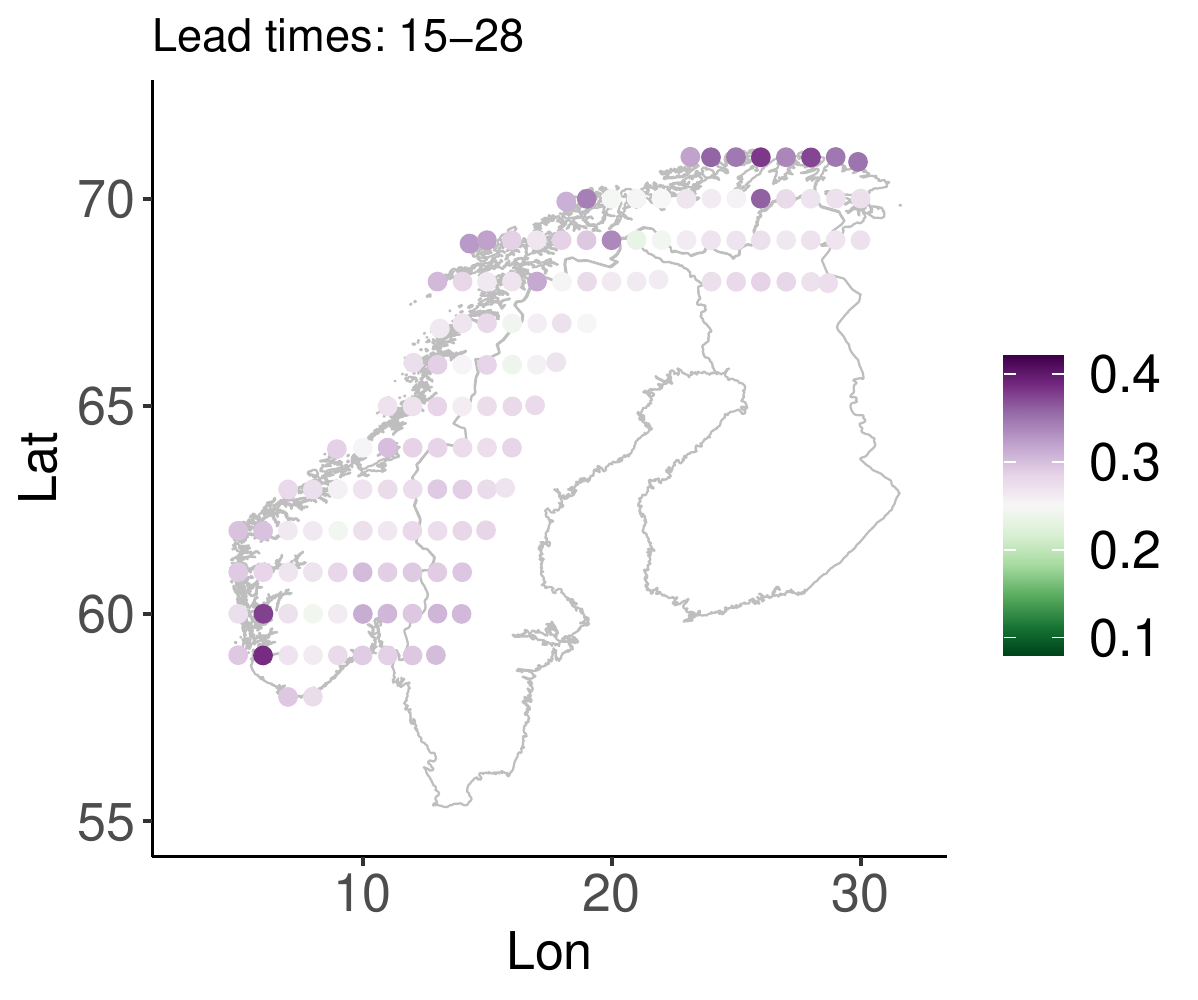}
		\caption{Raw: 15-28 days}
		\label{fig:PIT_error_cds15}
	\end{subfigure}
			\begin{subfigure}[b]{0.32\textwidth}
		\includegraphics[width=1\linewidth]{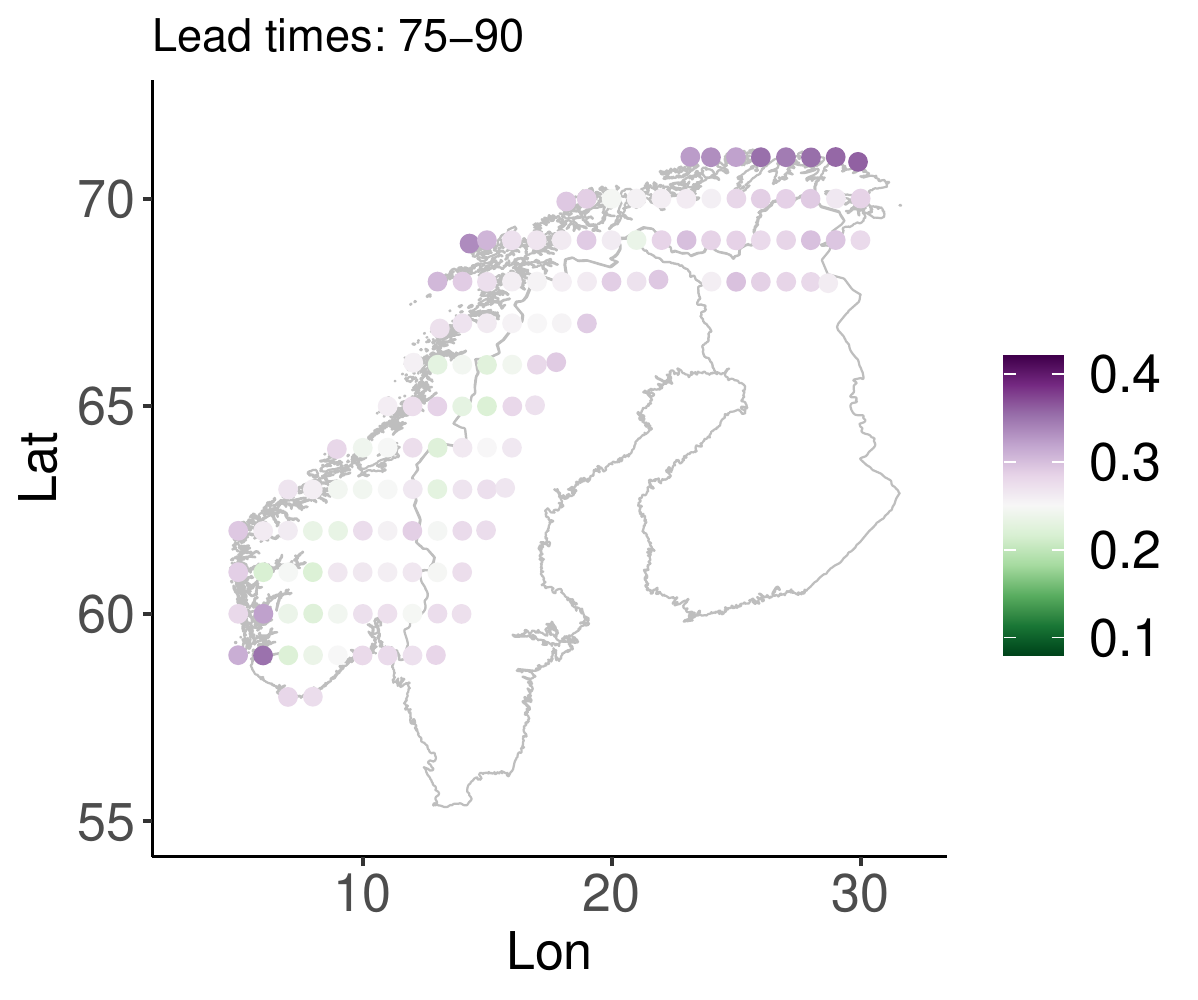}
		\caption{Raw: 75-90 days}
		\label{fig:PIT_error_cds75}
	\end{subfigure}
	\begin{subfigure}[b]{0.32\textwidth}
		\includegraphics[width=1\linewidth]{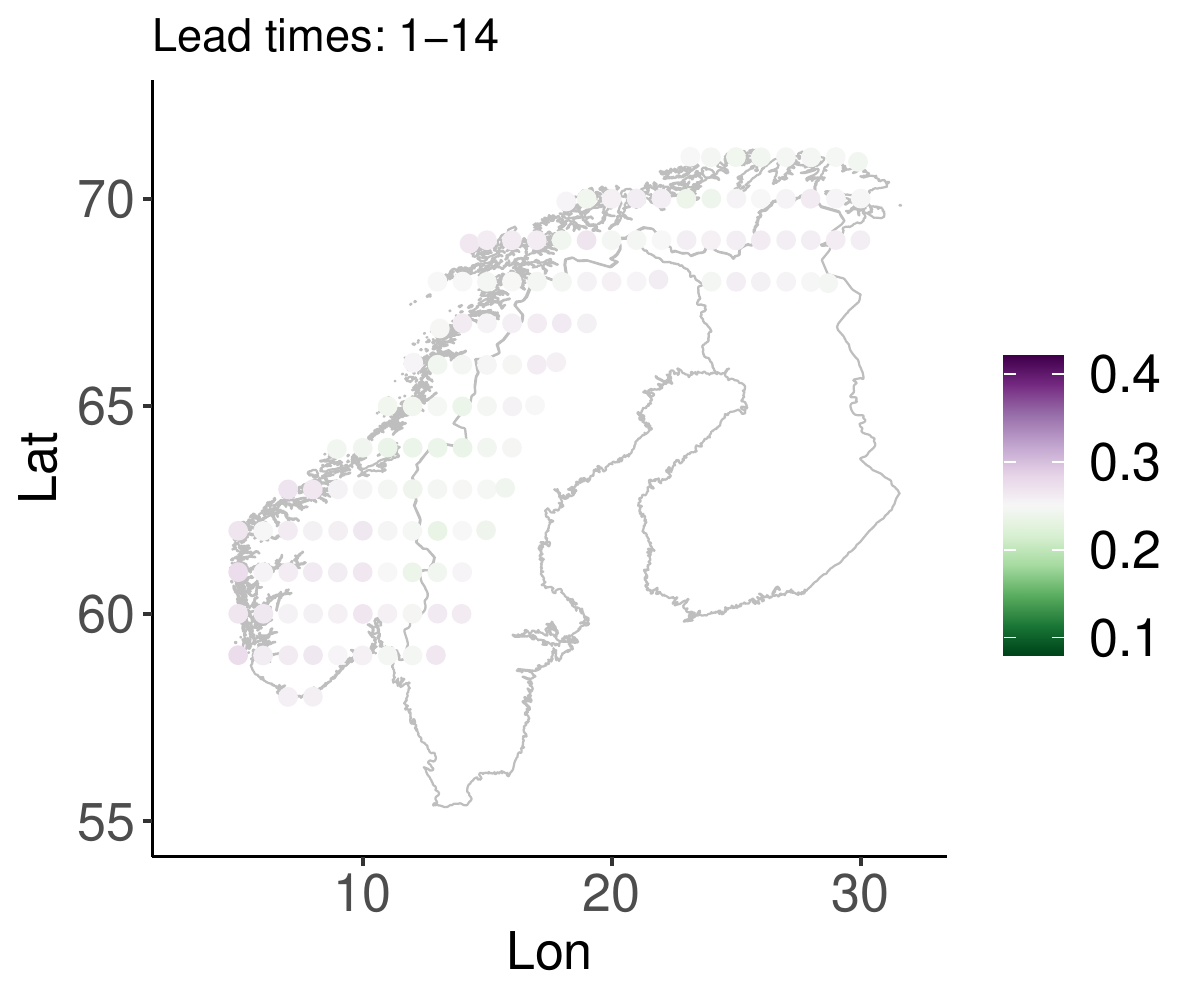}
		\caption{PP: 1-14 days}
		\label{fig:PIT_error_sfe1}
	\end{subfigure}
		\begin{subfigure}[b]{0.32\textwidth}
		\includegraphics[width=1\linewidth]{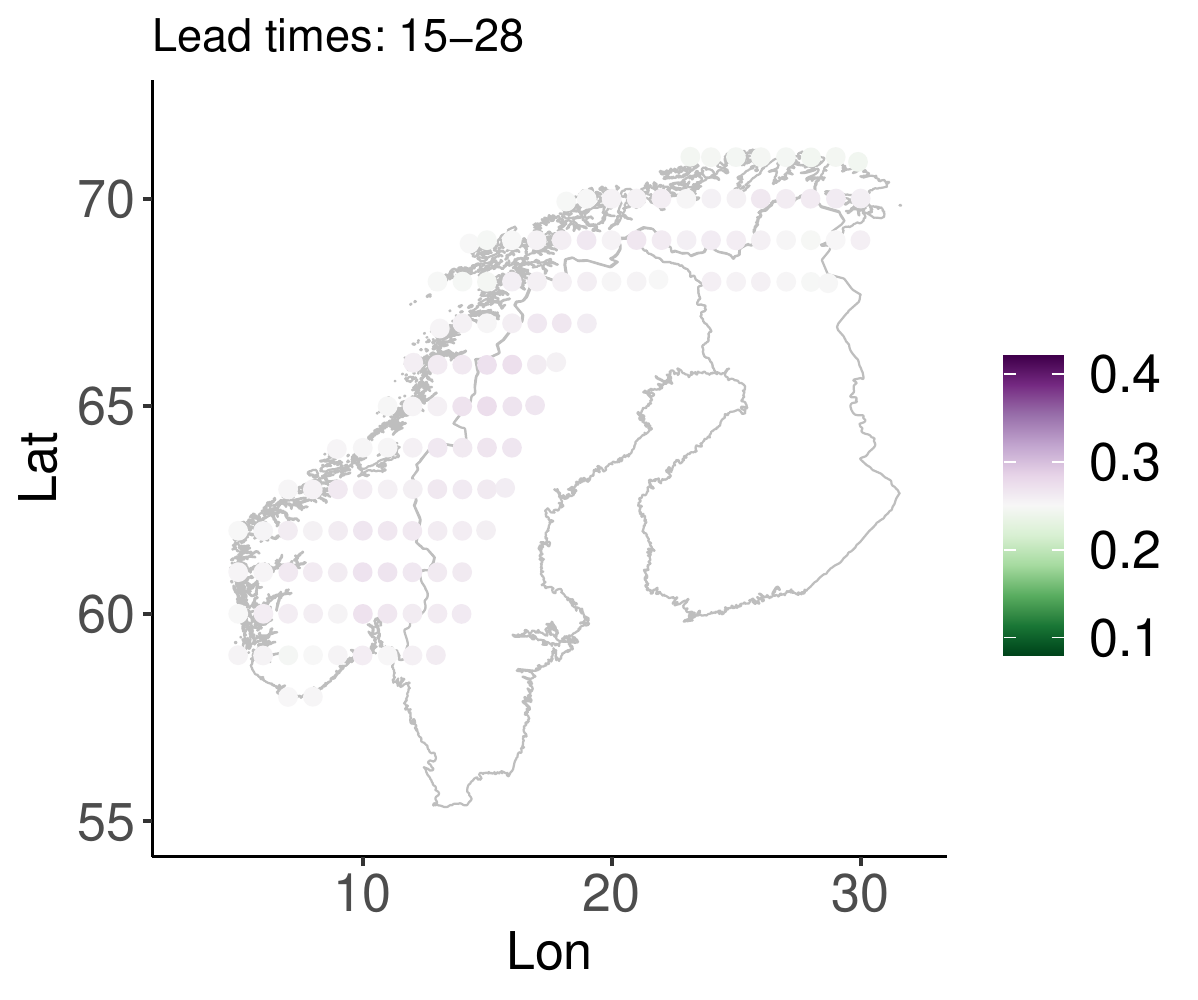}
		\caption{PP: 15-28 days}
		\label{fig:PIT_error_sfe15}
	\end{subfigure}
			\begin{subfigure}[b]{0.32\textwidth}
		\includegraphics[width=1\linewidth]{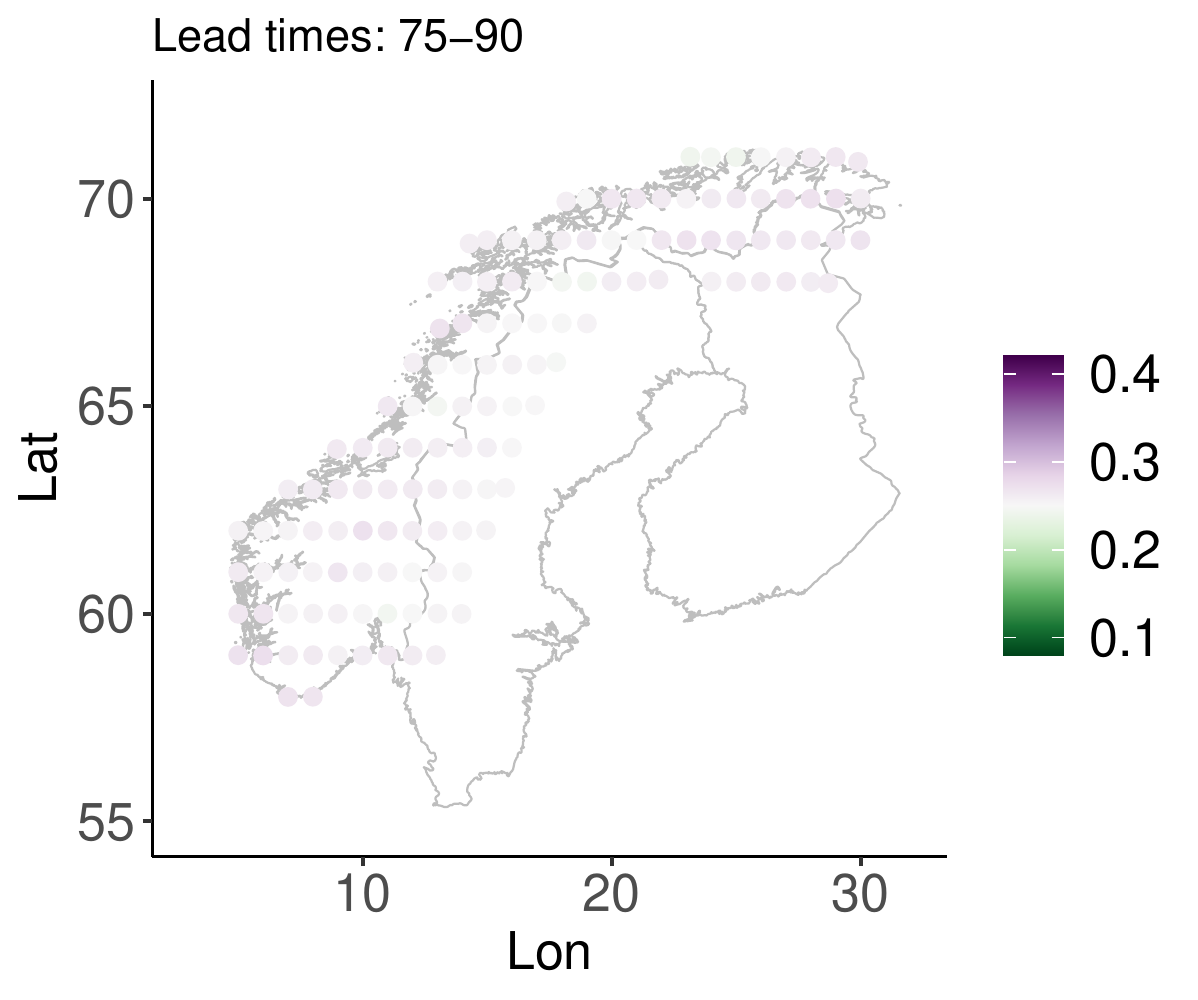}
		\caption{PP: 75-90 days}
		\label{fig:PIT_error_sfe75}
	\end{subfigure}
	\caption{\small Absolute deviance from 0.5 of observed standardized ranks for each location for daily temperature forecasts initialized on October 1 aggregated over 1993-2020, for daily lead times 1-14 (leftmost column), 15-28 (middle column) and 75-90 (rightmost column), for the raw (top row) and the postprocessed ensemble (bottom row). For a calibrated forecast, the mean absolute deviance should be 0.25.}\label{fig:PITerror}
\end{figure}

Figure~\ref{fig:PITerror}a-\ref{fig:PITerror}c show the mean rank deviation from 0.5, where a deviation of 0.25 indicates a calibrated forecast.  While the values for the raw daily temperature forecast are mostly larger than 0.25, the biases seen in Figure \ref{fig:PITbias} cause at least part of that increase and make it difficult to draw conclusions about additional dispersion errors in the ensemble. However, we can conclude from Figure \ref{fig:PITbias} that the post-processed ensemble is unbiased and neither under- nor overdispersive.  The post-processing algorithm hence seems to be able to translate a $1\times1^\circ$ daily temperature forecast into a calibrated local temperature forecast at a $1\times1$ km resolution.

\begin{figure}[h!]
	\centering 
	\begin{subfigure}[b]{0.32\textwidth}
		\includegraphics[width=1\linewidth]{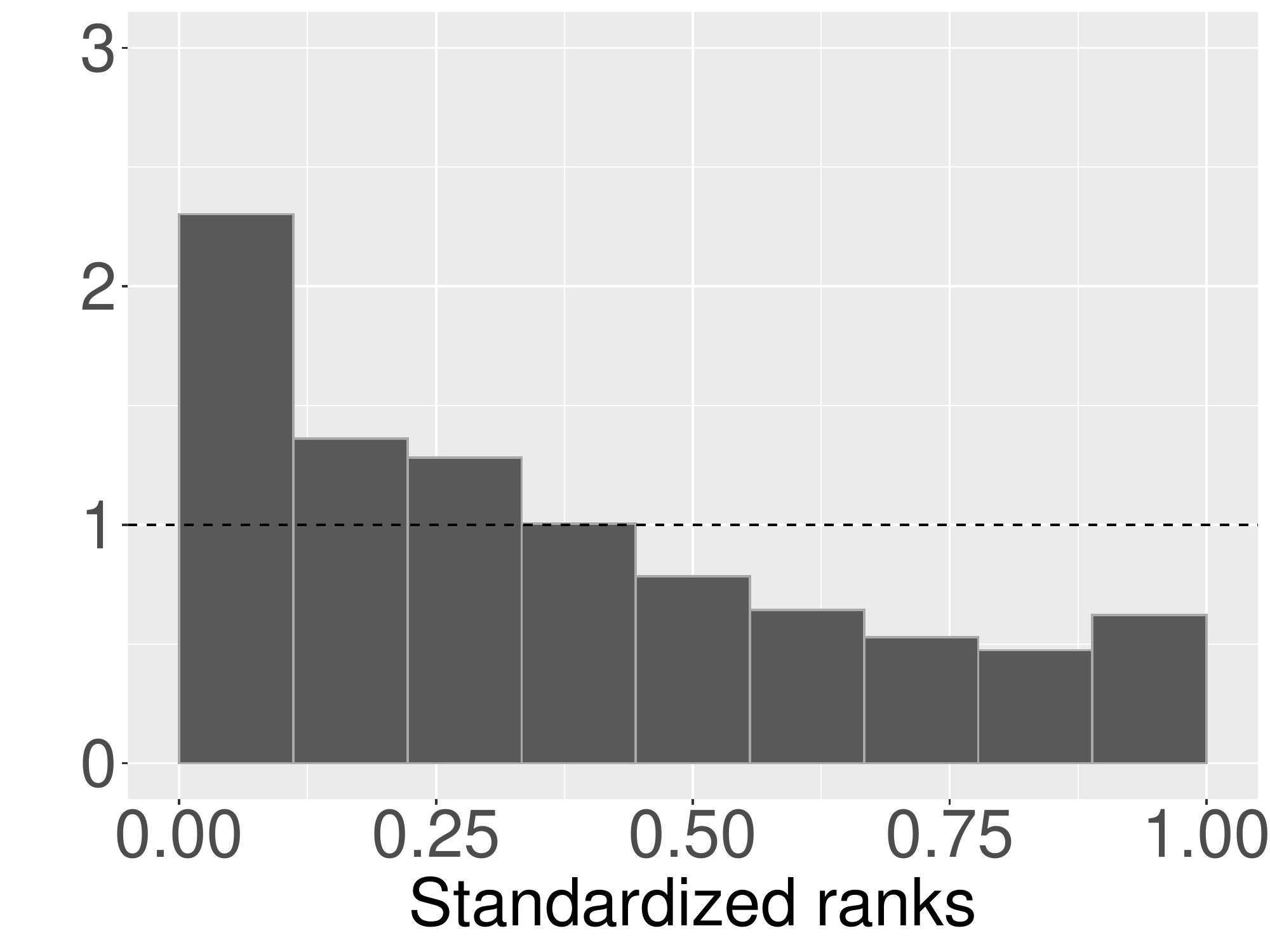}
		\caption{Raw }
		\label{fig:PITtimeraw}
	\end{subfigure}
		\begin{subfigure}[b]{0.32\textwidth}
		\includegraphics[width=1\linewidth]{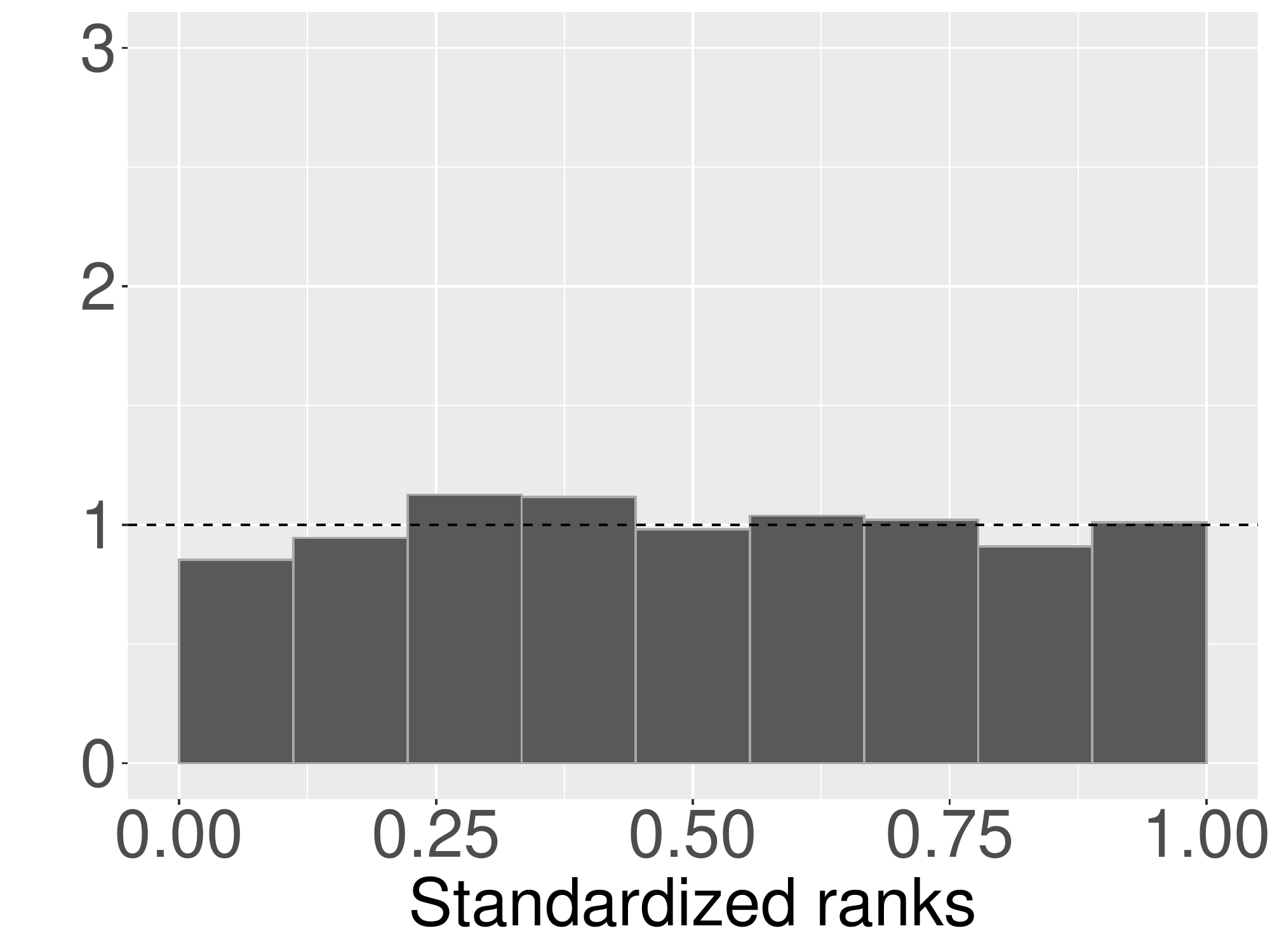}
		\caption{Post-processed}
		\label{fig:PITtimesfe}
	\end{subfigure}
	\caption{Rank histograms for the time to hard freeze forecasts for the raw (left) and the postprocessed ensemble (right) initialized on October 1, aggregated over the years 1993-2020 and the study locations shown in Figure~\ref{fig:seNorge}.}	\label{fig:PITtime}
\end{figure}

In Figure \ref{fig:PITtime}, rank histograms for the predictions of the time to hard freeze are presented. They show that, on average over all study locations, the raw ensemble tends to overestimate the time to hard freeze. However, the histogram does not show any signs of under- or overdispersion. Figure \ref{fig:PITtime} further shows that the rank histogram for the post-processed forecasts appears uniform indicating an overall calibrated forecast. Hence, the post-processing of the  $1\times1^\circ$ daily temperature forecasts also results in well-calibrated predictions for the time to hard freeze for our study locations.

\subsection{Evaluation of the time to hard freeze forecasts}
\subsubsection{Overall performance}\label{sec:Overall performance}
\begin{figure}[h!!]
	\centering
	\begin{subfigure}[b]{0.44\textwidth}
		\includegraphics[width=0.85\linewidth ,trim={0.2cm 0cm 2.5cm 0cm},clip]{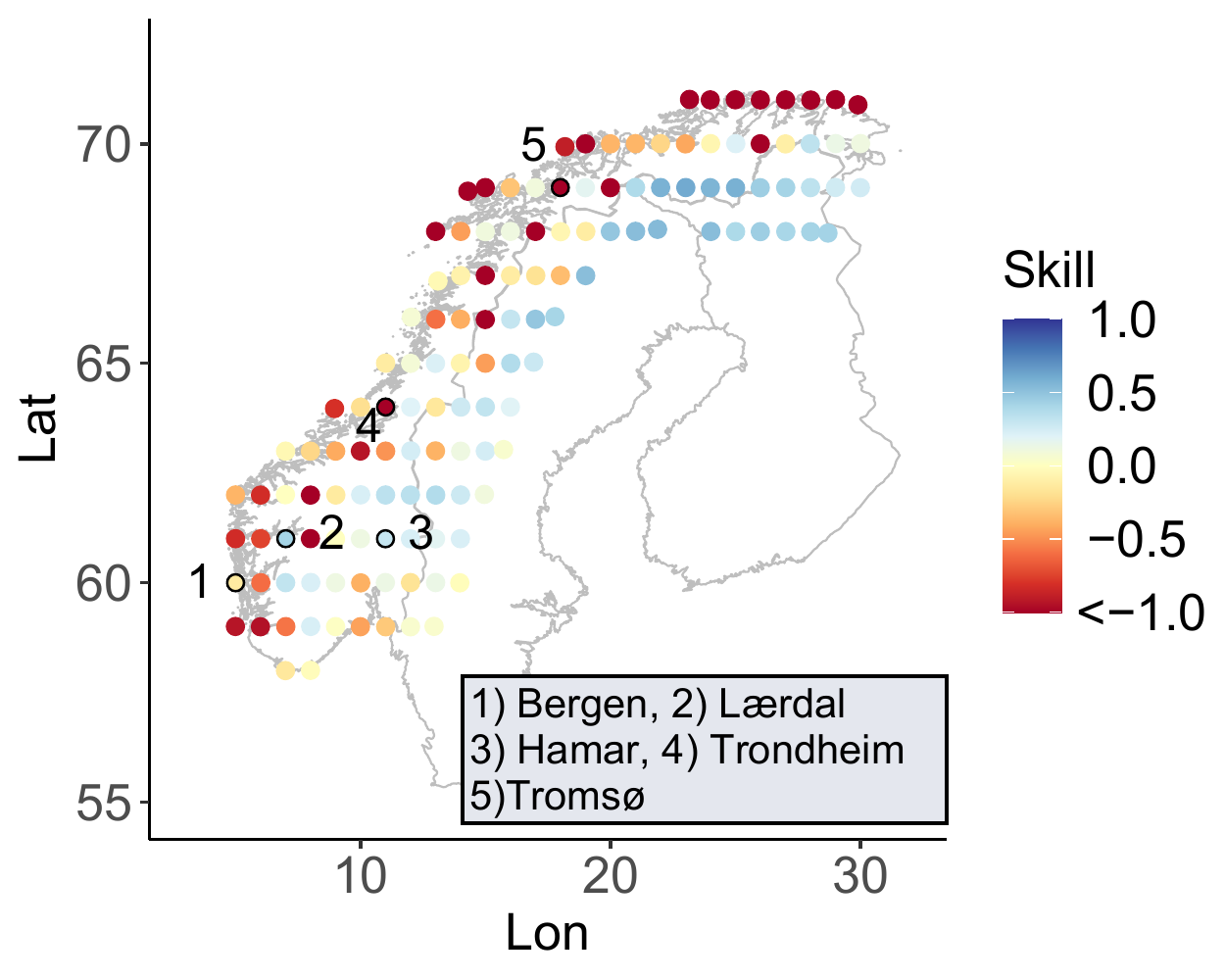}
		\caption{Raw}
		\label{fig:skillmap_cds}
	\end{subfigure}
	\begin{subfigure}[b]{0.44\textwidth}
		\includegraphics[width=1\textwidth,trim={0.6cm 0cm 0cm 0cm},clip]{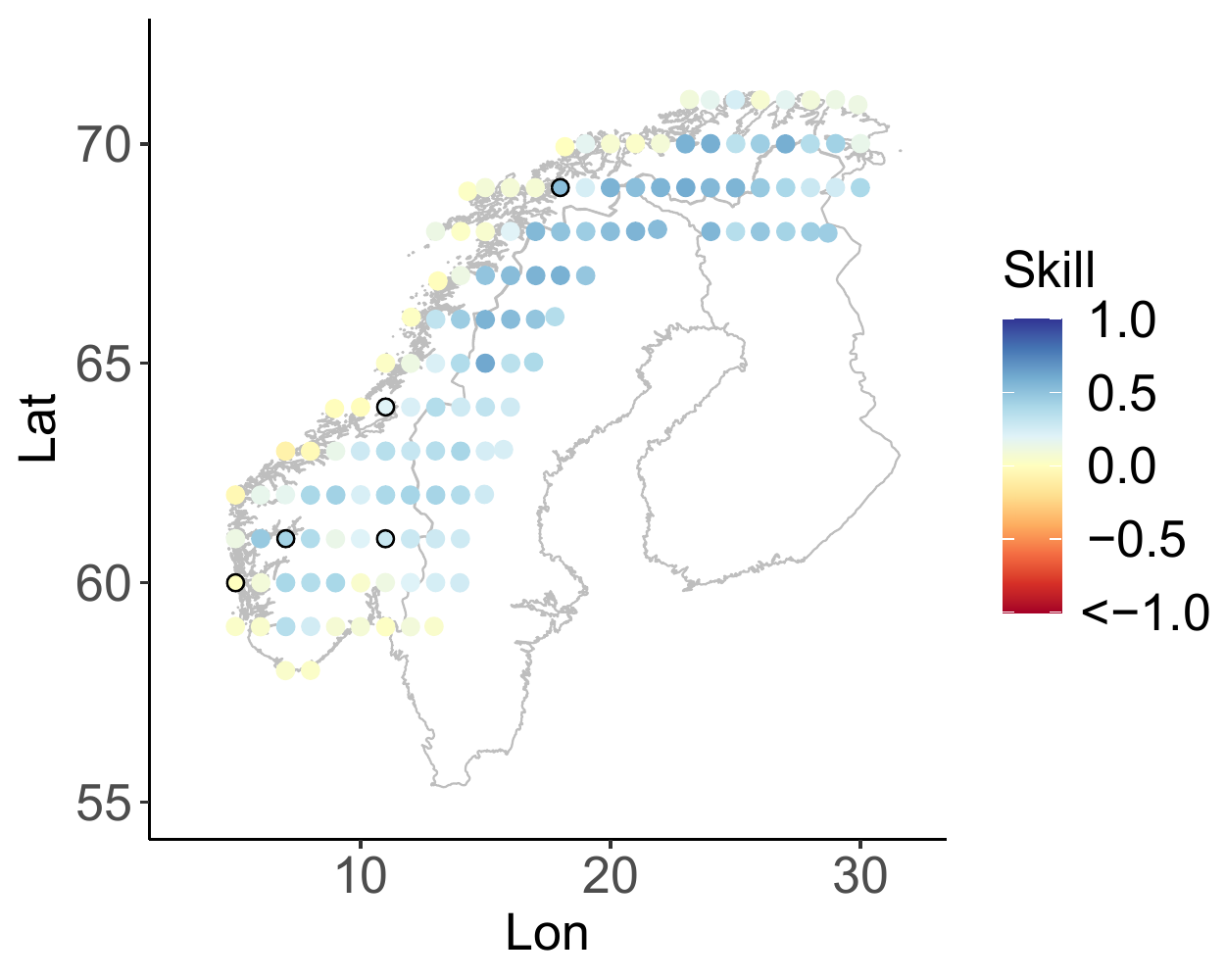}
		\caption{Post-processed}
		\label{fig:skillmap_sfe}
	\end{subfigure}
	\caption{\small Integrated Brier skill scores for time to hard freeze forecasts initialized on October 1 aggregated over 1993-2020 for the raw temperature forecasts $\S^{R}(t)$ and the post-processed temperature forecasts $\S^{P}(t)$ relative to the climatological forecast $\S^{C}(t)$. Five locations, close to the Norwegian cities/towns Bergen, Lærdal, Hamar, Trondheim and Tromsø, are marked with black circles for further reference. See also Table~\ref{tab:loctab}.}\label{fig:skill}
\end{figure}

In this section, we evaluate the overall performance of the proposed time to hard freeze KM forecasts. In Figure \ref{fig:skill}, the integrated Brier skill scores for all study locations are presented, both for the raw forecast  $\hat{\mathbb{S}}_{sy}^R(t)$ and the post-processed forecast $\hat{\mathbb{S}}^P_{sy}(t)$, where they are compared against climatology. The results show that $\hat{\mathbb{S}}^R_{sy}(t)$  yields negative skill scores at many grid points, particularly in  coastal areas, while in regions far away from the coast, the skill score is positive. In coastal areas, the frost in general comes later than in nearby non-coastal areas and is thus more difficult to predict with the October 1 initialization. The low spatial resolution of the forecasts can also be particularly challenging here, and is likely responsible for some of the biases noted above, especially as grid cells close to the coast overlap partially sea and partially land. This may be intensified by the fact that we have combined NWP products from different spatial grids. Furthermore, the west coast of Norway is quite mountainous, yielding rapidly changing temperatures fields.

Figure \ref{fig:skillmap_sfe} shows that the post-processing procedure increases the skill in areas where $\hat{\mathbb{S}}^R_{sy}(t)$ performed poorly. The post-processed model $\hat{\mathbb{S}}^{P}_{sy}(t)$ yields positive skill scores at most study locations, with near-zero skill scores near the coast. Consequently, the post-processed model performs just as well as a climatology forecast in coastal regions, and consistently outperforms climatology at most grid points.

\begin{figure}
	\centering
	\begin{subfigure}[b]{0.6\textwidth}
		\includegraphics[width=1\linewidth]{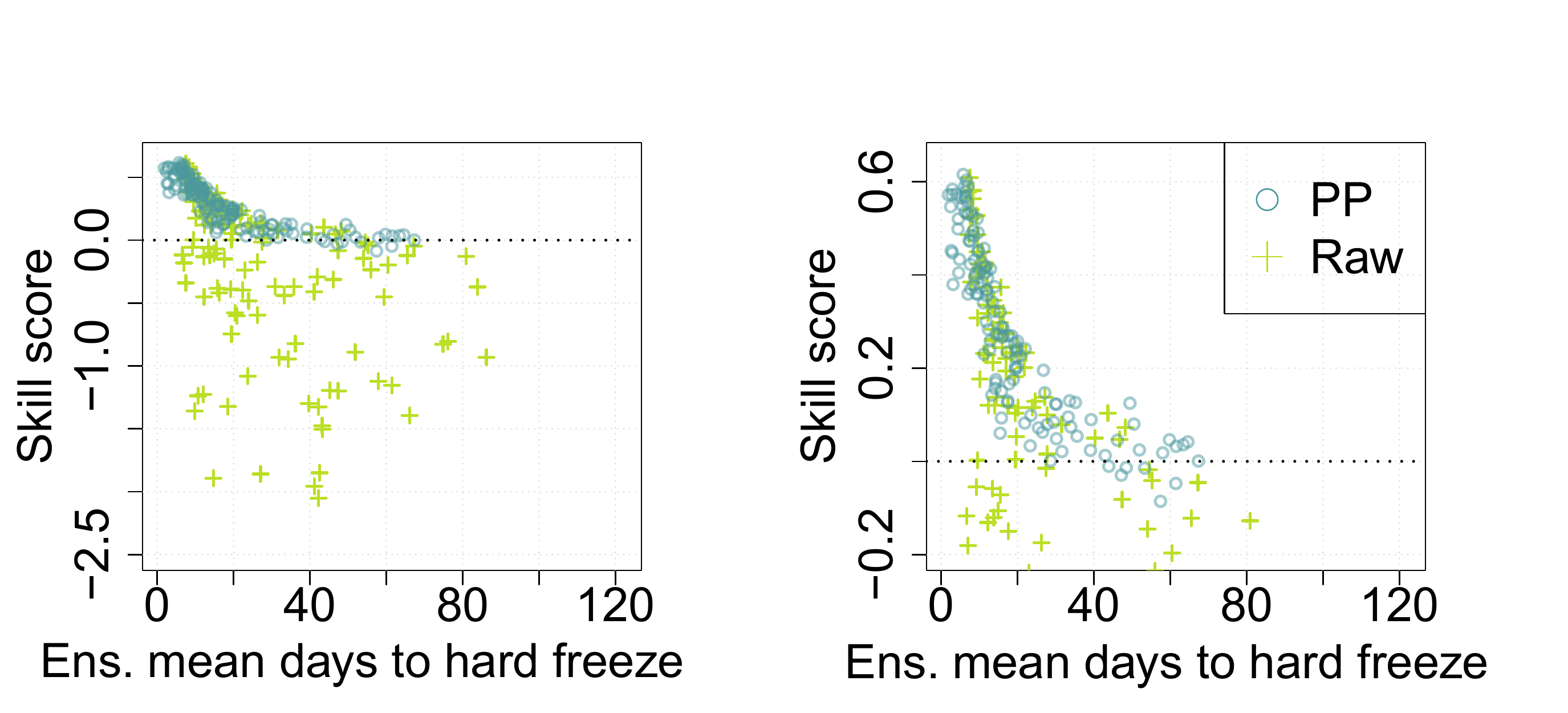}
		\caption{Forecasts initialized on October 1}
		\label{fig:skilloct}
	\end{subfigure}\\
	\begin{subfigure}[b]{0.6\textwidth}
		\includegraphics[width=1\textwidth]{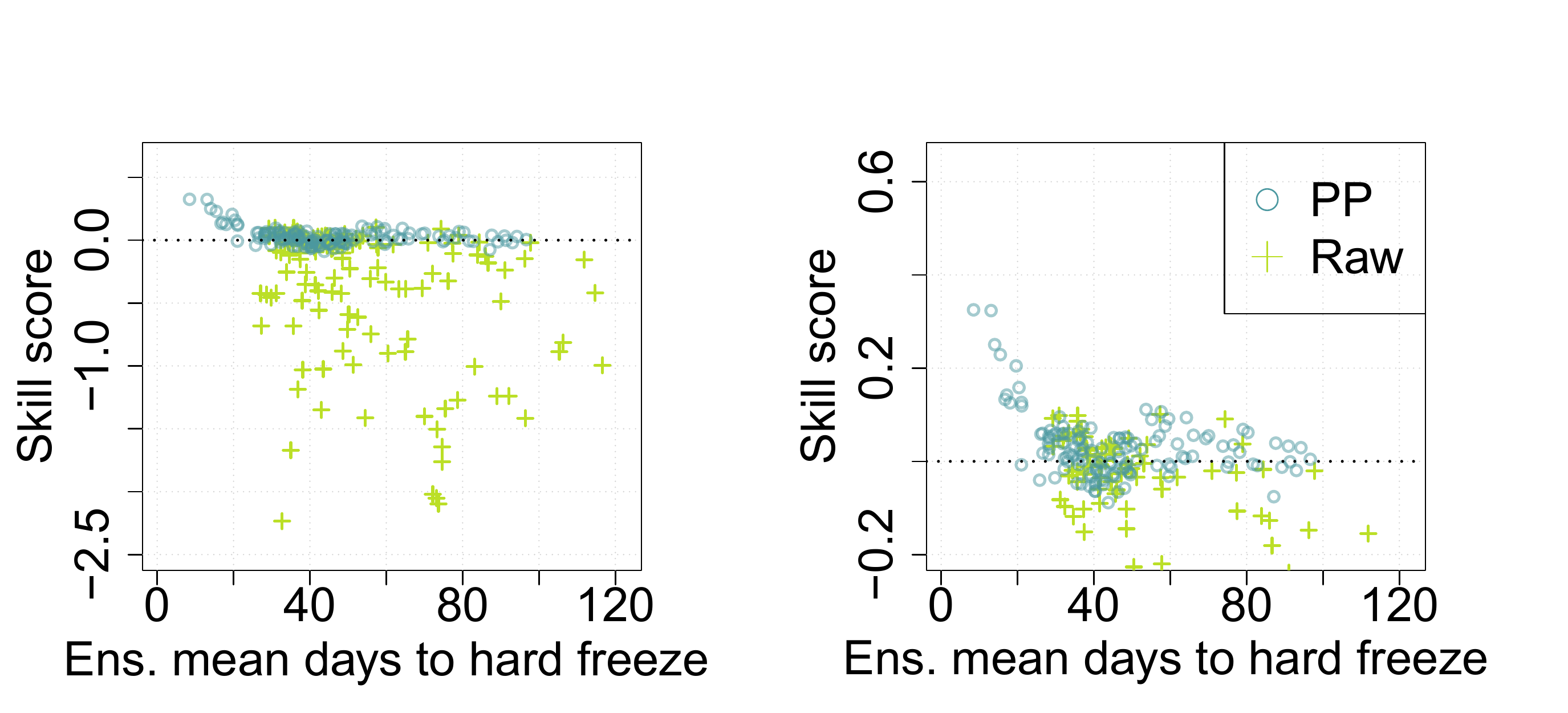}
		\caption{Forecasts initialized on September 1}
		\label{fig:skillsept}
	\end{subfigure}
	\caption{\small Top row: The location-specific skill scores in Figure~\ref{fig:skill} plotted as a function of the locations' mean predicted days to hard freeze. For the raw forecast, the mean predicted days are calculated from the raw ensemble mean for 1993-2020 (crosses), and for the post-processed forecast it is calculated from the post-processed ensemble mean (dots). The right plot is a zoomed-in version of the left plot. Bottom row: Corresponding results for forecasts initialized on September 1.}\label{fig:skill_vs_mean}
\end{figure}

Figure \ref{fig:skilloct} presents the skill score for each of the 135 study locations as a function of the mean predicted days to hard freeze from October 1. Here, the mean predicted days to hard freeze are obtained from the historical ensemble means for 1993-2020 for the raw or the post-processed ensemble, respectively. The figure shows that the skill score of the post-processed forecasts is positive for almost all study locations and that all locations where the average predicted time to hard freeze is less than 40 days have a positive skill score.  The skill of the forecasts decrease with increasing mean predicted time to hard freeze, as expected. However, for many of the locations where the mean predicted time to hard freeze is 15-30 days, the skill score is still quite high (0.10-0.20). 

\begin{figure}
	\centering
	\begin{subfigure}[b]{0.32\textwidth}
		\includegraphics[width=1\linewidth]{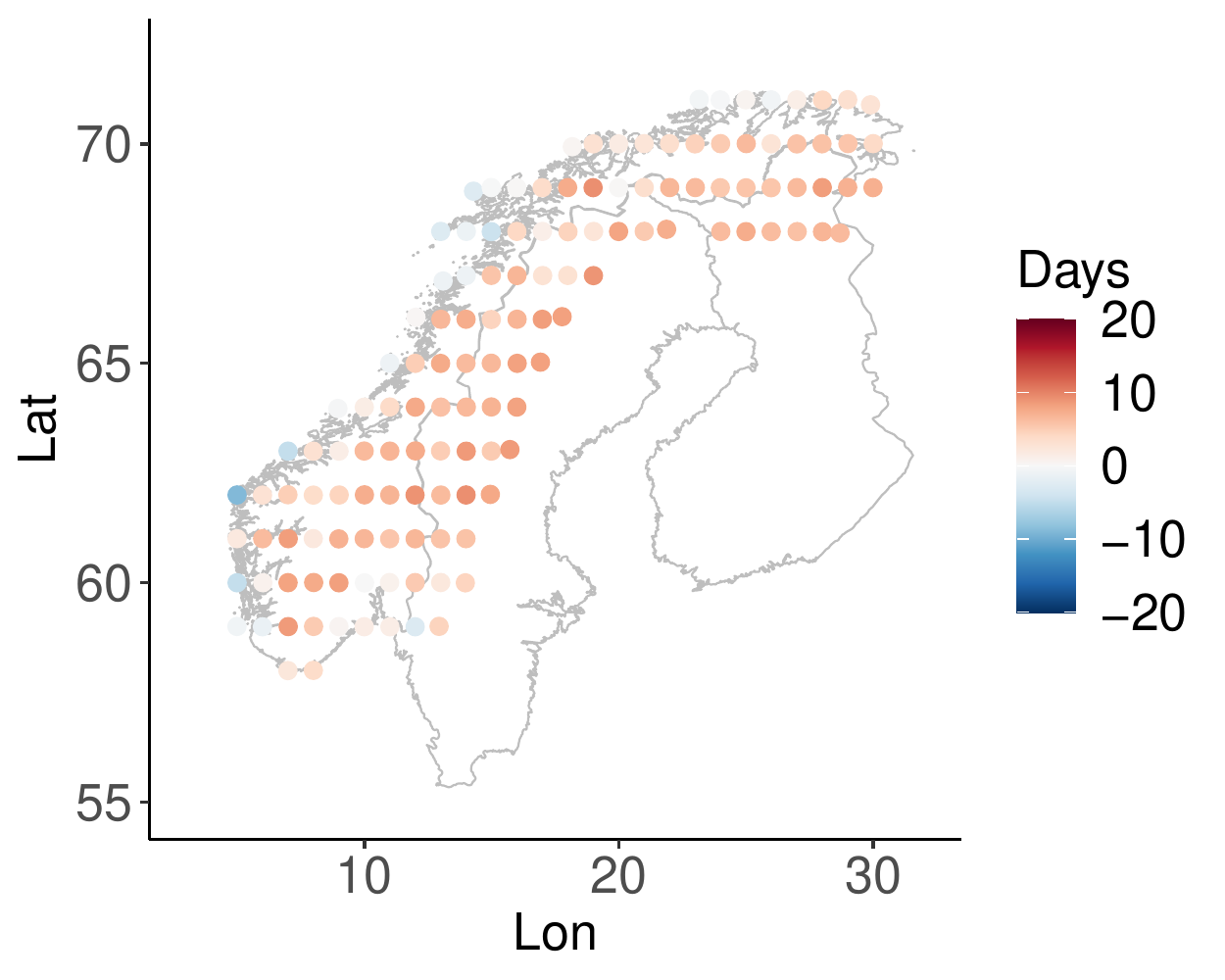}
		\caption{2005: PP - Clim.}
		\label{fig:trendmap2005sfe}
	\end{subfigure}
	\begin{subfigure}[b]{0.32\textwidth}
		\includegraphics[width=1\linewidth]{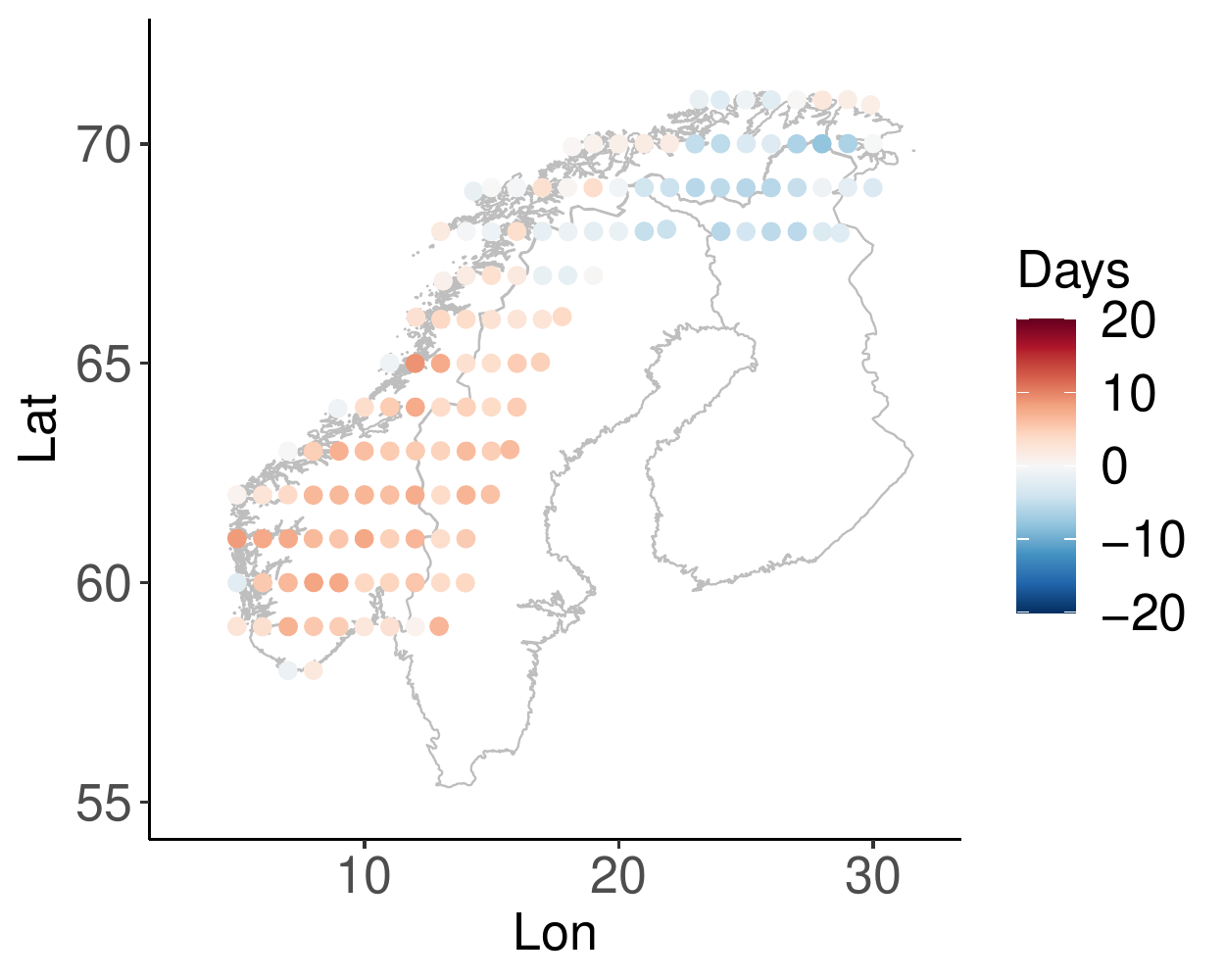}
		\caption{2006: PP - Clim.}
		\label{fig:trendmap2006sfe}
	\end{subfigure}
	\begin{subfigure}[b]{0.32\textwidth}
		\includegraphics[width=1\linewidth]{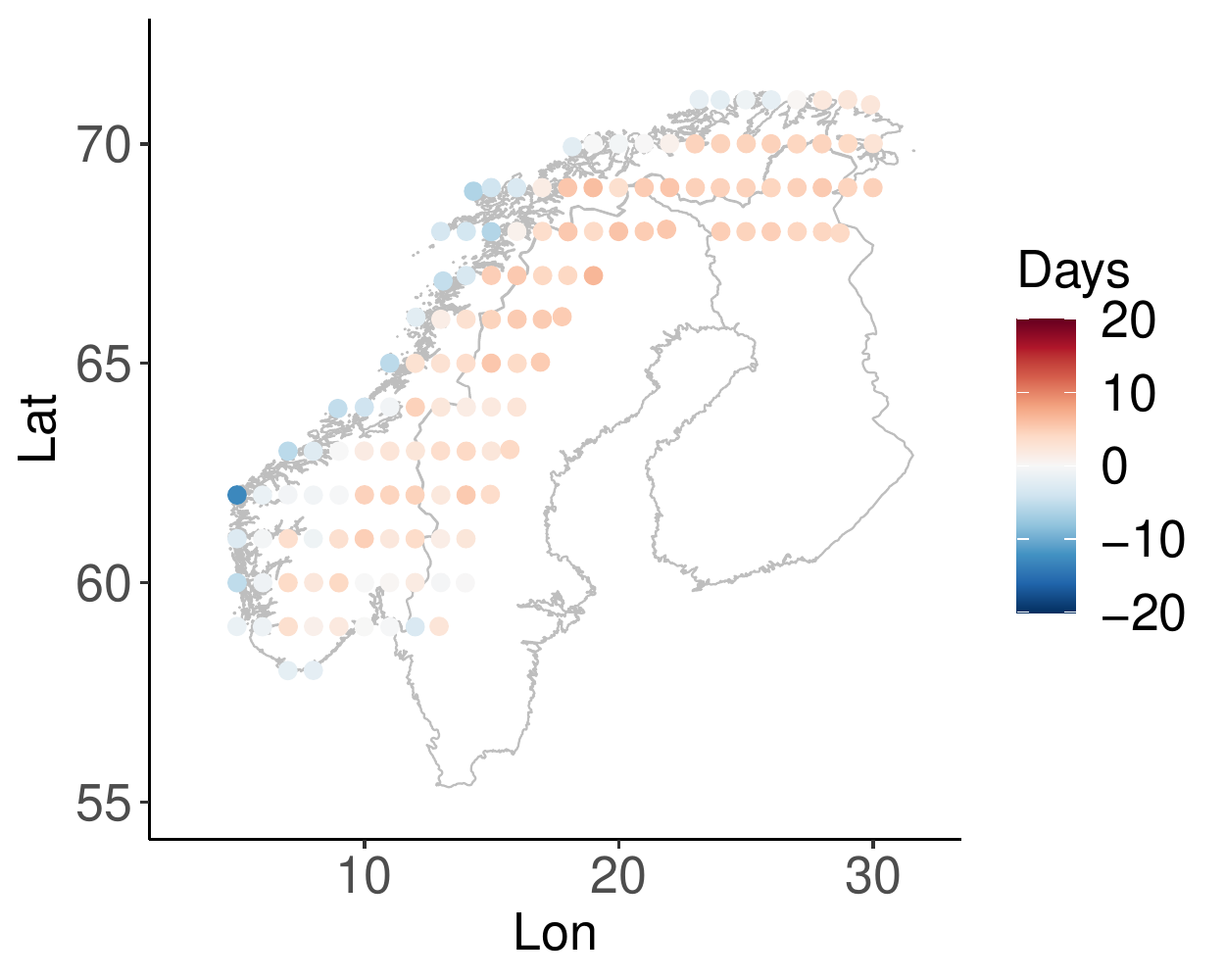}
		\caption{2007: PP - Clim.}
		\label{fig:trendmap2007sfe}
	\end{subfigure}
	\begin{subfigure}[b]{0.32\textwidth}
		\includegraphics[width=1\textwidth]{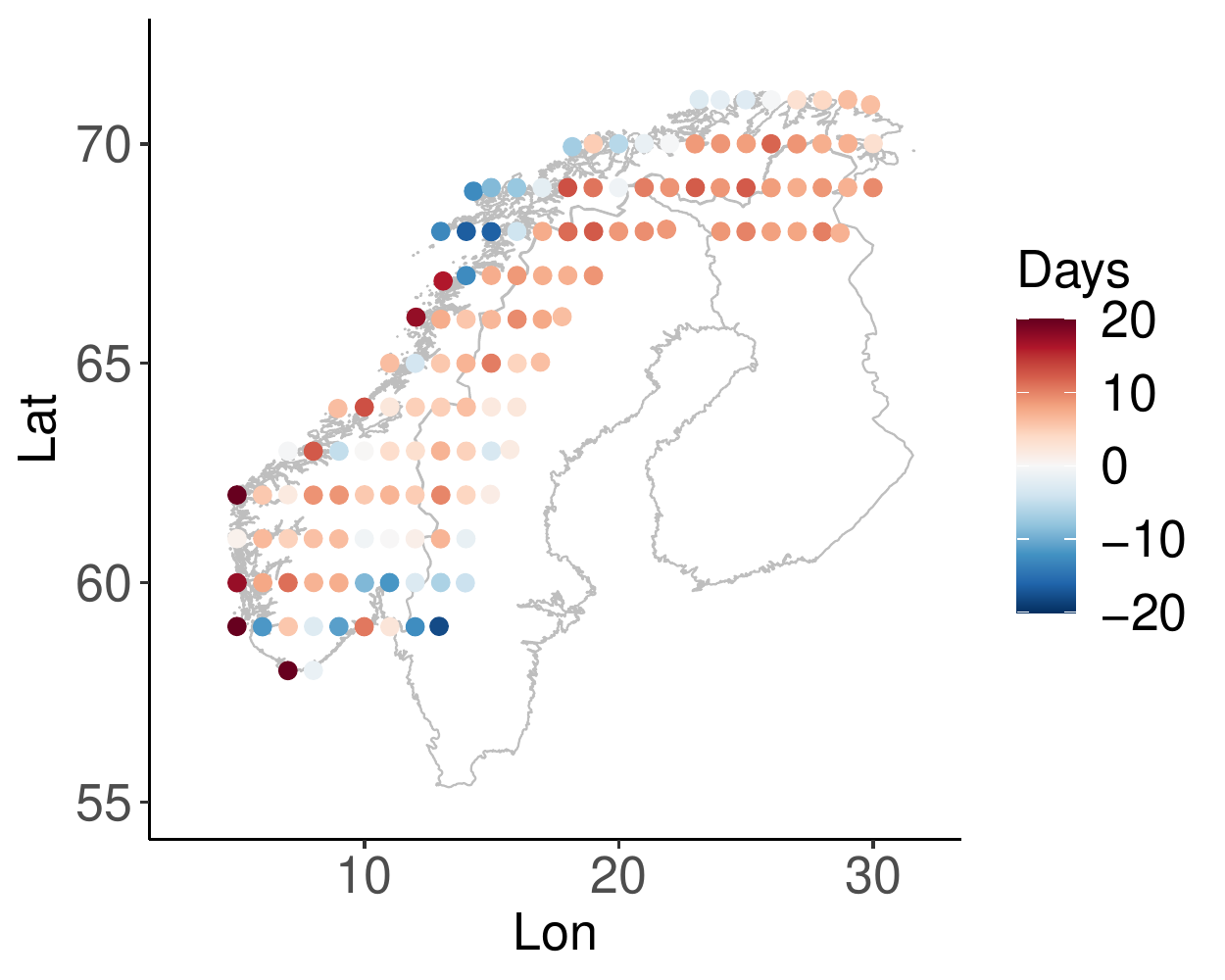}
		\caption{2005: Obs. - Clim.}
		\label{fig:trendmap2005obs}
	\end{subfigure}
	\begin{subfigure}[b]{0.32\textwidth}
		\includegraphics[width=1\textwidth]{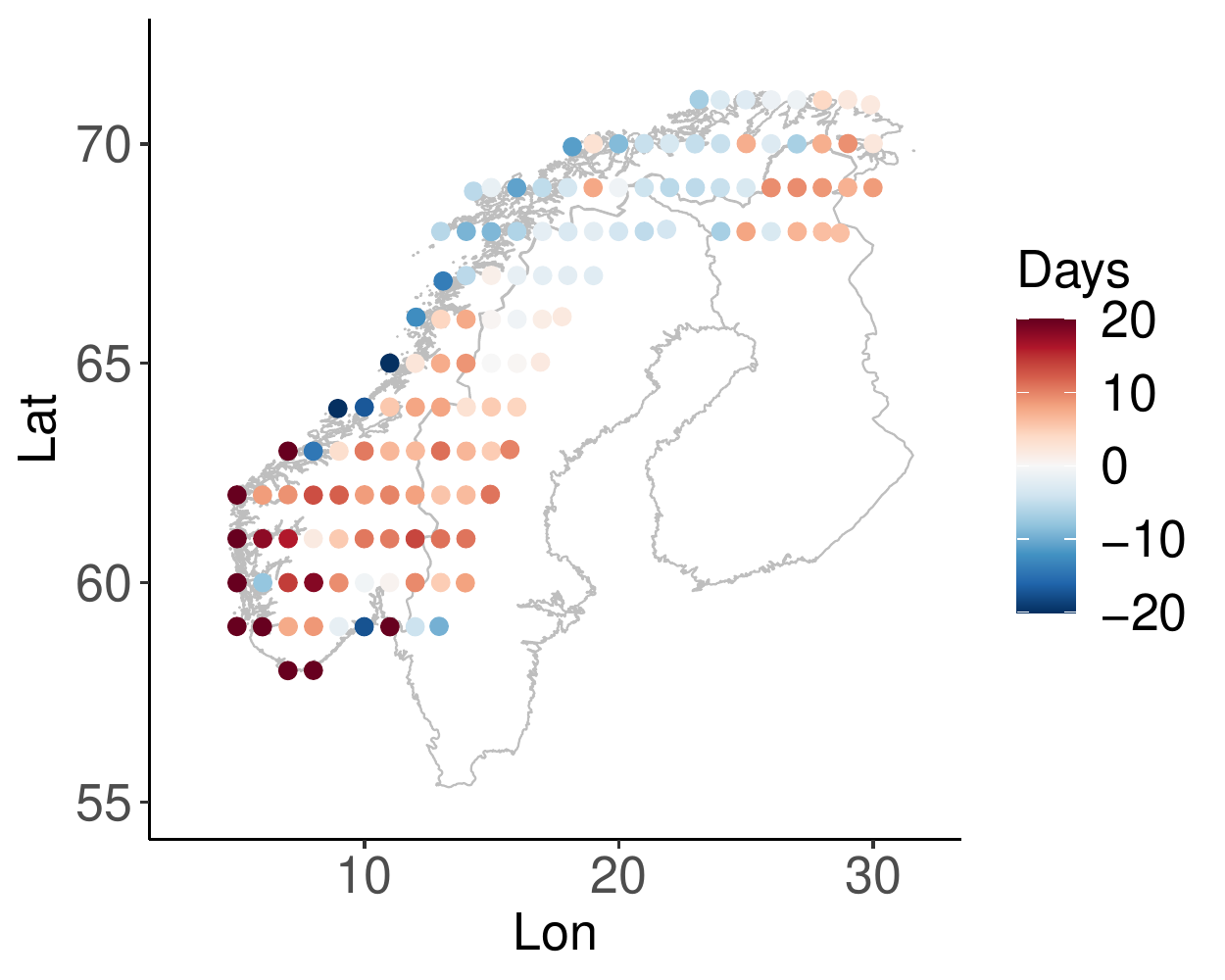}
		\caption{2006: Obs. - Clim.}
		\label{fig:trendmap2006obs}
	\end{subfigure}
	\begin{subfigure}[b]{0.32\textwidth}
		\includegraphics[width=1\textwidth]{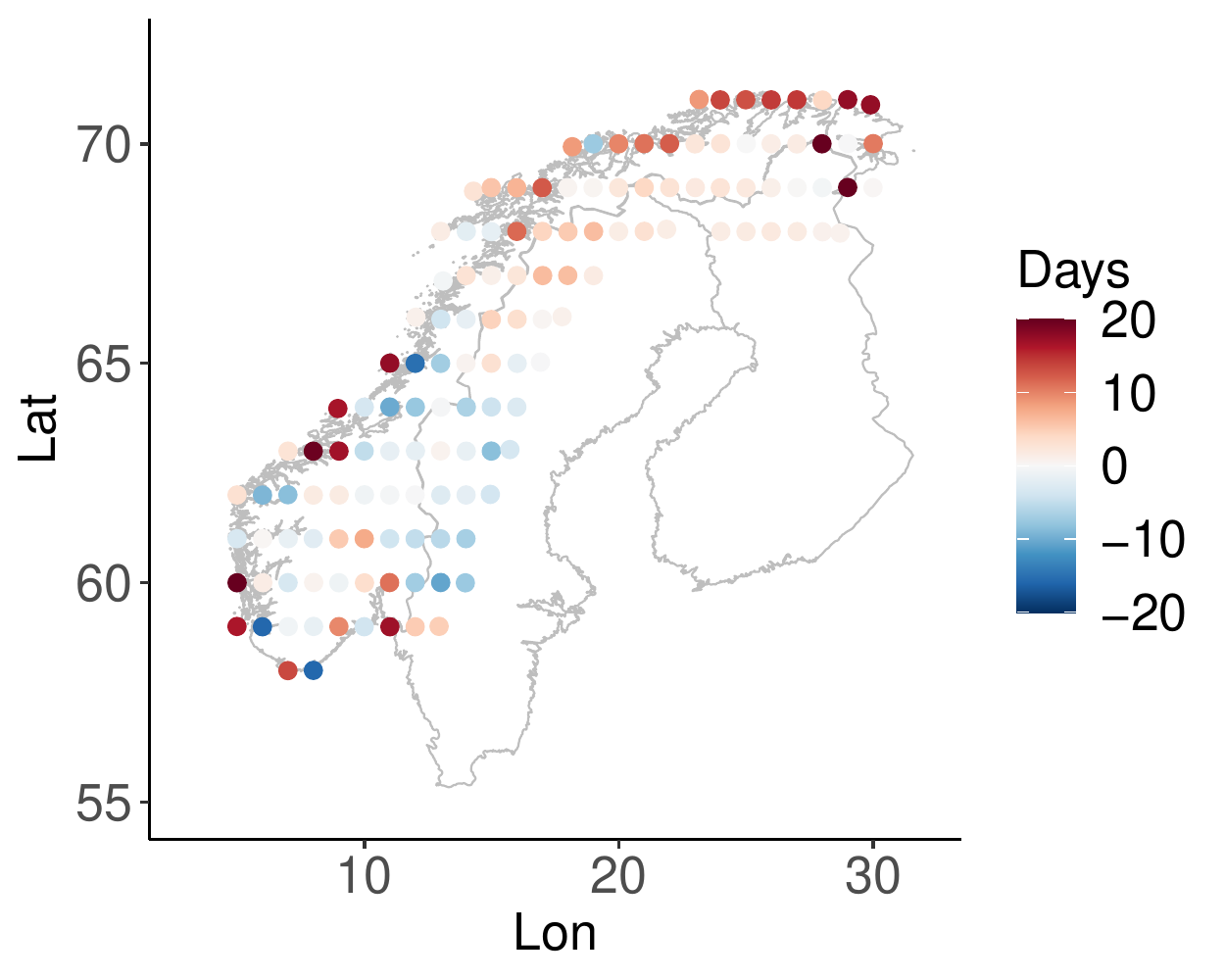}
		\caption{2007: Obs. - Clim.}
		\label{fig:trendmap2007obs}
	\end{subfigure}
	\caption{\small Top row: The deviation of the post-processed ensemble mean number of days to hard freeze in three given years from the climatological average over 1993-2020. Bottom row: Observed anomaly in days to hard freeze in the same three years. Positive values indicate that hard freeze was predicted or observed to come later than on average compared to the reference period.}\label{fig:trendmaps}
\end{figure}

\begin{figure}
	\centering
	\includegraphics[width=0.4\linewidth,trim={0cm 0.6cm 1.2cm 2.5cm},clip]{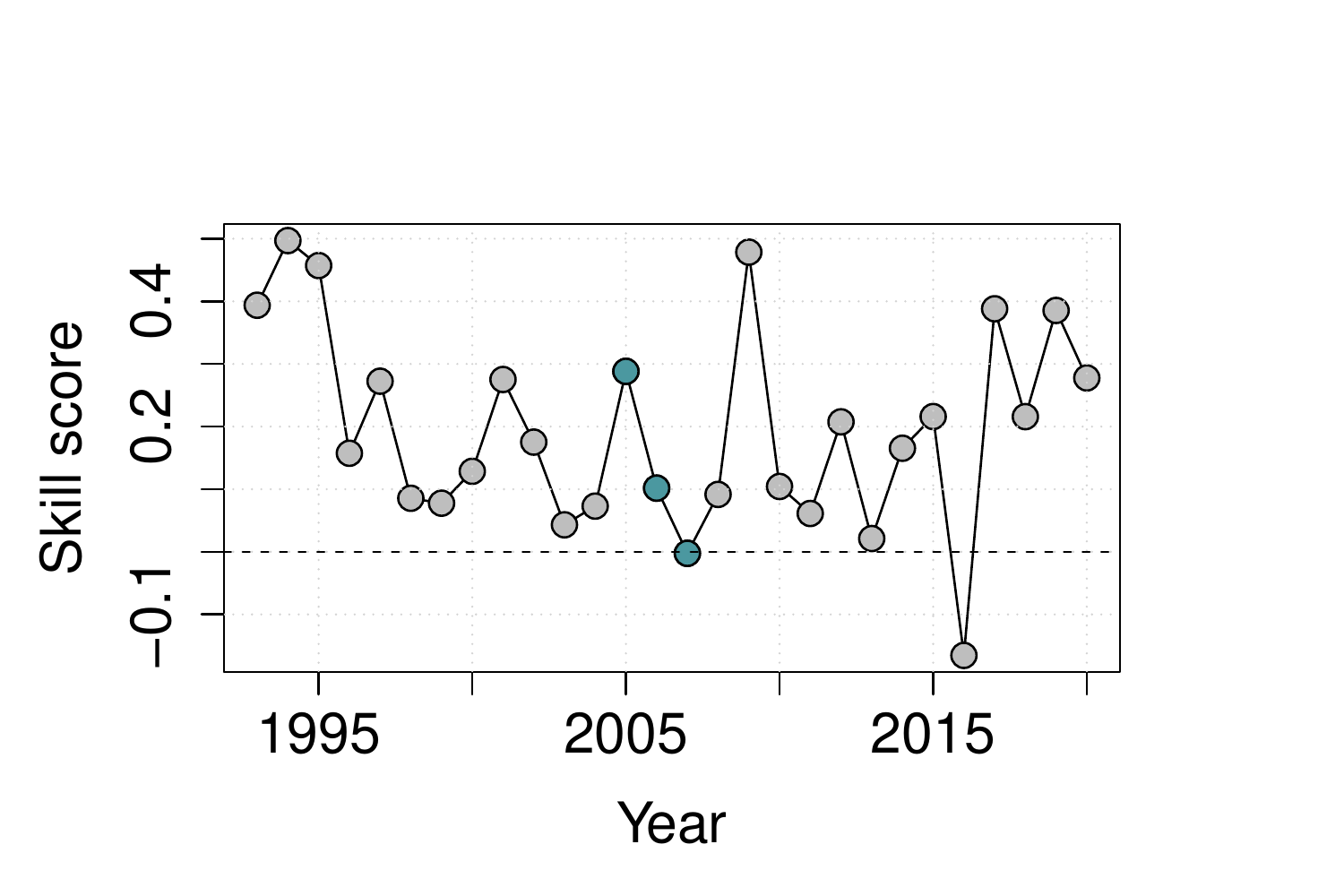}
	\caption{\small Year-specific skill scores for the post-processed forecast (1993-2020). The example years from Figure \ref{fig:trendmaps} are marked.} \label{fig:annualskill}
\end{figure}

For reference, we also include the skill scores we obtain when we use forecasts with initialization date September 1 over the same period 1993-2020, see Figure \ref{fig:skillsept}. The first occurrence of hard freeze from September 1 is usually further away from the initial date than for a counter that starts on October 1. However, the results from the September forecasts are comparable to the results from the October forecasts: For locations where hard freeze is expected within 30 days, the skill score tends to be positive and often high (0-0.35) for the post-processed forecasts. When the predicted time to hard freeze is more than 30 days, the quality of the forecasts is more variable with skill scores distributed around 0.

In Figures \ref{fig:trendmap2005sfe}-\ref{fig:trendmap2007sfe}, we show the difference between the ensemble mean days to hard freeze for our post-processed ensemble and the historically observed mean days to hard freeze (1993-2020) for three selected years: 2005, 2006 and 2007. This is a way to illustrate whether the forecasts predict hard freeze to come earlier or later than usual in different parts of the country.  For example in 2006, the NWP systems predicted the hard freeze to come later than usual in the southern parts of Norway and earlier than usual in the northern parts. In Figures \ref{fig:trendmap2005obs}-\ref{fig:trendmap2007obs}, we visualize what actually happened in 2005-2007 by showing the difference between the observed days to hard freeze and the historically observed mean days to hard freeze (1993-2020). The figures of 2005 and 2006 reveal that while the forecasts were not able to exactly predict the number of days to hard freeze at every single location, they were able to capture the large, general trends. Fall 2005 was indeed warmer than usual for most locations, and in 2006 the hard freeze came late in southern Norway and early in northern Norway as predicted. There are also some years where the overall performance is not as good, as year 2007 show. Comparing skill scores over years shows that 2007 is ranked as the year with the second lowest skill score for the post-processed forecast, with skill score 0. In comparison, the skill scores for years 2005 and 2006 are 0.29 and 0.11 respectively. See Figure \ref{fig:annualskill} for skill scores as a function of year for the post-processed forecasts.

We find more years comparable to the 2005 and 2006 examples in our reference period 1993-2020. In the next subsection we will show some examples of how survival curves can be used as a tool to evaluate whether hard freeze comes early or late locally, in a more probabilistic manner.

\subsubsection{Time to hard freeze forecasts at five selected locations}
We now present example time to hard freeze forecasts for five locations in Norway expressed probabilistically through the KM estimator of the survival curve. The selected locations are close to Bergen, Hamar, Lærdal, Tromsø and Trondheim and are marked in Figure \ref{fig:skillmap_cds}. See Table \ref{tab:loctab} for more information about each location. We also refer to the files attached to the paper, where observations and forecasts for these five locations are included together with code for calculating and plotting the KM estimator for 1993-2020.


\begin{table}[h!]
\caption{Location characteristics and integrated Brier skill scores aggregated over 1993-2020 compared to climatology for the locations marked in Figure \ref{fig:skillmap_cds}. Mean dtf indicates the historical mean days to hard freeze from October 1 over 1993-2020, while sd dtf indicates the corresponding standard deviation based on the seNorge data product.}
\centering
\begin{tabular}{lrrrrrrrr}
\hline 
  Location        & Lon & Lat & seNorge elev. & Mean dtf & Sd dtf &  & Skill  $\S^{R}(t)$  & Skill  $\S^{P}(t)$\\
  &  &  & \multicolumn{1}{c}{(m a.s.l.)} & \multicolumn{1}{c}{(days)} & \multicolumn{1}{c}{(days)} & &  \\ \hline
Trondheim & 11  & 64  & 386             & 20                 & 13               &  & -1.19    & 0.20    \\
Hamar     & 11  & 61  & 535              & 16                 & 10               &  & 0.30     & 0.29    \\
Bergen    & 5   & 60  & 0                & 71                 & 18               &  & -0.13         & 0.00  \\
Lærdal    & 7   & 61  & 1050              & 10                 & 9                &  & 0.42     & 0.43    \\
Tromsø    & 18  & 69  & 391              & 8                  & 6                &  &  -1.08    & 0.52  \\
\hline
\end{tabular}\label{tab:loctab}
\end{table}

\begin{figure}
	\centering
	\includegraphics[width=0.6\linewidth,trim={1cm 5cm 1cm 5cm},clip]{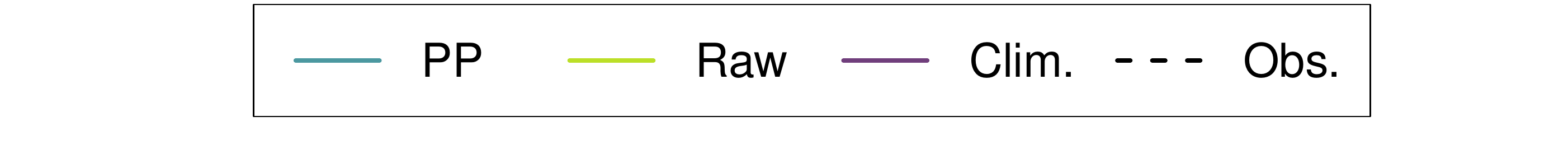}
	\includegraphics[width=0.8\linewidth]{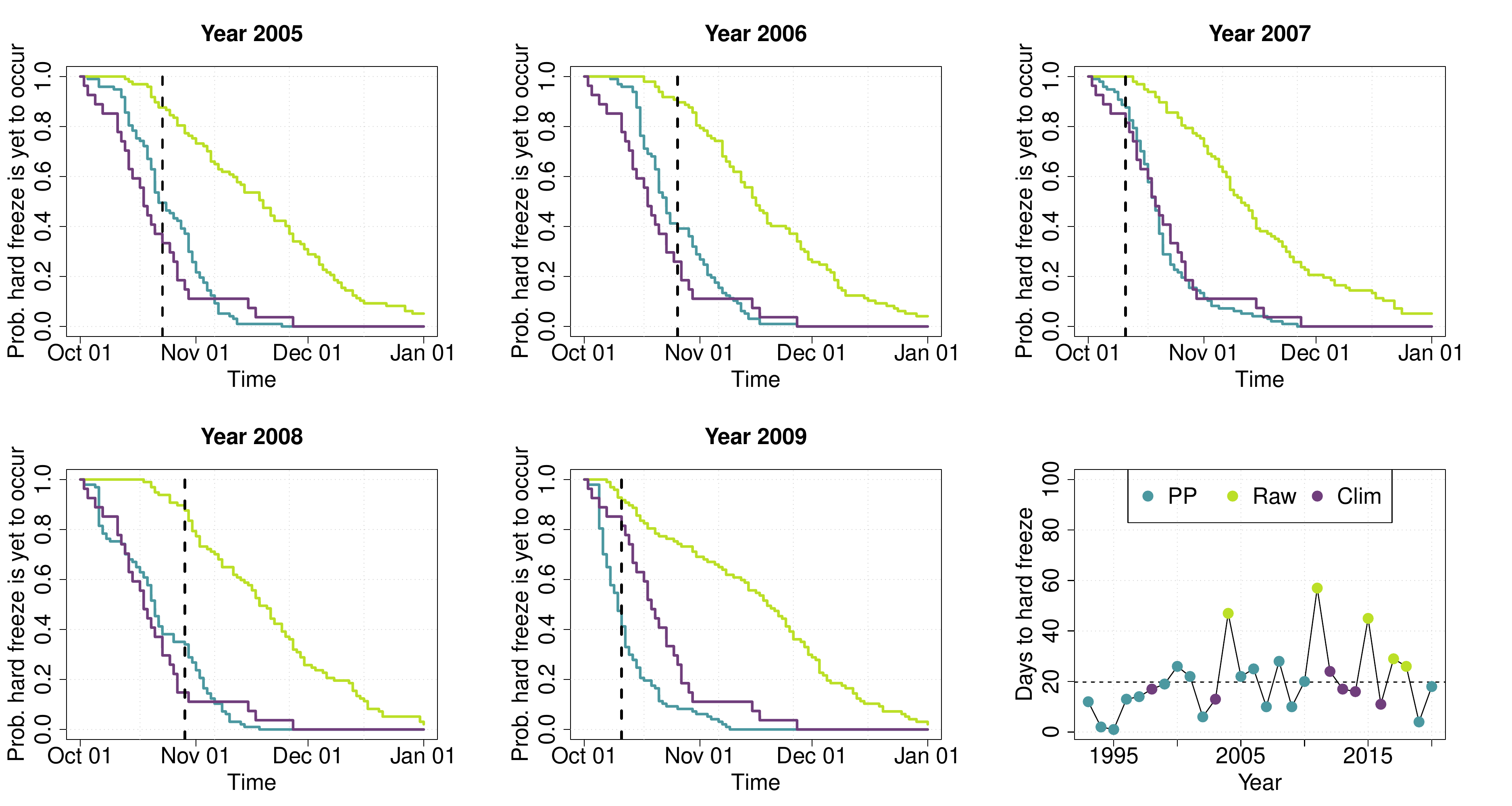}
	\caption{\small Survival curves for days to hard freeze based on the KM estimator for five selected years at a location close to Trondheim, at longitude 11$^\circ$ and latitude 64$^\circ$, with the observed first day of hard freeze indicated with a dashed line. The lower rightmost plot shows the observed number of days to hard freeze for all years in the study period with each observation colored according to the forecast that performed best in that year based on the integrated Brier score.}\label{fig:kmex9}
\end{figure}

Figure \ref{fig:kmex9} presents $\Shat{P}, \Shat{R}$ and $\Shat{C}$ for five example years for the location at longitude 11$^\circ$ and latitude 64$^\circ$, near the Norwegian city Trondheim. This is a region with a high density of farms and where a hard freeze forecast could be useful. We see that  $\Shat{P}$ has high skill, and performs considerably better than $\Shat{R}$. According to Figure \ref{fig:kmex9}, the raw  ensemble often overestimates the time to hard freeze by several weeks and gives an uncertainty that is too large. The forecasts given by the post-processed ensembles are sharper and for many of the study years they correspond quite well to what we observe: The probability that hard freeze has not occurred is often around 0.5 around the time when the first hard freeze actually arrived (dashed line). See years 2005, 2006, 2008 and 2009 in Figure \ref{fig:kmex9}. Also note how comparing  $\Shat{P}$ and $\Shat{C}$ gives information on whether hard freeze can be expected to come earlier or later than usual. In 2009, for example, the post-processed survival curve in general drops faster than the climatological survival curve, suggesting  that  hard freeze can be expected earlier than usual.  This was also what actually happened this year. The last panel in Figure \ref{fig:kmex9} reveals that $\Shat{P}$ gave the best forecast for 17 out of 28 study years. The overall skill scores for $\Shat{R}$ and  $\Shat{P}$ are -1.19 and 0.20 respectively, as summarized in Table \ref{tab:loctab}.

\begin{figure}
	\centering
	\includegraphics[width=0.6\linewidth,trim={1cm 5cm 1cm 5cm},clip]{legend2}
	\includegraphics[width=0.8\linewidth]{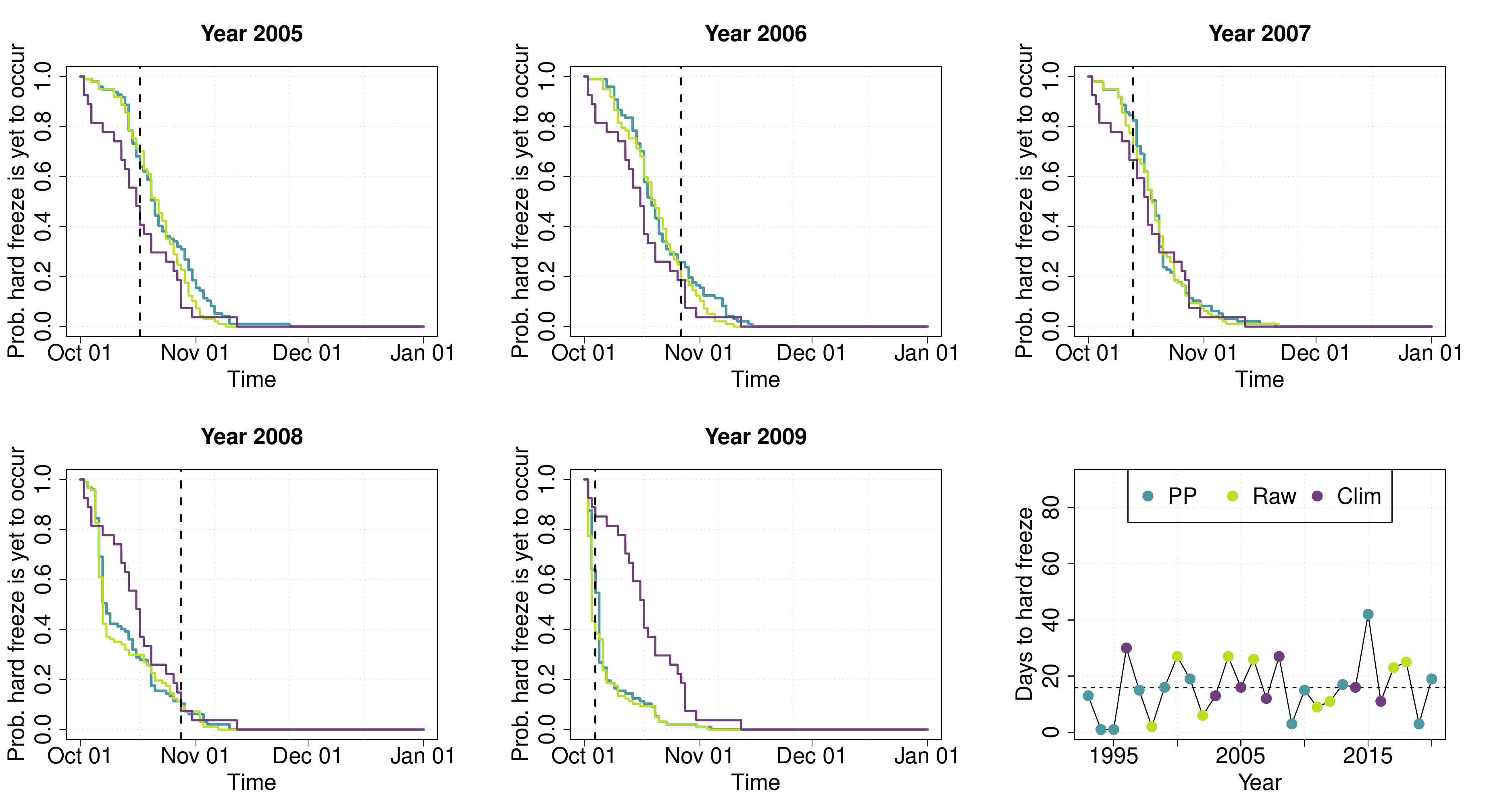}
	\caption{\small  As Figure \ref{fig:kmex9}, but for a location in the Hamar area, at longitude 11$^\circ$ and latitude 61$^\circ$.}\label{fig:kmex3}
\end{figure}

\begin{figure}
	\centering
	\includegraphics[width=0.6\linewidth,trim={1cm 5cm 1cm 5cm},clip]{legend2}
	\includegraphics[width=0.8\linewidth]{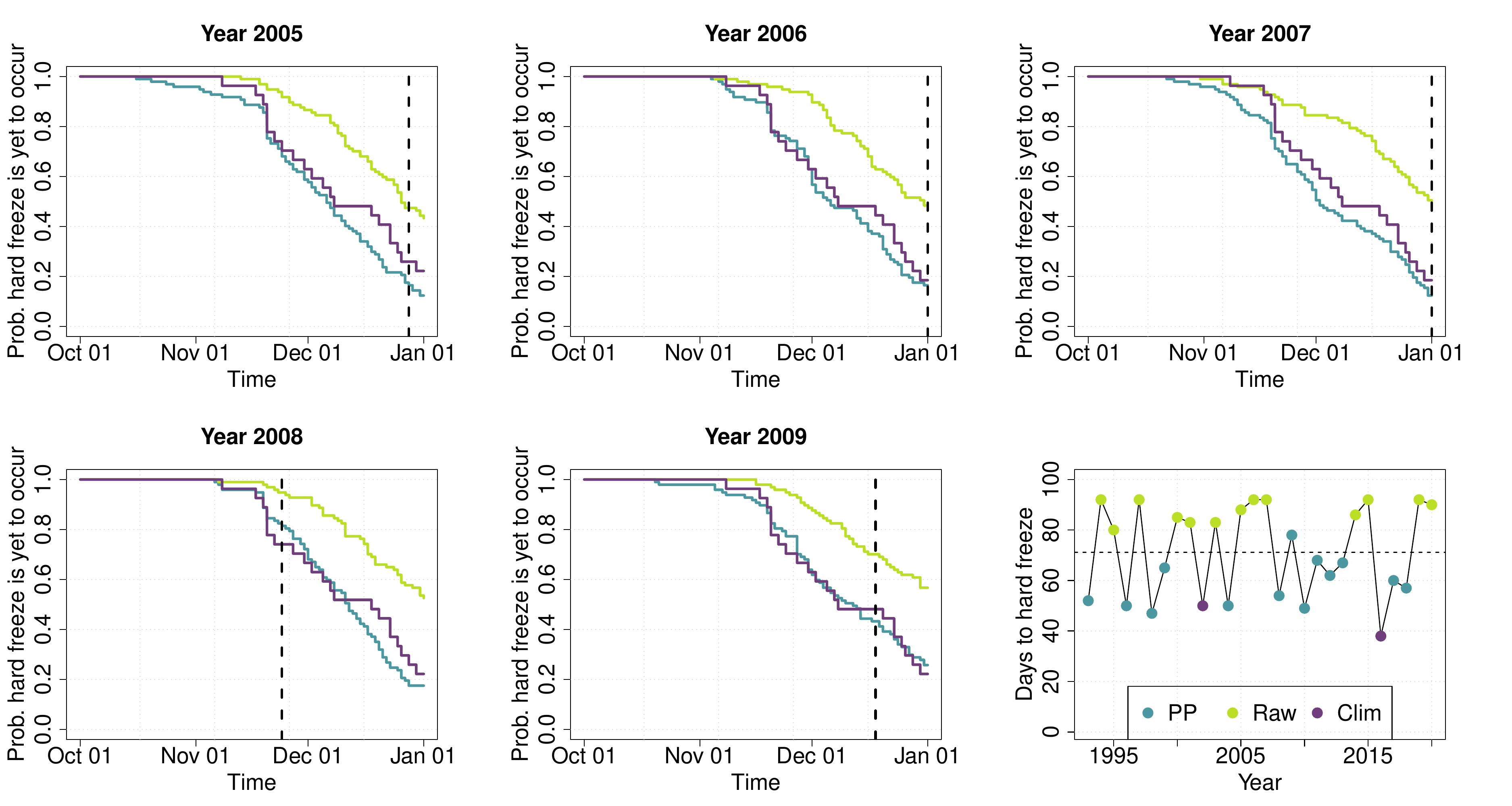}
	\caption{\small  As Figure \ref{fig:kmex9}, but for a location south of Bergen, at longitude 5$^\circ$ and latitude 60$^\circ$.}\label{fig:kmex10}
\end{figure}

We next consider survial curves for the location in the Hamar area, at longitude 11$^\circ$ and latitude 61$^\circ$, see Figure  \ref{fig:kmex3}. The Hamar region is another area of Norway where we find many farms and where a forecast for the days to hard freeze could be valuable. The hard freeze comes 16 days after October 1st on average according to the seNorge data product. We see that $\Shat{P}$  and  $\Shat{R}$ give more similar forecasts here compared to what we saw for Trondheim. Both $\Shat{P}$  and  $\Shat{R}$ seem to capture the probability of hard freeze occurrence quite well, with probabilities around 0.5 in the time periods where hard freeze occurred historically. This can e.g. be seen for years 2005, 2007 and 2009 in Figure \ref{fig:kmex3}. According to Table \ref{tab:loctab}, the overall skill scores of $\Shat{P}$  and  $\Shat{R}$ are 0.29 and 0.30 respectively. We see similar trends as in Hamar for many of the locations in the interior parts of southern Norway and Sweden. The year-to-year variability in the number of days to hard freeze is in general low (Figure \ref{fig:sdmap}) and the forecast skill is high (Figure \ref{fig:skill}).

Figure \ref{fig:kmex10} presents the survival curves for a location where the forecasts are less skillful. The selected location is at longitude 5$^\circ$ and latitude $60^\circ$, which is on the  west coast of Norway, south of Bergen. The mean number of days to hard freeze, from October 1, is 71 days, and Figure \ref{fig:kmex10} shows that the variability in the predicted time to hard freeze is also large. The shapes of $\Shat{P}$  and  $\Shat{R}$ tell us that the next hard freeze can come almost any time between day 14 and day 92 for some of the example years. The number of days to hard freeze varies a lot historically too, which we see from the shape of the baseline model $\Shat{C}$. For some of the historical years, hard freeze is not observed before the end of our study period on December 31. In general, $\Shat{R}$ predicts that hard freeze comes later than  $\Shat{C}$. The overall skill scores of $\Shat{R}$ and $\Shat{P}$ are -0.13 and 0.00 respectively (Table \ref{tab:loctab}), which means that $\Shat{P}$ on average is approximately as accurate as the climatological model $\Shat{C}$. We see large deviance between the raw and post-processed forecasts for many of the locations along the Norwegian coast. Further investigation reveals that the raw forecast can both under- and overestimate the time to hard freeze. This was also indicated by the results for the daily temperature forecasts shown in Figure \ref{fig:PITbias}.

\begin{figure}
	\centering
	\includegraphics[width=0.6\linewidth,trim={1cm 5cm 1cm 5cm},clip]{legend2}
	\begin{subfigure}[b]{0.48\textwidth}\centering
		\includegraphics[width=1\linewidth]{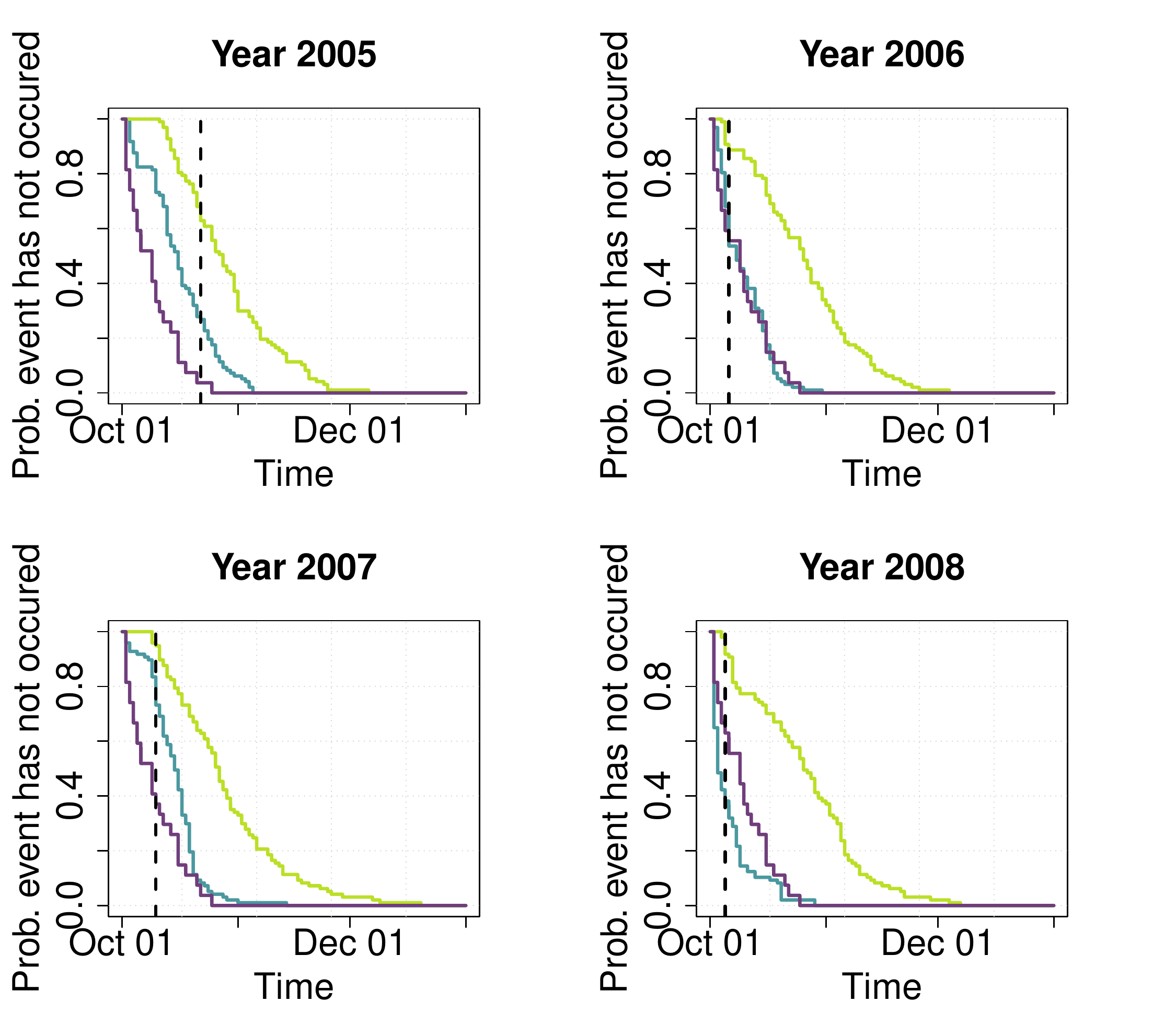}
		\caption{\small Tromsø.}
		\label{fig:km_ex4}
	\end{subfigure}\hspace{2mm}
	\begin{subfigure}[b]{0.48\textwidth}\centering
		\includegraphics[width=1\textwidth]{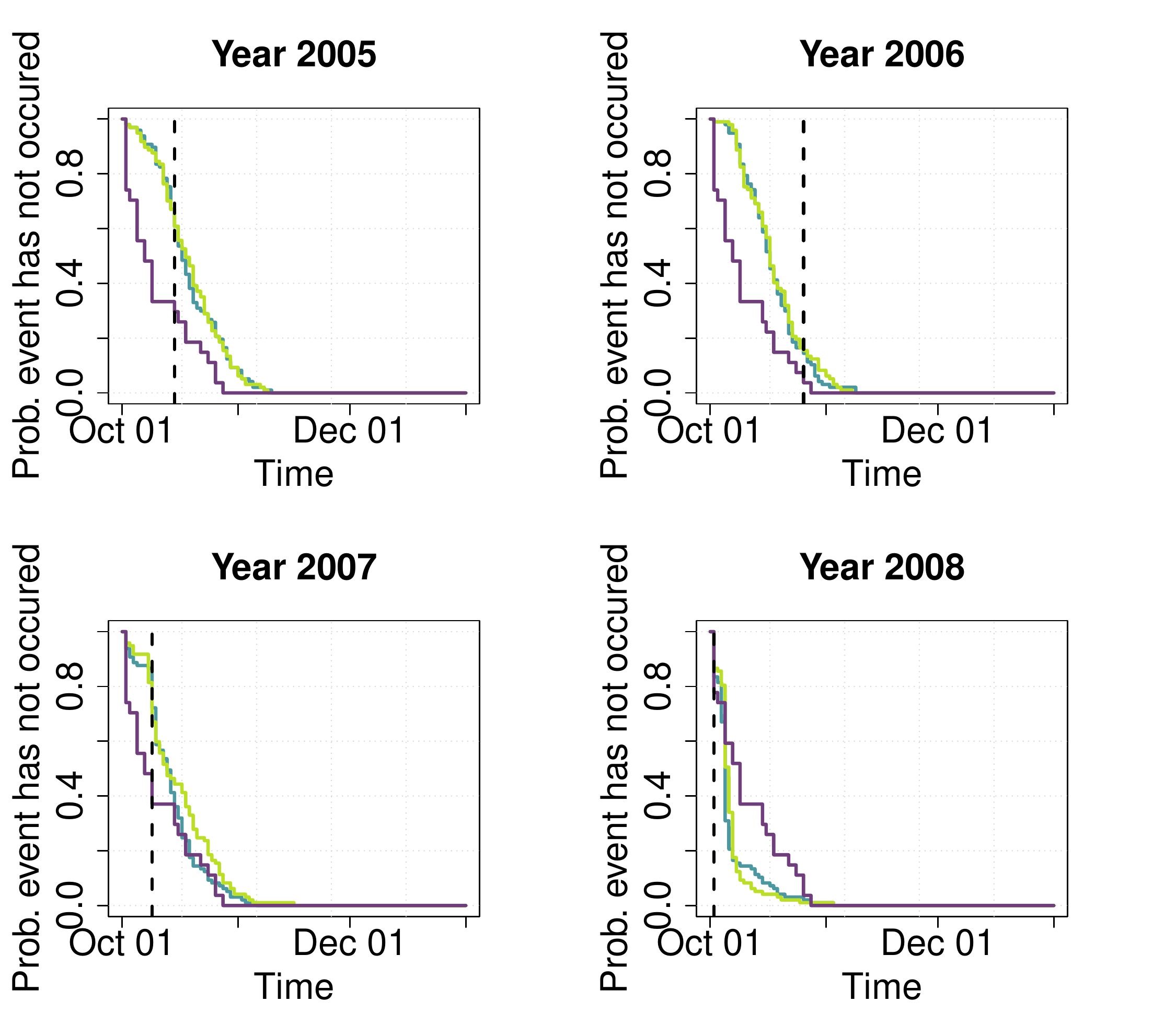}
		\caption{\small Lærdal.}
		\label{fig:days_ex5}
	\end{subfigure}
	\caption{\small  As Figure \ref{fig:kmex9},  but for a location in the Tromsø area (longitude 18$^\circ$, latitude 69$^\circ$) and close to Lærdal (longitude 7$^\circ$, latitude 61$^\circ$).}\label{fig:kmcold}
\end{figure}

In Figure \ref{fig:kmcold}, survival curves for the Tromsø and Lærdal area are presented. Tromsø is located far north in Norway, while the location close to Lærdal is in the interior of southern Norway, at elevation  1050 m a.s.l.  The skill of the post-processed forecast  $\Shat{P}$ is high for both locations, with skill scores 0.52 and 0.43, respectively. Most years,  $\Shat{C}$ and/or $\Shat{P}$ have probabilities around 0.5 around the time when the first hard freeze after October 1 was observed. The forecasts also often correctly identify whether the hard freeze comes earlier or later than usual. See e.g.  $\Shat{P}$ for Tromsø in 2005 and for Lærdal in 2006. 

The high skill for Lærdal and Tromsø can be explained by the early hard freeze, on average 8-10 days after October 1.  In some years, hard freeze has already occurred by the initialization date. This happens in Tromsø in 2006 and 2008, and at Lærdal in 2008 (Figure \ref{fig:kmcold}). The trends we see for Tromsø and Lærdal can also be seen for other locations north in the study area and in the mountainous areas in southern Norway. A difference between Tromsø and Lærdal is that for Lærdal, the raw forecast performs approximately as well as the post-procssed forecast, see Table \ref{tab:loctab}. Around Tromsø, the raw forecast gives an overall skill score around -1. This can be explained by that Tromsø is close to the coast. In coastal areas the skill of the raw forecast is poor in general as Figure \ref{fig:skillmap_cds} shows. The results for $\Shat{P}$ for Tromsø (and also for  e.g. Bergen and Trondheim) however illustrates that the post-processing algorithm is able to adjust the coarse  NWP forecasts such that they become skillful locally.

 \begin{figure}[h!!]
	\centering
	\includegraphics[width=0.7\linewidth]{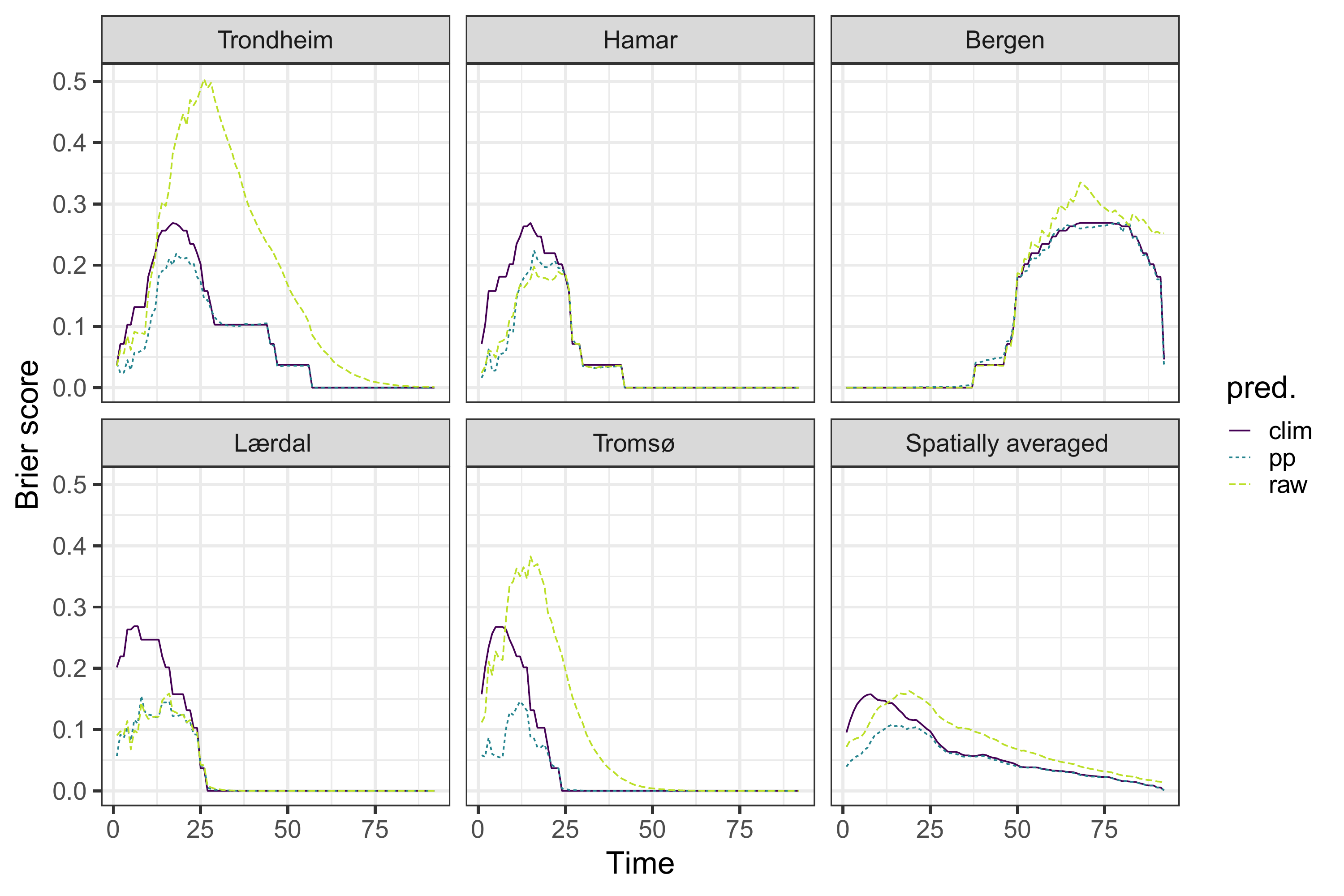}
	\caption{Mean Brier scores over the study period 1993-2020 at the five example locations indicated in Figure \ref{fig:skillmap_cds} and spatially averaged over all study locations, given as functions of lead time for the three time to freeze forecasts initialized on October 1. }\label{fig:BrierScores}
\end{figure}

While the skill scores considered in Section \ref{sec:Overall performance} assess the overall fit of the predicted time to hard freeze, we can gain some insights into how the error varies with the prediction time horizon by plotting the average Brier score $BS(t)$ as a function of $t$. Figure \ref{fig:BrierScores} shows the Brier scores for the five example locations considered in this Section, as well as $BS(t)$ averaged across all considered grid points shown in Figure \ref{fig:seNorge}.

The post-processed predictions consistently perform best across all examples and lead times, with near-climatological skill at long lead times. This is quite remarkable, since the post-processing technique is rather simple and was applied to surface temperature forecasts, not to the prediction target, i.e. the time to hard freeze.
 At first glance, it might seem counter-intuitive that the prediction error decreases after a certain lead time. However this is simply due to the fact that we predict whether hard freeze has occurred at any time up to time $t$, not whether it occurred at time $t$. 
 At long lead times the score converges to zero, because the climatological uncertainty of the predicted event goes to zero, which can be formalized by the score decompositions used in \citet{MurphyWinkler1987} and \citet{Broecker2009}.
  In particular, the time horizon when the score of the climatological model drops to zero highlights the latest occurrence of the first hard freeze present in our data at that location. For Bergen, the climatological model has a Brier score of 0 up to $t = 38$, highlighting that hard freeze never occurred before this time in our data.


\section{Discussion}\label{sec:discussion}
In this paper we propose to use techniques from survival analysis in combination with long-range numerical weather predictions to generate probabilistic forecasts for the time to the next hard freeze from an initialization date. The forecasts were evaluated for a region in Fennoscandia that covers Norway and parts of Sweden, Finland and Russia. The time to the next hard freeze is an example of an agroclimatic indicator that is highly relevant for agricultural operations planning. However, it cannot be derived from currently available seasonal forecast products which typically come in the form of averages over one or several months. This kind of aggregation is usually necessary for predictions to be skillful at a seasonal time scale. A priori, it is thus not clear whether model output from long-range NWP systems contains information that can be used to make skillful prediction of agroclimatic indicators like the time to the first hard freeze. The results presented in Section \ref{sec:results} suggest that at locations where the climatological average time to the first hard freeze is less than 40 days, skillful probabilistic predictions can indeed be obtained with our framework of post-processed, multi-model ensemble weather predictions.

The current time to hard freeze predictions are just one example of a wide range of meteorological time-to-event indicators that may fit well within the general survival time analysis framework. Other examples include the time to next drought, the time to the onset of the rain season, or the time to the next sudden stratospheric warming. Here, the general idea is to treat each ensemble member as one would treat an individual in a classical survival analysis setting. As far as the authors know, this is a new way to apply survival time models. Output from several NWPs can be combined to achieve a sufficiently large sample size for statistical analyses at separate locations. The survival curve examples in Section~\ref{sec:results} illustrate how the shape of the survival function provide an intuitive way of presenting the uncertainty of a time-to-event forecast and the probability of an event occurring earlier or later than usual.

Now that the applicability of survival models in long range weather forecasting is demonstrated, there exists a large collection of methods that can be used to build more advanced models for the time to hard freeze or other meteorological events of interest. A natural extension of the hard freeze analysis, is to replace the non-parametric KM estimator by a parametric or semi-parametric survival function. This can be achieved by using a proportional hazards model. In a proportional hazards model, the time to an event $t$ is modeled through a hazard rate $h(t)$ that expresses the probability of experiencing the event during the next instant of time after time $t$, given that the event has not already occurred. The hazard rate is given as a function of explanatory variables that are multiplicative with respect to a baseline hazard function $h_0(t)$, for example as $h(t|\boldsymbol{x}_i)=h_0(t)\times \exp(b_1 x_{i1} + ,...,+b_p x_{ip})$, where $\boldsymbol{x}_i=(x_{i1},...,x_{ip})$  is a vector of $p$ covariates associated with individual (or ensemble member) $i$, and $h_0(t)$ is a parametric or non-parametric model. The above form of  $h(t|\boldsymbol{x}_i)$ is known as a Cox proportional hazards model \citep{Cox1972} and has been used in a variety of applications \citep[e.g.][]{BankCox,KUMAR1994177}.

In a hard freeze forecasting context, a proportional hazard model can be applied to build a joint model for all study locations using explanatory variables like longitude, latitude and elevation. From the parametric relationship, predictions for any location can be derived. If we want to take the model a step further, one could add random effects to the exponential part of $h(t|\boldsymbol{x}_i)$, e.g. a Gaussian random field that captures spatial and/or temporal correlation \citep{SaraSurvival,spatialSurvival4,spatialSurvival2,INLASurv}. A fundamental assumption in the proportional hazard model is that the hazard ratio of two individuals is independent over time. If this is inappropriate, one can consider a Cox model with time dependent regression coefficients \citep{nonPropHazard} or an accelerated failure time model that assumes that the effect of a covariate increases or decreases during the study period by some constant \citep{AFT1,AFT2}. There also exist survival models where the hazard rate is estimated by using neural networks or other machine learning techniques \citep{Faraggi,deepSurvival}. 

In the hard freeze study, we use the ensemble members from long-range NWP systems as input to our survival  model. These were easily accessible and provided around 100 ensemble members per location and year.  A disadvantage with using the CDS seasonal forecasts, is that they are only produced once a month and published around 12 days after their initialization date. This makes it challenging to operationalize the forecast, particularly as our results indicate that the forecasts mainly have high skill in areas where the first hard freeze comes less then 4 weeks after the initialization date. As an alternative to the seasonal forecasts, one could use subseasonal forecasts, e.g. the extended-range forecasts from ECMWF that have lead time 0-46 days and are published twice a week. The drawback is that their hindcasts only have 11 ensemble members, which hampers model validation. Furthermore, with a maximum lead time of 46 days we also lose the opportunity to study the future daily temperature to a point where it has dropped to zero at almost every location in Norway. The last is feasible with the seasonal forecasts with initialization date of September 1 or October 1.

The overlapping release of forecast products, the varying length of their lead times, and the often substantially different ensemble sizes of each product poses a critical operational challenge in general, and for hard freeze modeling in particular. Substantial effort will now be placed on the problem of data fusion, where a long-range forecast is updated on a daily basis to incorporate new observations and new forecasts across all time scales. One of the motivating reasons for choosing the survival time framework was that it is highly amenable to incorporate these concepts.  In the future, we envision a system which dynamically updates survival time forecasts on the basis of observations and updates to forecasting systems.

When making forecasts for the agricultural sector, it is important to also consider the most reasonable definition of the relevant meteorological event for the particular application. In the case of hard freeze, the definition should depend on the type of crop we are interested in and the temperatures that are harmful for that specific culture. Potatoes protected by the soil can e.g. handle a lower temperature than berries or fruit. For some crops it might be the minimum temperature during a day that matters, while for others it could be the mean temperature over several days. Luckily, any such definition fits into the survival time framework outlined in this paper. The models can also easily be rephrased to consider the time to the last hard freeze or frost before spring.

\section*{Acknowledgements}
The authors thank Ellen-Margrethe Hovland, Trine Thanh Ha and Stian Bj{\o}rntvedt for helpful discussions. This work was supported by the Research Council of Norway through grants 309562 "Climate Futures" and 270733 "Seasonal Forecasting Engine".


{\footnotesize
	\bibliographystyle{abbrvnat}
	\bibliography{sample.bib}}	

\begin{thebibliography}{61}
\providecommand{\natexlab}[1]{#1}
\providecommand{\url}[1]{\texttt{#1}}
\expandafter\ifx\csname urlstyle\endcsname\relax
  \providecommand{\doi}[1]{doi: #1}\else
  \providecommand{\doi}{doi: \begingroup \urlstyle{rm}\Url}\fi

\bibitem[An-Vo et~al.(2019)An-Vo, Mushtaq, Reardon-Smith, Kouadio, Attard,
  Cobon, and Stone]{An-Vo&2019}
D.-A. An-Vo, S.~Mushtaq, K.~Reardon-Smith, L.~Kouadio, S.~Attard, D.~Cobon, and
  R.~Stone.
\newblock Value of seasonal forecasting for sugarcane farm irrigation planning.
\newblock \emph{European journal of agronomy}, 104:\penalty0 37--48, 2019.

\bibitem[Angel et~al.(2017)Angel, Widhalm, Todey, Massey, and
  Biehl]{Angel&2017}
J.~R. Angel, M.~Widhalm, D.~Todey, R.~Massey, and L.~Biehl.
\newblock The {U2U} corn growing degree day tool: Tracking corn growth across
  the {US Corn Belt}.
\newblock \emph{Climate Risk Management}, 15:\penalty0 73--81, 2017.

\bibitem[Aswi et~al.(2020)Aswi, Cramb, Duncan, Hu, White, and
  Mengersen]{spatialSurvival4}
A.~Aswi, S.~Cramb, E.~Duncan, W.~Hu, G.~White, and K.~Mengersen.
\newblock Bayesian spatial survival models for hospitalisation of dengue: A
  case study of wahidin hospital in makassar, indonesia.
\newblock \emph{International Journal of Environmental Research and Public
  Health}, 17\penalty0 (3), 2020.
\newblock ISSN 1660-4601.
\newblock \doi{10.3390/ijerph17030878}.

\bibitem[Ata~Tutkun and Tekin(2007)]{nonPropHazard}
N.~Ata~Tutkun and M.~Tekin.
\newblock {C}ox regression models with nonproportional hazards applied to lung
  cancer survival data.
\newblock \emph{Hacettepe Journal of Mathematics and Statistics}, 36:\penalty0
  157--167, 01 2007.

\bibitem[Belu\v{s}i\'c et~al.(2020)Belu\v{s}i\'c, de~Vries, Dobler, Landgren,
  Lind, Lindstedt, Pedersen, S\'anchez-Perrino, Toivonen, van Ulft, Wang,
  Andrae, Batrak, Kjellstr\"om, Lenderink, Nikulin, Pietik\"ainen,
  Rodr\'{\i}guez-Camino, Samuelsson, van Meijgaard, and Wu]{seNorgeApplic}
D.~Belu\v{s}i\'c, H.~de~Vries, A.~Dobler, O.~Landgren, P.~Lind, D.~Lindstedt,
  R.~A. Pedersen, J.~C. S\'anchez-Perrino, E.~Toivonen, B.~van Ulft, F.~Wang,
  U.~Andrae, Y.~Batrak, E.~Kjellstr\"om, G.~Lenderink, G.~Nikulin, J.-P.
  Pietik\"ainen, E.~Rodr\'{\i}guez-Camino, P.~Samuelsson, E.~van Meijgaard, and
  M.~Wu.
\newblock Hclim38: a flexible regional climate model applicable for different
  climate zones from coarse to convection-permitting scales.
\newblock \emph{Geoscientific Model Development}, 13\penalty0 (3):\penalty0
  1311--1333, 2020.
\newblock \doi{10.5194/gmd-13-1311-2020}.

\bibitem[Brier(1950)]{Brier1950}
G.~W. Brier.
\newblock Verification of forecasts expressed in terms of probability.
\newblock \emph{Monthly weather review}, 78\penalty0 (1):\penalty0 1--3, 1950.

\bibitem[Br{\"o}cker(2009)]{Broecker2009}
J.~Br{\"o}cker.
\newblock Reliability, sufficiency, and the decomposition of proper scores.
\newblock \emph{Quarterly Journal of the Royal Meteorological Society},
  135\penalty0 (643):\penalty0 1512--1519, 2009.

\bibitem[Cox(1972)]{Cox1972}
D.~R. Cox.
\newblock Regression models and life-tables.
\newblock \emph{Journal of the Royal Statistical Society: Series B
  (Methodological)}, 34\penalty0 (2):\penalty0 187--202, 1972.
\newblock \doi{10.1111/j.2517-6161.1972.tb00899.x}.

\bibitem[Dabernig et~al.(2017)Dabernig, Mayr, Messner, and
  Zeileis]{DabernigEA2017}
M.~Dabernig, G.~Mayr, J.~Messner, and A.~Zeileis.
\newblock Spatial ensemble post-processing with standardized anomalies.
\newblock \emph{Quarterly Journal of the Royal Meteorological Society},
  143:\penalty0 909--916, 2017.

\bibitem[Erlandsen et~al.(2021)Erlandsen, Beldring, Eisner, Hisdal, Huang, and
  Tallaksen]{Erlandsen}
H.~B. Erlandsen, S.~Beldring, S.~Eisner, H.~Hisdal, S.~Huang, and L.~M.
  Tallaksen.
\newblock {Constraining the HBV model for robust water balance assessments in a
  cold climate}.
\newblock \emph{Hydrology Research}, 52\penalty0 (2):\penalty0 356--372, 01
  2021.
\newblock ISSN 0029-1277.
\newblock \doi{10.2166/nh.2021.132}.

\bibitem[Faraggi and Simon(1995)]{Faraggi}
D.~Faraggi and R.~Simon.
\newblock A neural network model for survival data.
\newblock \emph{Statistics in Medicine}, 14\penalty0 (1):\penalty0 73--82,
  1995.
\newblock \doi{10.1002/sim.4780140108}.

\bibitem[Faruk(2018)]{AFT1}
A.~Faruk.
\newblock The comparison of proportional hazards and accelerated failure time
  models in analyzing the first birth interval survival data.
\newblock \emph{Journal of Physics: Conference Series}, 974:\penalty0 012008,
  03 2018.
\newblock \doi{10.1088/1742-6596/974/1/012008}.

\bibitem[Fischer and Connor(2018)]{FischerConnor2018}
R.~Fischer and D.~Connor.
\newblock Issues for cropping and agricultural science in the next 20 years.
\newblock \emph{Field Crops Research}, 222:\penalty0 121--142, 2018.

\bibitem[Gandin and Hardin(1965)]{Gandin}
L.~Gandin and R.~Hardin.
\newblock Objective analysis of meteorological fields.
\newblock \emph{Israel program for scientific translations, Jerusalem}, 242,
  1965.

\bibitem[Glahn and Lowry(1972)]{Glahn+Lowry1972}
H.~R. Glahn and D.~A. Lowry.
\newblock The use of model output statistics {(MOS)} in objective weather
  forecasting.
\newblock \emph{Journal of Applied Meteorology}, 11:\penalty0 1203--1211, 1972.

\bibitem[Gómez-Rubio(2020)]{INLASurv}
V.~Gómez-Rubio.
\newblock Bayesian inference with inla.
\newblock \emph{Chapman \& Hall/CRC Press}, Boca Raton, FL, 2020.

\bibitem[Gneiting et~al.(2005)Gneiting, Raftery, Westveld, and
  Goldman]{GneitingEA2005}
T.~Gneiting, A.~E. Raftery, A.~H. Westveld, and T.~Goldman.
\newblock Calibrated probabilistic forecasting using ensemble model output
  statistics and minimum crps estimation.
\newblock \emph{Monthly Weather Review}, 133:\penalty0 1098--1118, 2005.

\bibitem[Gneiting et~al.(2007)Gneiting, Balabdaoui, and Raftery]{Gneiting&2007}
T.~Gneiting, F.~Balabdaoui, and A.~E. Raftery.
\newblock Probabilistic forecasts, calibration and sharpness.
\newblock \emph{Journal of the Royal Statistical Society: Series B (Statistical
  Methodology)}, 69\penalty0 (2):\penalty0 243--268, 2007.

\bibitem[Hamill(2001)]{Hamill2001}
T.~M. Hamill.
\newblock Interpretation of rank histograms for verifying ensemble forecasts.
\newblock \emph{Monthly Weather Review}, 129\penalty0 (3):\penalty0 550--560,
  2001.

\bibitem[Hammer et~al.(1996)Hammer, Holzworth, and Stone]{Hammer&1996}
G.~Hammer, D.~Holzworth, and R.~Stone.
\newblock The value of skill in seasonal climate forecasting to wheat crop
  management in a region with high climatic variability.
\newblock \emph{Australian Journal of Agricultural Research}, 47\penalty0
  (5):\penalty0 717--737, 1996.

\bibitem[Handeland et~al.(2021)Handeland, Tunheim, Madslien, Vikøren,
  Viljugrein, Mossing, Børve, Strand, and Hamnes]{HANDELAND2021214}
K.~Handeland, K.~Tunheim, K.~Madslien, T.~Vikøren, H.~Viljugrein, A.~Mossing,
  I.~Børve, O.~Strand, and I.~S. Hamnes.
\newblock High winter loads of {Oestrid} larvae and {Elaphostrongylus
  Rangiferi} are associated with emaciation in wild reindeer calves.
\newblock \emph{International Journal for Parasitology: Parasites and
  Wildlife}, 15:\penalty0 214--224, 2021.
\newblock ISSN 2213-2244.
\newblock \doi{10.1016/j.ijppaw.2021.05.008}.

\bibitem[Heinrich(2021)]{Heinrich2021}
C.~Heinrich.
\newblock On the number of bins in a rank histogram.
\newblock \emph{Quarterly Journal of the Royal Meteorological Society},
  147\penalty0 (734):\penalty0 544--556, 2021.

\bibitem[Heinrich et~al.(2021)Heinrich, Hellton, Lenkoski, and
  Thorarinsdottir]{Heinrich&2021}
C.~Heinrich, K.~H. Hellton, A.~Lenkoski, and T.~L. Thorarinsdottir.
\newblock Multivariate postprocessing methods for high-dimensional seasonal
  weather forecasts.
\newblock \emph{Journal of the American Statistical Association}, 116\penalty0
  (535):\penalty0 1048--1059, 2021.

\bibitem[Hemri et~al.(2014)Hemri, Scheuerer, Pappenberger, Bogner, and
  Haiden]{HemriEA2014}
S.~Hemri, M.~Scheuerer, F.~Pappenberger, K.~Bogner, and T.~Haiden.
\newblock Trends in the predictive performance of raw ensemble weather
  forecasts.
\newblock \emph{Geophysical Research Letters}, 41:\penalty0 9197--9205, 2014.

\bibitem[Hemri et~al.(2020)Hemri, Bhend, Liniger, Manzanas, Siegert,
  Stephenson, Guti{\'e}rrez, Brookshaw, and Doblas-Reyes]{hemri_et_2020}
S.~Hemri, J.~Bhend, M.~A. Liniger, R.~Manzanas, S.~Siegert, D.~B. Stephenson,
  J.~M. Guti{\'e}rrez, A.~Brookshaw, and F.~J. Doblas-Reyes.
\newblock How to create an operational multi-model of seasonal forecasts?
\newblock \emph{Climate Dynamics}, 55\penalty0 (5):\penalty0 1141--1157, 2020.

\bibitem[Hersbach(2000)]{Hersbach2000}
H.~Hersbach.
\newblock Decomposition of the continuous ranked probability score for ensemble
  prediction systems.
\newblock \emph{Weather and Forecasting}, 15\penalty0 (5):\penalty0 559--570,
  2000.

\bibitem[J.(1992)]{AFT2}
W.~L. J.
\newblock The accelerated failure time model: a useful alternative to the {C}ox
  regression model in survival analysis.
\newblock \emph{Statistics in medicine}, 11 (14-15):\penalty0 1871–1879,
  1992.
\newblock \doi{10.1002/sim.4780111409}.

\bibitem[Kalbfleisch and Prentice(2011)]{kalbfleisch_ross_2011}
J.~D. Kalbfleisch and R.~L. Prentice.
\newblock \emph{The statistical analysis of failure time data}, volume 360.
\newblock John Wiley \& Sons, 2011.

\bibitem[Kalnay(2003)]{Kalnay}
E.~Kalnay.
\newblock \emph{Atmospheric modeling, data assimilation and predictability}.
\newblock Cambridge university press, 2003.

\bibitem[Kaplan and Meier(1958)]{kaplan_meier_1958}
E.~L. Kaplan and P.~Meier.
\newblock Nonparametric estimation from incomplete observations.
\newblock \emph{Journal of the {A}merican statistical association}, 53\penalty0
  (282):\penalty0 457--481, 1958.

\bibitem[Katzman et~al.(2018)Katzman, Shaham, Cloninger, Bates, Jiang, and
  Kluger]{deepSurvival}
J.~L. Katzman, U.~Shaham, A.~Cloninger, J.~Bates, T.~Jiang, and Y.~Kluger.
\newblock {DeepSurv}: personalized treatment recommender system using a {C}ox
  proportional hazards deep neural network.
\newblock \emph{{BMC} medical research methodology}, 18\penalty0 (1):\penalty0
  1--12, 2018.

\bibitem[Keller et~al.(2021)Keller, Rajczak, Bhend, Spirig, Hemri, Liniger, and
  Wernli]{KellerEA2021}
R.~Keller, J.~Rajczak, J.~Bhend, C.~Spirig, S.~Hemri, M.~A. Liniger, and
  H.~Wernli.
\newblock Seamless multimodel postprocessing for air temperature forecasts in
  complex topography.
\newblock \emph{Weather and Forecasting}, 36\penalty0 (3):\penalty0 1031 --
  1042, 2021.
\newblock \doi{10.1175/WAF-D-20-0141.1}.

\bibitem[Klemm and McPherson(2017)]{KlemmMcPherson2017}
T.~Klemm and R.~A. McPherson.
\newblock The development of seasonal climate forecasting for agricultural
  producers.
\newblock \emph{Agricultural and forest meteorology}, 232:\penalty0 384--399,
  2017.

\bibitem[Kukal and Irmak(2018)]{Kukal&2018}
M.~S. Kukal and S.~Irmak.
\newblock {US} agro-climate in 20th century: growing degree days, first and
  last frost, growing season length, and impacts on crop yields.
\newblock \emph{Scientific reports}, 8\penalty0 (1):\penalty0 1--14, 2018.

\bibitem[Kumar and Klefsjö(1994)]{KUMAR1994177}
D.~Kumar and B.~Klefsjö.
\newblock Proportional hazards model: a review.
\newblock \emph{Reliability Engineering \& System Safety}, 44\penalty0
  (2):\penalty0 177--188, 1994.
\newblock ISSN 0951-8320.
\newblock \doi{10.1016/0951-8320(94)90010-8}.

\bibitem[Kunkel et~al.(2004)Kunkel, Easterling, Hubbard, and
  Redmond]{Kunkel&2004}
K.~E. Kunkel, D.~R. Easterling, K.~Hubbard, and K.~Redmond.
\newblock Temporal variations in frost-free season in the {United States}:
  1895--2000.
\newblock \emph{Geophysical Research Letters}, 31\penalty0 (3), 2004.

\bibitem[Lane et~al.(1986)Lane, Looney, and Wansley]{BankCox}
W.~R. Lane, S.~W. Looney, and J.~W. Wansley.
\newblock An application of the {C}ox proportional hazards model to bank
  failure.
\newblock \emph{Journal of Banking \& Finance}, 10\penalty0 (4):\penalty0
  511--531, 1986.
\newblock ISSN 0378-4266.
\newblock \doi{10.1016/S0378-4266(86)80003-6}.

\bibitem[Lawrence(2020)]{Lawrence}
D.~Lawrence.
\newblock Uncertainty introduced by flood frequency analysis in projections for
  changes in flood magnitudes under a future climate in {N}orway.
\newblock \emph{Journal of Hydrology: Regional Studies}, 28:\penalty0 100675,
  04 2020.
\newblock \doi{10.1016/j.ejrh.2020.100675}.

\bibitem[Lehmann et~al.(2020)Lehmann, Kretschmer, Schauberger, and
  Wechsung]{Lehmann&2020}
J.~Lehmann, M.~Kretschmer, B.~Schauberger, and F.~Wechsung.
\newblock Potential for early forecast of {M}oroccan wheat yields based on
  climatic drivers.
\newblock \emph{Geophysical Research Letters}, 47\penalty0 (12):\penalty0
  e2020GL087516, 2020.

\bibitem[Lerch et~al.(2020)Lerch, Baran, M\"oller, Gro{\ss}, Schefzik, Hemri,
  and Graeter]{LerchEA2020}
S.~Lerch, S.~Baran, A.~M\"oller, J.~Gro{\ss}, R.~Schefzik, S.~Hemri, and
  M.~Graeter.
\newblock Simulation-based comparison of multivariate ensemble post-processing
  methods.
\newblock \emph{Nonlinear Processes in Geophysics}, 27\penalty0 (2):\penalty0
  349--371, 2020.
\newblock \doi{10.5194/npg-27-349-2020}.

\bibitem[Liu et~al.(2008)Liu, Henderson, and Xu]{Liu&2008}
B.~Liu, M.~Henderson, and M.~Xu.
\newblock Spatiotemporal change in {C}hina's frost days and frost-free season,
  1955--2000.
\newblock \emph{Journal of Geophysical Research: Atmospheres}, 113\penalty0
  (D12), 2008.

\bibitem[Lussana et~al.(2018)Lussana, Tveito, and Uboldi]{Lussana2018}
C.~Lussana, O.~Tveito, and F.~Uboldi.
\newblock Three-dimensional spatial interpolation of two-meter temperature over
  {N}orway.
\newblock \emph{Quarterly Journal of the Royal Meteorological Society}, 144, 01
  2018.
\newblock \doi{10.1002/qj.3208}.

\bibitem[Lussana et~al.(2019)Lussana, Tveito, Dobler, and Tunheim]{Lussana2019}
C.~Lussana, O.~E. Tveito, A.~Dobler, and K.~Tunheim.
\newblock se{N}orge\_2018, daily precipitation, and temperature datasets over
  {N}orway.
\newblock \emph{Earth System Science Data}, 11\penalty0 (4):\penalty0
  1531--1551, 2019.
\newblock \doi{10.5194/essd-11-1531-2019}.

\bibitem[Martino et~al.(2011)Martino, Akerkar, and Rue]{SaraSurvival}
S.~Martino, R.~Akerkar, and H.~Rue.
\newblock Approximate {B}ayesian inference for survival models.
\newblock \emph{Scandinavian Journal of Statistics}, 38\penalty0 (3):\penalty0
  514--528, 2011.
\newblock ISSN 03036898, 14679469.

\bibitem[Messner et~al.(2017)Messner, Mayr, and Zeileis]{MessnerEA2017}
J.~W. Messner, G.~J. Mayr, and A.~Zeileis.
\newblock Nonhomogeneous boosting for predictor selection in ensemble
  postprocessing.
\newblock \emph{Monthly Weather Review}, 145\penalty0 (1):\penalty0 137 -- 147,
  2017.
\newblock \doi{10.1175/MWR-D-16-0088.1}.

\bibitem[Mogensen et~al.(2012)Mogensen, Ishwaran, and Gerds]{Mogensen&2012}
U.~B. Mogensen, H.~Ishwaran, and T.~A. Gerds.
\newblock Evaluating random forests for survival analysis using prediction
  error curves.
\newblock \emph{Journal of statistical software}, 50\penalty0 (11):\penalty0 1,
  2012.

\bibitem[Murphy and Winkler(1987)]{MurphyWinkler1987}
A.~H. Murphy and R.~L. Winkler.
\newblock A general framework for forecast verification.
\newblock \emph{Monthly weather review}, 115\penalty0 (7):\penalty0 1330--1338,
  1987.

\bibitem[Rasp and Lerch(2018)]{Rasp+Lerch2018}
S.~Rasp and S.~Lerch.
\newblock Neural networks for postprocessing ensemble weather forecasts.
\newblock \emph{Monthly Weather Review}, 146\penalty0 (11):\penalty0
  3885--3900, 2018.

\bibitem[Robertson and Vitart(2018)]{Robertson&2018}
A.~Robertson and F.~Vitart.
\newblock \emph{Sub-seasonal to seasonal prediction: The gap between weather
  and climate forecasting}.
\newblock Elsevier, 2018.

\bibitem[Roebber and Crockett(2019)]{Roebber+Crockett2019}
P.~J. Roebber and J.~Crockett.
\newblock Using a coevolutionary postprocessor to improve skill for both
  forecasts of surface temperature and nowcasts of convection occurrence.
\newblock \emph{Monthly Weather Review}, 147\penalty0 (11):\penalty0 4241 --
  4259, 2019.
\newblock \doi{10.1175/MWR-D-19-0063.1}.

\bibitem[Schwartz et~al.(2006)Schwartz, Ahas, and Aasa]{Schwartz&2006}
M.~D. Schwartz, R.~Ahas, and A.~Aasa.
\newblock Onset of spring starting earlier across the {Northern Hemisphere}.
\newblock \emph{Global Change Biology}, 12\penalty0 (2):\penalty0 343--351,
  2006.

\bibitem[Taillardat et~al.(2016)Taillardat, Mestre, Zamo, and
  Naveau]{TaillardatEA2016}
M.~Taillardat, O.~Mestre, M.~Zamo, and P.~Naveau.
\newblock Calibrated ensemble forecasts using quantile regression forests and
  ensemble model output statistics.
\newblock \emph{Monthly Weather Review}, 144:\penalty0 2375--2393, 2016.

\bibitem[{Van Schaeybroeck} and Vannitsem(2015)]{VanSchaeybroeck+Vannitsem2015}
B.~{Van Schaeybroeck} and S.~Vannitsem.
\newblock Ensemble post-processing using member-by-member approaches:
  Theoretical aspects.
\newblock \emph{Quarterly Journal of the Royal Meteorological Society},
  141:\penalty0 807--818, 2015.

\bibitem[Van~Schaeybroeck and Vannitsem(2018)]{VanSchaeybroeck&2018}
B.~Van~Schaeybroeck and S.~Vannitsem.
\newblock Postprocessing of long-range forecasts.
\newblock In \emph{Statistical Postprocessing of Ensemble Forecasts}, pages
  267--290. Elsevier, 2018.

\bibitem[van Straaten et~al.(2020)van Straaten, Whan, Coumou, van~den Hurk, and
  Schmeits]{vanStraaten&2020}
C.~van Straaten, K.~Whan, D.~Coumou, B.~van~den Hurk, and M.~Schmeits.
\newblock The influence of aggregation and statistical post-processing on the
  subseasonal predictability of {E}uropean temperatures.
\newblock \emph{Quarterly Journal of the Royal Meteorological Society},
  146\penalty0 (731):\penalty0 2654--2670, 2020.

\bibitem[Weigel et~al.(2009)Weigel, Liniger, and Appenzeller]{Weigel&2009}
A.~P. Weigel, M.~A. Liniger, and C.~Appenzeller.
\newblock Seasonal ensemble forecasts: Are recalibrated single models better
  than multimodels?
\newblock \emph{Monthly Weather Review}, 137\penalty0 (4):\penalty0 1460--1479,
  2009.

\bibitem[Weltzin et~al.(2020)Weltzin, Betancourt, Cook, Crimmins, Enquist,
  Gerst, Gross, Henebry, Hufft, Kenney, et~al.]{Weltzin&2020}
J.~F. Weltzin, J.~L. Betancourt, B.~I. Cook, T.~M. Crimmins, C.~A. Enquist,
  M.~D. Gerst, J.~E. Gross, G.~M. Henebry, R.~A. Hufft, M.~A. Kenney, et~al.
\newblock Seasonality of biological and physical systems as indicators of
  climatic variation and change.
\newblock \emph{Climatic Change}, 163\penalty0 (4):\penalty0 1755--1771, 2020.

\bibitem[Wilks(2011)]{Wilks2011}
D.~S. Wilks.
\newblock \emph{Statistical methods in the atmospheric sciences}, volume 100.
\newblock Academic press, 2011.

\bibitem[Woldemeskel et~al.(2018)Woldemeskel, McInerney, Lerat, Thyer,
  Kavetski, Shin, Tuteja, and Kuczera]{Woldemeskel&2018}
F.~Woldemeskel, D.~McInerney, J.~Lerat, M.~Thyer, D.~Kavetski, D.~Shin,
  N.~Tuteja, and G.~Kuczera.
\newblock Evaluating post-processing approaches for monthly and seasonal
  streamflow forecasts.
\newblock \emph{Hydrology and Earth System Sciences}, 22\penalty0
  (12):\penalty0 6257--6278, 2018.

\bibitem[Zhang et~al.(2014)Zhang, Xu, Li, Cai, and An]{Zhang&2014}
D.~Zhang, W.~Xu, J.~Li, Z.~Cai, and D.~An.
\newblock Frost-free season lengthening and its potential cause in the {Tibetan
  Plateau} from 1960 to 2010.
\newblock \emph{Theoretical and applied climatology}, 115\penalty0
  (3):\penalty0 441--450, 2014.

\bibitem[Zhou et~al.(2020)Zhou, Hanson, and Zhang]{spatialSurvival2}
H.~Zhou, T.~Hanson, and J.~Zhang.
\newblock {spBayesSurv: Fitting Bayesian Spatial Survival Models Using R}.
\newblock \emph{Journal of Statistical Software}, 92\penalty0 (9):\penalty0
  1–33, 2020.
\newblock \doi{10.18637/jss.v092.i09}.

\end{thebibliography}
	

\end{document}